\documentclass[10pt,letterpaper]{article}
\pdfoutput=1

\usepackage[margin=1in]{geometry}
\usepackage[utf8]{inputenc}
\usepackage[T1]{fontenc}
\usepackage[nopatch=footnote]{microtype}
\usepackage{amsfonts}
\usepackage{amssymb}
\usepackage{amsmath}
\usepackage{amsthm}
\usepackage{booktabs}
\usepackage[inline]{enumitem}
\usepackage[hidelinks,bookmarks,bookmarksopen,bookmarksnumbered]{hyperref}
\usepackage[capitalize]{cleveref}
\usepackage{mathtools}
\usepackage{tikz}
\usepackage[font=small,labelfont=bf]{caption}
\usepackage{titlesec}
\usepackage{thmtools}
\usepackage{wrapfig}
\usepackage{tablefootnote}
\usepackage{thm-restate}
\usepackage{soul}
\usepackage{framed}
\usepackage{tikz}
\usepackage{subcaption}
\usepackage[draft]{fixme}
\usepackage[norefs,nocites,nomsgs]{refcheck}  %
\usepackage{setspace}
\usepackage{array}
\usepackage{multirow}

\usetikzlibrary{trees}
\usetikzlibrary{arrows}
\usetikzlibrary{arrows.meta,arrows}
\usetikzlibrary{decorations.pathreplacing}
\usetikzlibrary{positioning}
\usetikzlibrary{calc}

\input{settings}

\newcommand{\bigO}{\mathcal{O}}

\DeclareMathOperator*{\argmin}{arg\,min}

\newcommand{\dd}{\mathinner{.\,.}}
\newcommand{\probname}[1]{\text{\sc #1}}

\newcommand{\Z}{\mathbb{Z}}
\newcommand{\Zz}{\Z_{\ge 0}}

\newcommand{\Zp}{\Z_{>0}}

\newcommand{\Pat}{P}
\newcommand{\Text}{T}
\newcommand{\Textinf}{\Text^{\infty}}
\newcommand{\Textlen}{n}
\newcommand{\AlphabetSize}{\sigma}
\newcommand{\IntegerAlphabet}{[0 \dd \AlphabetSize)}
\newcommand{\BinaryAlphabet}{\{{\tt 0}, {\tt 1}\}}
\newcommand{\emptystring}{\varepsilon}

\newcommand{\SA}[1]{\mathrm{SA}_{#1}}
\newcommand{\ISA}[1]{\mathrm{SA}^{-1}_{#1}}
\newcommand{\LCP}[1]{\mathrm{LCP}_{#1}}
\newcommand{\PLCP}[1]{\mathrm{PLCP}_{#1}}
\newcommand{\BWT}[1]{\mathrm{BWT}_{#1}}
\newcommand{\LF}[1]{\mathrm{LF}_{#1}}
\newcommand{\ILF}[1]{\mathrm{LF}^{-1}_{#1}}
\newcommand{\PhiArray}[1]{\Phi_{#1}}
\newcommand{\InvPhiArray}[1]{\Phi^{-1}_{#1}}
\newcommand{\LPF}[1]{\mathrm{LPF}_{#1}}

\newcommand{\LCE}[3]{\mathrm{LCE}_{#1}(#2,#3)}
\newcommand{\RL}[1]{\mathrm{RL}(#1)}
\newcommand{\lcp}[2]{\mathrm{lcp}(#1,#2)}
\newcommand{\per}[1]{\mathrm{per}(#1)}
\newcommand{\revstr}[1]{\overline{#1}}
\newcommand{\bin}[2]{{\rm bin}_{#1}(#2)}
\newcommand{\ebin}[2]{{\rm ebin}_{#1}(#2)}
\newcommand{\SubstrCount}[2]{\mathrm{d}_{#1}(#2)}

\newcommand{\SubstringComplexitySym}{\delta}
\newcommand{\LZSizeSym}{z}

\newcommand{\RLBWTSizeSym}{r}

\newcommand{\SubstringComplexity}[1]{\SubstringComplexitySym(#1)}
\newcommand{\LZSize}[1]{\LZSizeSym(#1)}

\newcommand{\RLBWTSize}[1]{\RLBWTSizeSym(#1)}

\newcommand{\OccTwo}[2]{\mathrm{Occ}(#1, #2)}
\newcommand{\RangeBegTwo}[2]{\mathrm{RangeBeg}(#1, #2)}
\newcommand{\RangeEndTwo}[2]{\mathrm{RangeEnd}(#1, #2)}

\newcommand{\Predecessor}[2]{\mathsf{pred}_{#1}(#2)}
\newcommand{\PredecessorColor}[2]{\mathsf{pred\mbox{-}color}_{#1}(#2)}

\newcommand{\TwoSidedRangeCount}[3]{\mathsf{range\mbox{-}count}_{#1}(#2, #3)}
\newcommand{\RangeSelect}[3]{\mathsf{range\mbox{-}select}_{#1}(#2, #3)}
\newcommand{\RMQ}[3]{\mathsf{rmq}_{#1}(#2, #3)}

\newcommand{\Rhs}[2]{\mathrm{rhs}_{#1}(#2)}
\newcommand{\Exp}[2]{\mathrm{exp}_{#1}(#2)}
\newcommand{\Lang}[1]{L(#1)}

\newcommand{\zero}{{\tt 0}}
\newcommand{\one}{{\tt 1}}

\begin{document}

\title{Tight Lower Bounds for Central String Queries\\ in Compressed Space}

\author{
  \large Dominik Kempa\thanks{Partially funded by the
  NSF CAREER Award 2337891 and the Simons Foundation
  Junior Faculty Fellowship.}\\[-0.3ex]
  \normalsize Department of Computer Science,\\[-0.3ex]
  \normalsize Stony Brook University,\\[-0.3ex]
  \normalsize Stony Brook, NY, USA\\[-0.3ex]
  \normalsize \texttt{kempa@cs.stonybrook.edu}
  \and
  \large Tomasz Kociumaka\\[-0.3ex]
  \normalsize Max Planck Institute for Informatics,\\[-0.3ex]
  \normalsize Saarland Informatics Campus,\\[-0.3ex]
  \normalsize Saarbrücken, Germany\\[-0.3ex]
  \normalsize \texttt{tomasz.kociumaka@mpi-inf.mpg.de}
}

\date{\vspace{-0.5cm}}
\maketitle

\begin{abstract}
  In this work, we study limits of compressed data structures, i.e.,
  data structures that support various queries on the input text $\Text \in \Sigma^{\Textlen}$
  in space proportional to the size of $\Text$ in compressed form.
  On the \emph{upper bound} side, currently nearly all fundamental
  queries can be efficiently supported in $\bigO(\SubstringComplexity{\Text}
  \log^{\bigO(1)} \Textlen)$ space (where
  $\SubstringComplexity{\Text}$ is the \emph{substring complexity} --
  a strong measure of compressibility that lower-bounds the
  optimal achievable space to represent the text [Kociumaka, Navarro,
    Prezza, IEEE Trans. Inf. Theory 2023]); this includes
  queries like random access, longest common extension,
  suffix array, longest common prefix array, and many others. In
  contrast, \emph{lower bounds} for compressed data structures remained
  elusive, and currently they are known only for the basic random
  access problem. This work addresses this important gap and develops
  \ul{tight lower bounds} for nearly all other fundamental queries:
  \begin{itemize}
  \item First, we prove that several core string-processing queries,
    including suffix array (SA) queries, inverse suffix array
    ($\text{SA}^{-1}$) queries, longest common prefix (LCP) array
    queries, and longest common extension (LCE) queries, all require
    $\Omega\big(\tfrac{\log \Textlen}{\log \log \Textlen}\big)$ time
    within $\bigO(\SubstringComplexity{\Text} \log^{\bigO(1)}
    \Textlen)$ space. All our lower bounds are asymptotically tight,
    i.e., we match the performance of known upper bounds.

  \item We then extend our framework to include other common
    queries that are currently supported in $\bigO(\log \log
    \Textlen)$ time and $\bigO(\SubstringComplexity{\Text}
    \log^{\bigO(1)} \Textlen)$ space. This includes Burrows--Wheeler
    Transform (BWT) queries, permuted longest common prefix array
    (PLCP) queries, Last-to-First (LF) queries, inverse LF
    ($\text{LF}^{-1}$) queries, lexicographical predecessor ($\Phi$)
    queries, and inverse lexicographical predecessor ($\Phi^{-1}$)
    queries, and again we prove that, within this space, they all
    \emph{require} $\Omega(\log \log \Textlen)$ time, obtaining again
    a set of asymptotically tight lower bounds.
  \end{itemize}

  All our lower bounds hold even for texts over a binary alphabet.
  Our results establish a clean dichotomy: the optimal time complexity to
  support central string queries in compressed space is either
  $\Theta(\log \Textlen / \log \log \Textlen)$ or $\Theta(\log \log
  \Textlen)$. This completes the theoretical foundation of compressed
  indexing, closing a crucial gap between upper and lower bounds,
  and providing a clear target for future data structures: seeking
  either the optimal time in the smallest space, or fastest time in
  the optimal space (both of which are now known) for central string queries.
\end{abstract}

\thispagestyle{empty}
\clearpage
\setcounter{page}{1}

\thispagestyle{empty}

\begin{table}
  \vspace{5cm}
  \centering
  \begin{tabular}{lccc}
    \toprule
    \textbf{Query} & \textbf{Time Complexity} & \textbf{Lower Bound} & \textbf{Upper Bound} \\
    \midrule
    $\mathrm{T}$ &
      \multirow{5}{*}{$\Theta(\tfrac{\log \Textlen}{\log \log \Textlen})$} &
      \cite{VerbinY13} &
      \cite{blocktree}\\
    $\mathrm{LCP}$ &&
      \cref{sec:lcp-lower-bound} &
      \cite{Gagie2020}\footnoteref{rindex-note}\\
    $\mathrm{SA}$ &&
      \cref{sec:sa-lower-bound} &
      \cite{Gagie2020}\tablefootnote{\label{rindex-note}These results
        are obtained through a cosmetic modification to the
        structures by Gagie, Navarro, and Prezza~\cite[Theorems~5.4, 5.7, and 5.8]{Gagie2020} (originally
        using $\bigO(\RLBWTSize{\Text} \log \Textlen)$ space and supporting
        the queries in $\bigO(\log \Textlen)$ time), resulting in
        $\bigO(\RLBWTSize{\Text} \log^{1+\epsilon} \Textlen)$ space usage
        and $\bigO(\tfrac{\log \Textlen}{\log \log \Textlen})$ query time; see \cref{th:rindex,rm:rindex}.
        Combining with the upper bound
        $\RLBWTSize{\Text} = \bigO(\SubstringComplexity{\Text} \log^2
        \Textlen)$~\cite{resolution} (see \cref{th:rlbwt-size,rm:rlbwt-size}), this results in
        suffix array, inverse suffix array, and LCP array queries in
        $\bigO(\SubstringComplexity{\Text} \log^{4+\epsilon} \Textlen)
        = \bigO(\SubstringComplexity{\Text} \log^{\bigO(1)} \Textlen)$ space and
        $\bigO(\tfrac{\log \Textlen}{\log \log\Textlen})$ time.}\\
    $\mathrm{SA}^{-1}$ &&
      \cref{sec:isa-lower-bound} &
      \cite{Gagie2020}\footnoteref{rindex-note}\\
    $\mathrm{LCE}$ &&
      \cref{sec:lce-lower-bound} &
      \cite{Gagie2020}\tablefootnote{A data structure using
      $\bigO(\LZSize{\Text} \log \Textlen)$ (resp.\ $\bigO(\SubstringComplexity{\Text} \log \Textlen)$) space and answering
      LCE queries in $\bigO(\log \Textlen)$ time was presented
      in~\cite{tomohiro-lce} (resp.~\cite{collapsing}). A similar result, achieved via more general LCP RMQ queries
      (see \cref{sec:lcp-rmq-problem-def}), was presented in~\cite{Gagie2020}, resulting
      in LCE queries using $\bigO(\RLBWTSize{\Text} \log \Textlen)$
      space and $\bigO(\log \Textlen)$ time. In \cref{sec:lcp-rmq}, we explain in detail the modifications to the
      technique of Gagie, Navarro, and Prezza~\cite{Gagie2020} needed to
      achieve LCP RMQ queries in $\bigO(\RLBWTSize{\Text} \log^{2+\epsilon} \Textlen)$ space
      and $\bigO(\tfrac{\log \Textlen}{\log \log \Textlen})$ time.
      By applying the
      upper bound $\RLBWTSize{\Text} = \bigO(\SubstringComplexity{\Text} \log^2 \Textlen)$~\cite{resolution}
      (see \cref{th:rlbwt-size,rm:rlbwt-size}), this results in LCE queries
      using $\bigO(\SubstringComplexity{\Text} \log^{4+\epsilon} \Textlen)
      = \bigO(\SubstringComplexity{\Text} \log^{\bigO(1)} \Textlen)$ space and
      $\bigO(\tfrac{\log \Textlen}{\log \log \Textlen})$ time;
      see \cref{sec:lce-upper-bound}.}\\
    \midrule
    $\mathrm{BWT}$ &
      \multirow{6}{*}{$\Theta(\log \log \Textlen)$} &
      \cref{sec:bwt-lower-bound} &
      \cite{Gagie2020} \\
    $\mathrm{PLCP}$ &&
      \cref{sec:plcp-lower-bound} &
      \cite{Gagie2020} \\
    $\mathrm{LF}$ &&
      \cref{sec:lf-lower-bound} &
      \cite{Gagie2020} \\
    $\mathrm{LF}^{-1}$ &&
      \cref{sec:ilf-lower-bound} &
      \cref{sec:ilf-upper-bound} \\
    $\Phi$ &&
      \cref{sec:phi-lower-bound} &
      \cite{Gagie2020} \\
    $\Phi^{-1}$ &&
      \cref{sec:iphi-lower-bound} &
      \cite{Gagie2020} \\
    \bottomrule
  \end{tabular}
  \caption{Query time lower and upper bounds for fundamental string queries in
    $\bigO(\SubstringComplexity{\Text} \log^{\bigO(1)} \Textlen)$
    space for a text $\Text \in \Sigma^{\Textlen}$. All lower bounds
    hold even when $\Sigma = \BinaryAlphabet$.}\label{tab:query-bounds}
\end{table}

\clearpage
\pagenumbering{arabic}
\setcounter{page}{1}

\section{Introduction}\label{sec:intro}

Compressed indexing~\cite{NavarroIndexes,NavarroMeasures} is a recent
paradigm in the design of data structures that combines elements of
data compression, information theory, and combinatorial pattern
matching to build data structures whose space usage is proportional to
the size of the input text (string, sequence) in compressed form. For
example, given a text $\Text \in \Sigma^{\Textlen}$ and a compression
algorithm $C : \Sigma^{*} \rightarrow \hat{\Sigma}^{*}$, a compressed
index supporting \emph{random access} to $\Text$ will typically use
$\bigO(|C(\Text)| \cdot \log^{\bigO(1)} \Textlen)$ space and, given
any position $i \in [1 \dd \Textlen]$, return the symbol $\Text[i]$ in
$\bigO(\log^{\bigO(1)} \Textlen)$ time.

In the past two decades, interest in compressed computation has spiked
dramatically, as datasets in many applications have grown to such
unwieldy sizes that they essentially cannot be handled otherwise. This
includes source code storage systems (like GitHub), versioned
databases (like Wikipedia), and computational biology, where some
datasets already exceed petabytes~\cite{stephens2015big}, with
estimates predicting growth to exabytes in the near
future~\cite{estimate}. These datasets are extremely
repetitive/redundant, with some starting at 99.9\%
redundancy~\cite{Przeworski2000} and increasing with dataset
size~\cite{Gagie2020}. Crucially, \emph{misrepresenting} any portion
of this data is often infeasible -- the presence of a disease could
hinge on a single DNA mutation, and a bug might result from a single
typo.

On the \emph{upper bound} front, compressed indexing supports a rich
repertoire of queries:
\begin{itemize}

\item Compressed indexes supporting random access were among the
  first~\cite{Rytter03,charikar}. They achieve
  $\bigO(\SubstringComplexity{\Text} \log^{\bigO(1)} \Textlen)$
  space\footnote{The value $\SubstringComplexity{\Text}$ is a
    fundamental measure of repetitiveness~\cite{delta}, closely related
    to other core measures (such as the sizes of LZ77 factorization
    and run-length BWT compression). We discuss them all in more
    detail shortly.} and query time $\bigO(\log
  \Textlen)$. Subsequent works reduced space by a polylogarithmic
  factor~\cite{balancing,BLRSRW15,KempaS22,delta} or query time to
  $\bigO(\tfrac{\log \Textlen}{\log \log
    \Textlen})$~\cite{BelazzouguiCPT15,blocktree,balancing}. These
  indexes are typically used to simply replace the text.

\item Compressed indexes also support more complex longest common
  extension (LCE) queries (see \cref{sec:prelim}) using
  $\bigO(\SubstringComplexity{\Text} \log^{\bigO(1)} \Textlen)$ space
  and $\bigO(\log \Textlen)$
  time~\cite{tomohiro-lce,Gagie2020,collapsing}. LCE queries are
  central in many applications, including approximate pattern
  matching~\cite{LandauV88,KociumakaNW24}, BWT
  compression~\cite{resolution}, and internal pattern
  matching~\cite{KociumakaRRW24}.

\item Compressed indexes supporting suffix array (SA) queries (which
  return $\SA{\Text}[i]$, given any $i \in [1 \dd \Textlen]$), inverse
  SA queries, and LCP array queries
  (\cref{def:sa,def:isa,def:lcp-array}) are among the most
  powerful~\cite{Gagie2020,collapsing}. They use
  $\bigO(\SubstringComplexity{\Text} \log^{\bigO(1)} \Textlen)$ space,
  with query times ranging from $\bigO(\log
  \Textlen)$~\cite{Gagie2020} to $\bigO(\log^{\bigO(1)}
  \Textlen)$~\cite{collapsing}. SA/LCP queries solve myriad problems,
  including longest common
  substring~\cite{gusfield,Charalampopoulos21}, LZ77
  factorization~\cite{CrochemoreIS08,OhlebuschG11,KempaP13,GotoB13,sublinearlz,PolicritiP18},
  maximal exact matches (MEMs)~\cite{gusfield}, maximal unique matches
  (MUMs)~\cite{kurtz2004versatile}, shortest unique
  substring~\cite{IlieS11}, minimal absent word~\cite{BartonHMP14},
  sequence alignment~\cite{bowtie2,bwa}, and many
  others~\cite{gusfield,bwtbook}.

\item Compressed indexes also support faster algorithms for
  certain query types. Lexicographical predecessor $\Phi$
  (\cref{def:phi-array}), LF mapping (\cref{def:lf}), BWT
  (\cref{def:bwt}), and PLCP queries (\cref{def:plcp-array}) can all
  be supported in $\bigO(\log \log \Textlen)$ time and
  $\bigO(\SubstringComplexity{\Text} \log^{\bigO(1)} \Textlen)$
  space~\cite{Gagie2020,NishimotoT21}.\footnote{\cite{NishimotoT21}
  can support LF and inverse lexicographical predecessor queries (see
  \cref{sec:lf-problem-def,sec:iphi-problem-def}) faster than
  $\bigO(\log \log \Textlen)$ if the query algorithm is given
  additional information (in addition to the index $i$) as input.
  This variant is useful since such a scenario occurs in some
  applications (e.g., pattern matching). The most general LF and
  inverse lexicographical predecessor queries (which are given only
  the index $i$) are supported in $\bigO(\log \log \Textlen)$ time
  by~\cite{Gagie2020}.} Their applications include pattern
  listing~\cite{Gagie2020}, suffix tree queries~\cite{cst,Gagie2020},
  and pattern counting~\cite{FerraginaM00,GrossiV00,Gagie2020}.

\end{itemize}

We refer to~\cite{NavarroIndexes,NavarroMeasures} for a further
overview of \emph{upper bound} results in compressed indexing. At
this point, it is natural to ask: what can we say about compressed
indexing on the \emph{lower bound} front? The two fundamental
questions are: (1) What is the \emph{smallest achievable index space}
that still supports efficient queries (e.g., in $\bigO(\log^{\bigO(1)}
\Textlen)$ time)? Equally important is the complementary question:
(2) What is the \emph{best achievable query time} within near-optimal
space (say, off by at most a polylogarithmic factor from the best
possible)?

The first question has been largely resolved in recent years. Each of
the indexes mentioned above achieves a specific space bound. For
example, the indexes in~\cite{Gagie2020,NishimotoT21} use
$\bigO(\RLBWTSize{\Text})$ or $\bigO(\RLBWTSize{\Text} \log \Textlen)$
space (where $\RLBWTSize{\Text}$ is the size of the run-length
Burrows--Wheeler transform~\cite{bwt} of~$\Text$), while others use
$\bigO(\SubstringComplexity{\Text} \log \Textlen)$
space~\cite{delta,collapsing} (where $\SubstringComplexity{\Text}$ is
the substring complexity of $\Text$~\cite{delta}),
$\bigO(\LZSize{\Text} \log \Textlen)$
space~\cite{Rytter03,charikar,balancing,blocktree,tomohiro-lce} (where
$\LZSize{\Text}$ is the LZ77 size~\cite{LZ77}), or $\bigO(z_e(\Text))$
space~\cite{KempaS22} (where $z_e(\Text)$ is the LZ-End parsing
size~\cite{kreft2010navarro}). However, in a series of
works~\cite{Rytter03,charikar,attractors,GNPlatin18,resolution,KempaS22,delta},
\emph{all} of these (and several other) repetitiveness measures and
compression sizes have been shown to be within $\log^{\bigO(1)}
\Textlen$ factors of each other in the worst case. Moreover,
$\bigO(\SubstringComplexity{\Text} \log \tfrac{\Textlen \log
  \AlphabetSize}{\SubstringComplexity{\Text} \log \Textlen}) =
\bigO(\SubstringComplexity{\Text} \log \Textlen)$\footnote{Henceforth,
  $\AlphabetSize$ denotes the alphabet size.} has been proven to be
the optimal space~\cite{delta} and subsequently shown to be attainable
for a range of queries (even for the most powerful queries like SA,
inverse SA, LCP, and LCE queries)~\cite{collapsing},\footnote{LCP
  array queries are not directly addressed in~\cite{collapsing}, but
  they follow as an immediate consequence. For every $\Text \in
  \Sigma^{\Textlen}$ and $i \in [2 \dd \Textlen]$, it holds
  $\LCP{\Text}[i] = \LCE{\Text}{\SA{\Text}[i]}{\SA{\Text}[i-1]}$ (see
  \cref{sec:prelim}). In other words, if we can support SA and LCE
  queries, then we can also support LCP array queries in the same time
  (up to a constant factor).} without slowing down query times beyond
$\log^{\bigO(1)} \Textlen$. For these reasons, it is reasonable to
simplify space considerations and assume that by \emph{compressed
  space} we mean space $\bigO(\SubstringComplexity{\Text}
\log^{\bigO(1)} \Textlen)$.

The second question -- regarding the limits of \emph{query time} -- has
seen significantly less progress. To date, the only query among those
mentioned above whose complexity is well understood is the most basic
\emph{random access} query: Verbin and Yu~\cite{VerbinY13}
demonstrated that in $\bigO(\SubstringComplexity{\Text}
\log^{\bigO(1)} \Textlen)$ space, random access queries require
$\Omega(\tfrac{\log \Textlen}{\log \log \Textlen})$ time, which
matches the complexity of the best upper bounds. Currently, no lower
bounds are known for any other fundamental queries (SA,
$\text{SA}^{-1}$, LCP, LCE, BWT, PLCP, LF, $\text{LF}^{-1}$, $\Phi$,
$\Phi^{-1}$). Given their central role and wealth of applications,
we thus ask:

\begin{center}
  \emph{What are the lower bounds for fundamental string queries in
    compressed space?}
\end{center}

\paragraph{Our Results}

We develop a series of new lower bounds that establish a clean
dichotomy of fundamental string queries in
$\bigO(\SubstringComplexity{\Text} \log^{\bigO(1)} \Textlen)$ space
into two categories: those with time complexity
$\Theta\big(\tfrac{\log \Textlen}{\log \log \Textlen}\big)$ and those
with time $\Theta(\log \log \Textlen)$ (cf.\ \cref{tab:query-bounds}).
In more detail, our results are as follows:

First, in \cref{sec:lcp},~\ref{sec:sa},~\ref{sec:isa},
and~\ref{sec:lce}, we prove that core string processing queries,
including suffix array (SA), inverse suffix array ($\text{SA}^{-1}$),
longest common prefix (LCP) array, and longest common extension (LCE)
queries, require $\Omega\big(\tfrac{\log \Textlen}{\log \log
  \Textlen}\big)$ time in $\bigO(\SubstringComplexity{\Text}
\log^{\bigO(1)} \Textlen)$ space. These lower bounds are tight up to a
constant factor, matching the performance of the $r$-index of Gagie,
Navarro, and Prezza~\cite{Gagie2020}, which supports these queries in
$\bigO(\RLBWTSize{\Text} \log \Textlen) =
\bigO(\SubstringComplexity{\Text} \log^{\bigO(1)} \Textlen)$ space and
$\bigO(\log \Textlen)$ time and, with minor modifications, achieves
$\bigO\big(\tfrac{\log \Textlen}{\log \log \Textlen}\big)$ time and
$\bigO(\SubstringComplexity{\Text} \log^{\bigO(1)} \Textlen)$ space;
see \cref{tab:query-bounds}. Our lower bounds
(\cref{th:lcp-lower-bound,th:sa-lower-bound,%
  th:isa-lower-bound,th:lce-lower-bound}) hold even when the input
text $\Text \in \Sigma^{\Textlen}$ is over a binary alphabet $\Sigma =
\BinaryAlphabet$. In a single theorem, they can be summarized
as follows:

\begin{theorem}\label{th:main-result-slow-queries}
  There is no data structure that, for every text $\Text \in
  \BinaryAlphabet^{\Textlen}$, uses $\bigO(\SubstringComplexity{\Text}
  \log^{\bigO(1)} \Textlen)$ space and answers any of the following
  queries:
  \begin{itemize}
  \item longest common prefix (LCP) array queries (\cref{sec:lcp}),
  \item suffix array (SA) queries (\cref{sec:sa}),
  \item inverse suffix array ($\text{SA}^{-1}$) queries (\cref{sec:isa}), and
  \item longest common extension (LCE) queries (\cref{sec:lce}),
  \end{itemize}
  in $o\big(\tfrac{\log \Textlen}{\log \log \Textlen}\big)$ time.
\end{theorem}

Together with previous
works~\cite{Rytter03,charikar,attractors,GNPlatin18,resolution,KempaS22,delta}
(which established relations between repetitiveness measures as well
as the optimal \emph{space} bounds for compressed indexing), our
results (concerning the optimal \emph{query time} bounds of compressed
indexes) establish the two frontiers of optimization for future
trade-offs, which can focus on either achieving the optimal query
  time in the smallest possible space, or the fastest possible query
within the optimal space (and everything in between), \emph{both}
of which are now known for central string queries.

\paragraph{Lower Bounds for Faster Queries}

Our second contribution is extending our set of lower bounds to cover
fundamental string queries that admit faster query times. This
includes central queries such as BWT access, LF, PLCP, and
lexicographical predecessor ($\Phi$) queries (see \cref{sec:prelim}),
all of which can be supported in $\bigO(\log \log \Textlen)$ time
using $\bigO(\RLBWTSize{\Text} \log \Textlen) =
\bigO(\SubstringComplexity{\Text} \log^{\bigO(1)} \Textlen)$
space~\cite{Gagie2020}.

The $\Theta(\log \log \Textlen)$ runtime of these queries stems from
the use of predecessor data structures, and until now, it was not
known whether this was inherent or could be improved by further
algorithmic advances. We show that, for all these (and related)
problems, we can reverse the reductions and answer predecessor or
colored predecessor queries (see
\cref{sec:prelim-predecessor,sec:prelim-colored-predecessor}), which,
in our regime of parameters, require $\Omega(\log \log \Textlen)$
time~\cite{PatrascuT06}.

All our lower bounds again hold even when the input text
$\Text \in \Sigma^{\Textlen}$ is over a binary alphabet
$\Sigma = \BinaryAlphabet$.
Our results
(\cref{th:bwt-lower-bound,%
 th:plcp-lower-bound,%
 th:lf-lower-bound,%
 th:ilf-lower-bound,%
 th:phi-lower-bound,%
 th:iphi-lower-bound})
can be summarized in a single theorem as follows:

\begin{theorem}\label{th:main-result-fast-queries}
  There is no data structure that, for every text $\Text \in
  \BinaryAlphabet^{\Textlen}$, uses $\bigO(\SubstringComplexity{\Text}
  \log^{\bigO(1)} \Textlen)$ space and answers any of the following
  queries:
  \begin{itemize}
  \item Burrows--Wheeler Transform (BWT) queries (\cref{sec:bwt}),
  \item permuted longest common prefix (PLCP) array queries (\cref{sec:plcp}),
  \item last-to-first (LF) mapping queries (\cref{sec:lf}),
  \item inverse last-to-first ($\text{LF}^{-1}$) queries (\cref{sec:ilf}),
  \item lexicographical predecessor ($\Phi$) queries (\cref{sec:phi}), and
  \item inverse lexicographical predecessor ($\Phi^{-1}$) queries (\cref{sec:iphi}),
  \end{itemize}
  in $o(\log \log \Textlen)$ time.
\end{theorem}

\paragraph{Related Work}

In this work, we focus on compressed indexes capable of strong (up to
exponential) compression of the input text, which exploit redundancy
arising from repeated substrings. Such redundancy is quantified using
measures including substring complexity~\cite{delta}, string attractor
size~\cite{attractors}, LZ77 size~\cite{LZ77}, run-length BWT
size~\cite{bwt,Gagie2020}, grammar
compression~\cite{Rytter03,charikar}, LZ-End
size~\cite{kreft2010navarro,KempaS22}, and others;
see~\cite{NavarroMeasures} for an overview.

Beyond highly repetitive texts, prior work has also addressed the
\emph{compact} or \emph{entropy-bounded} setting, in which the text
$\Text \in \IntegerAlphabet^{\Textlen}$ is indexed using
$\bigO(\Textlen \log \AlphabetSize)$ bits of space (or, in some cases,
space proportional to the empirical entropy $H_k(\Text)$). This is
asymptotically precisely the space needed to represent any text $\Text
\in \IntegerAlphabet^{\Textlen}$ in the worst case and improves upon
classical indexes such as the suffix array~\cite{ManberM90} and the
suffix tree~\cite{Weiner73}, which require $\Omega(\Textlen \log
\Textlen)$ bits. The two main classes of space-efficient
indexes---those based on repetitiveness and those based on entropy
bounds---are largely incomparable, since entropy-based measures do not
account for large-scale repetitions~\cite{attractors}. As a result,
these approaches have been studied independently, using different
techniques.

Many of the queries considered in this paper have been studied in the
compact or entropy-bounded setting, including suffix array
queries~\cite{FerraginaM00,GrossiV00,breaking}, BWT and LF-mapping
queries~\cite{FerraginaM00}, lexicographic predecessor/successor
queries~\cite{GrossiV00}, longest common extension (LCE)
queries~\cite{sss}, and LCP/PLCP array queries~\cite{cst}. These
foundational compact indexes introduced several unidirectional
reductions, showing that, e.g., suffix array access can
be reduced to rank queries over the BWT~\cite{FerraginaM00}.

Recent work on compact indexes~\cite{breaking,sublinearlz,PrefixEquiv} introduced
an alternative framework based on \emph{prefix range queries}. In
particular,~\cite{breaking} showed that many central indexing queries
(such as SA and inverse SA queries) can be reduced to prefix select
and prefix special rank queries (two specific types of prefix range
queries). The resulting text indexes match the query time and space of
state-of-the-art indexes~\cite{FerraginaM00,GrossiV00} while also
supporting sublinear-time construction in optimal working
space. Subsequent work~\cite{PrefixEquiv} has further extended these
reductions, making them \emph{bidirectional}, i.e., establishing that,
up to a very small additive term in query time, prefix range queries
are in fact \emph{equivalent} to central string queries. This
provides a unified view of previous reduction-based techniques and
shows that, in the compact setting, many core string-processing queries
can be equivalently expressed in terms of much simpler prefix range
queries.

Reductions of this nature for \emph{algorithms} are less common. A
notable example is~\cite{hierarchy}, which develops a broad reduction
hierarchy including problems such as LZ77 factorization, BWT
construction, batched ISA/LPF queries, longest common factor, and
inversion counting, whose state-of-the-art algorithms achieve
$\bigO(\Textlen \sqrt{\log \Textlen})$ time on inputs of $\Textlen$
words (i.e., $\Textlen \log \Textlen$ bits).

\paragraph{Organization of the Paper}

First, in \cref{sec:prelim}, we introduce the notation and
definitions used in the paper. In \cref{sec:overview}, we then give an
overview of our lower bounds. In \cref{sec:slow-queries}, we
describe our lower bounds for queries with complexity
$\Theta\big(\tfrac{\log \Textlen}{\log \log \Textlen}\big)$, including
longest common prefix (LCP) array queries (\cref{sec:lcp}), suffix
array (SA) queries (\cref{sec:sa}), inverse suffix array
($\text{SA}^{-1}$) queries (\cref{sec:isa}), and longest common
extension (LCE) queries (\cref{sec:lce}). In \cref{sec:fast-queries},
we describe our lower bounds for queries with complexity $\Theta(\log
\log \Textlen)$, including Burrows--Wheeler Transform (BWT) queries
(\cref{sec:bwt}), permuted longest common prefix (PLCP) array queries
(\cref{sec:plcp}), last-to-first (LF) mapping queries (\cref{sec:lf}),
inverse last-to-first ($\text{LF}^{-1}$) queries (\cref{sec:ilf}),
lexicographical predecessor ($\Phi$) queries (\cref{sec:phi}), and
inverse lexicographical predecessor ($\Phi^{-1}$) queries
(\cref{sec:iphi}).

\section{Preliminaries}\label{sec:prelim}

\subsection{Basic Definitions}\label{sec:prelim-basic}

A \emph{string} is a finite sequence of characters from a given
\emph{alphabet} $\Sigma$.  The length of a string $S$ is denoted
$|S|$. For $i \in [1\dd |S|]$,\footnote{For $i,j\in \mathbb{Z}$, we let
  $[i \dd j] = \{k \in \Z : i \leq k \leq j\}$,
  $[i \dd j) = \{k \in \Z : i \leq k < j\}$, and
  $(i \dd j] = \{k \in \Z: i < k \leq j\}$.}
the $i$th character of $S$ is denoted $S[i]$. A~\emph{substring} or a
\emph{factor} of $S$ is a string of the form
$S[i \dd j) = S[i]S[{i+1}] \cdots S[{j-1}]$ for some
$1 \leq i \leq j \leq |S| + 1$.
Substrings of the form $S[1 \dd j)$ and $S[i \dd |S|{+}1)$ are called
\emph{prefixes} and \emph{suffixes}, respectively. We use
$\revstr{S}$ to denote the \emph{reverse} of $S$, i.e.,
$S[|S|]\cdots S[2]S[1]$.
We denote the \emph{concatenation} of two strings $U$ and
$V$, that is, the string $U[1]\cdots U[|U|]V[1]\cdots V[|V|]$, by $UV$
or $U\cdot V$.  Furthermore, $S^k = \bigodot_{i=1}^k S$ is the
concatenation of $k \in \Zz$ copies of $S$; note that $S^0 =
\emptystring$ is the \emph{empty string}. A nonempty string $S$ is
said to be \emph{primitive} if it cannot be written as $S = U^k$,
where $U\in \Sigma^+$ and $k \geq 2$.  An integer $p\in [1\dd |S|]$ is
a \emph{period} of $S$ if $S[i] = S[i + p]$ holds for every
$i \in [1 \dd |S|-p]$. We denote the smallest period of $S$ as
$\per{S}$.  For every $S \in \Sigma^{+}$, we define the infinite power
$S^{\infty}$ so that $S^{\infty}[i] = S[1 + (i-1) \bmod |S|]$ for
$i \in \Z$.  In particular, $S = S^{\infty}[1 \dd |S|]$.  By $\lcp{U}{V}$
we denote the length of the longest common prefix of $U$ and $V$. For
any string $S \in \Sigma^{*}$ and any $j_1, j_2 \in [1 \dd |S|]$, we
denote $\LCE{S}{j_1}{j_2} = \lcp{S[j_1 \dd |S|]}{S[j_2 \dd |S|]}$. We
use $\preceq$ to denote the order on $\Sigma$, extended to the
\emph{lexicographic} order on $\Sigma^*$ so that $U,V\in \Sigma^*$
satisfy $U \preceq V$ if and only if either
\begin{enumerate*}[label=(\alph*)]
  \item $U$ is a prefix of $V$, or
  \item $U[1 \dd i) = V[1 \dd i)$ and
    $U[i]\prec V[i]$ holds for some $i\in [1\dd \min(|U|,|V|)]$.
\end{enumerate*}
For any string $S = c_1^{\ell_1} c_2^{\ell_2} \cdots c_h^{\ell_h}$,
where $c_i \in \Sigma$ and $\ell_i > 0$ holds for $i \in [1 \dd h]$,
and $c_i \neq c_{i+1}$ holds for $i \in [1 \dd h)$, we define the
\emph{run-length encoding} of $S$ as a sequence
$\RL{S} = ((c_1, \ell_1), \ldots, (c_h, \ell_h))$.

\begin{definition}[Pattern occurrences and SA-interval]\label{def:occ}
  For any $\Pat \in \Sigma^{*}$ and $\Text \in \Sigma^*$, we define
  \vspace{-.7cm}

  \begin{align*}
    \OccTwo{\Pat}{\Text}
      &= \{j \in [1 \dd |\Text|] : j + |\Pat| \leq |\Text| + 1\text{ and }\Text[j \dd j + |\Pat|) = \Pat\},\\
    \RangeBegTwo{\Pat}{\Text}
      &= |\{j \in [1 \dd |\Text|] : \Text[j \dd |\Text|] \prec \Pat\}|,\\
    \RangeEndTwo{\Pat}{\Text}
      &= \RangeBegTwo{\Pat}{\Text} + |\OccTwo{\Pat}{\Text}|.
  \end{align*}
\end{definition}

\subsection{Suffix Arrays}\label{sec:prelim-sa}

\begin{figure}[t!]
  \centering
  \begin{tikzpicture}[yscale=0.35]

    \draw(1.9,0) node[right] {\scriptsize $\Text[\SA{\Text}[i]\dd \Textlen]$};
    \foreach \x [count=\i] in {a, aababa, aababababaababa,
        aba, abaababa, abaababababaababa, ababa, ababaababa,
        abababaababa, ababababaababa, ba, baababa,
        baababababaababa, baba, babaababa, babaababababaababa,
        bababaababa, babababaababa, bbabaababababaababa}
      \draw (1.9, -\i) node[right] {$\texttt{\x}$};

    \draw (-3.2,0) node{\scriptsize $i$};
    \foreach \i in {1,...,19}
      \draw (-3.2, -\i) node {\footnotesize $\i$};

    \draw(-2.3,0) node{\scriptsize $\SA{\Text}$};
    \foreach \x [count=\i] in {19,14,5,17,12,3,15,10,8,6,18,13,4,16,11,2,9,7,1}
      \draw (-2.3, -\i) node {$\x$};

    \draw (-1.4,0) node {\scriptsize $\LF{\Text}$};
    \foreach \x [count=\i] in {11,12,13,14,15,16,2,17,18,3,4,5,6,7,8,19,9,10,1}
      \draw (-1.4, -\i) node {\x};

    \draw (-0.5,0.05) node {\scriptsize $\ILF{\Text}$};
    \foreach \x [count=\i] in {19,7,10,11,12,13,14,15,17,18,1,2,3,4,5,6,8,9,16}
      \draw (-0.5, -\i) node {\x};

    \draw(0.4,0) node{\scriptsize $\LCP{\Text}$};
    \foreach \x [count=\i] in {0,1,6,1,3,8,3,5,5,7,0,2,7,2,4,9,4,6,1}
      \draw (0.4, -\i) node {$\x$};

    \draw (1.3,0) node{\scriptsize $\BWT{\Text}$};
    \foreach \x [count=\i] in {b,b,b,b,b,b,a,b,b,a,a,a,a,a,a,b,a,a,a}
      \draw (1.3, -\i) node {\texttt{\x}};

  \end{tikzpicture}

  \vspace{-1.3ex}
  \caption{A list of all sorted suffixes of $\Text =
    \texttt{bbabaababababaababa}$ along with
    the arrays
    $\SA{\Text}$ (\cref{def:sa}),
    $\LF{\Text}$ (\cref{def:lf}),
    $\ILF{\Text}$ (\cref{def:ilf}),
    $\LCP{\Text}$ (\cref{def:lcp-array}), and
    $\BWT{\Text}$ (\cref{def:bwt}).}\label{fig:lex-order}
\end{figure}

\begin{figure}[t!]
  \begin{alignat*}{20}
    i   &=        \,1  \,&&\,
                    2  \,&&\,
                    3  \,&&\,
                    4  \,&&\,
                    5  \,&&\,
                    6  \,&&\,
                    7  \,&&\,
                    8  \,&&\,
                    9  \,&&\,
                    10 \,&&\,
                    11 \,&&\,
                    12 \,&&\,
                    13 \,&&\,
                    14 \,&&\,
                    15 \,&&\,
                    16 \,&&\,
                    17 \,&&\,
                    18 \,&&\,
                    19 \\
    \Text &=        \,\texttt{b} \,&&\,
                    \texttt{b} \,&&\,
                    \texttt{b} \,&&\,
                    \texttt{b} \,&&\,
                    \texttt{b} \,&&\,
                    \texttt{b} \,&&\,
                    \texttt{a} \,&&\,
                    \texttt{b} \,&&\,
                    \texttt{b} \,&&\,
                    \texttt{a} \,&&\,
                    \texttt{a} \,&&\,
                    \texttt{a} \,&&\,
                    \texttt{a} \,&&\,
                    \texttt{a} \,&&\,
                    \texttt{a} \,&&\,
                    \texttt{b} \,&&\,
                    \texttt{a} \,&&\,
                    \texttt{a} \,&&\,
                    \texttt{a}\\
    \ISA{\Text} &= [19, \,&&\,
                    16, \,&&\,
                    6,  \,&&\,
                    13, \,&&\,
                    3,  \,&&\,
                    10, \,&&\,
                    18, \,&&\,
                    9,  \,&&\,
                    17, \,&&\,
                    8,  \,&&\,
                    15, \,&&\,
                    5,  \,&&\,
                    12, \,&&\,
                    2,  \,&&\,
                    7,  \,\,\,&&\,
                    14, \,&&\,
                    4,  \,\,\,&&\,
                    11, \,&&\,
                    1]\\
    \PhiArray{\Text} &= [7,  \,&&\,
                         11, \,&&\,
                         12, \,&&\,
                         13, \,&&\,
                         14, \,&&\,
                         8,  \,&&\,
                         9,  \,&&\,
                         10, \,&&\,
                         2,  \,&&\,
                         15, \,&&\,
                         16, \,&&\,
                         17, \,&&\,
                         18, \,&&\,
                         19, \,&&\,
                         3,  \,&&\,
                         4,  \,&&\,
                         5,  \,&&\,
                         6,  \,&&\,
                         1]\\
    \InvPhiArray{\Text} &= [19, \,&&\,
                            9,  \,&&\,
                            15, \,&&\,
                            16, \,&&\,
                            17, \,&&\,
                            18, \,&&\,
                            1,  \,&&\,
                            6,  \,&&\,
                            7,  \,&&\,
                            8,  \,&&\,
                            2,  \,&&\,
                            3,  \,&&\,
                            4,  \,&&\,
                            5,  \,&&\,
                            10, \,&&\,
                            11, \,&&\,
                            12, \,&&\,
                            13, \,&&\,
                            14]\\
    \PLCP{\Text} &= [1, \,&&\,
                     9, \,&&\,
                     8, \,&&\,
                     7, \,&&\,
                     6, \,&&\,
                     7, \,&&\,
                     6, \,&&\,
                     5, \,&&\,
                     4, \,&&\,
                     5, \,&&\,
                     4, \,&&\,
                     3, \,&&\,
                     2, \,&&\,
                     1, \,&&\,
                     3, \,&&\,
                     2, \,&&\,
                     1, \,&&\,
                     0, \,&&\,
                     0]
  \end{alignat*}
  \caption{The arrays
    $\ISA{\Text}$ (\cref{def:isa}),
    $\PhiArray{\Text}$ (\cref{def:phi-array}),
    $\InvPhiArray{\Text}$ (\cref{def:inv-phi-array}), and
    $\PLCP{\Text}$ (\cref{def:plcp-array})
    for the example string
    $\Text = \texttt{bbabaababababaababa}$
    (the same as in~\cref{fig:lex-order}).}\label{fig:text-order}
\end{figure}

\begin{definition}[Suffix array]\label{def:sa}
  For any string $\Text \in \Sigma^{\Textlen}$ of length
  $\Textlen \geq 1$, the \emph{suffix array} $\SA{\Text}[1 \dd \Textlen]$
  of $\Text$ is a permutation of $[1 \dd \Textlen]$ such that
  $\Text[\SA{\Text}[1] \dd \Textlen] \prec
  \Text[\SA{\Text}[2] \dd \Textlen] \prec \cdots \prec
  \Text[\SA{\Text}[\Textlen] \dd \Textlen]$, i.e., $\SA{\Text}[i]$ is
  the starting position of the lexicographically $i$th suffix of
  $\Text$.
  See \cref{fig:lex-order} for an example.
\end{definition}

\begin{remark}\label{rm:occ}
  Note that the two values $\RangeBegTwo{\Pat}{\Text}$ and
  $\RangeEndTwo{\Pat}{\Text}$ (\cref{def:occ}) are the endpoints of
  the so-called \emph{SA-interval} representing the occurrences of
  $\Pat$ in $\Text$, i.e.,
  \[
    \OccTwo{\Pat}{\Text} =
      \{\SA{\Text}[i] : i \in (\RangeBegTwo{\Pat}{\Text} \dd \RangeEndTwo{\Pat}{\Text}]\}
  \]
  holds for every $\Text \in \Sigma^{+}$ and $\Pat \in \Sigma^{*}$,
  including when $\Pat = \emptystring$ and when
  $\OccTwo{\Pat}{\Text} = \emptyset$.
\end{remark}

\begin{example}\label{ex:occ-and-ranges}
  For the example text $\Text$ in \cref{fig:lex-order} and pattern
  $\Pat = \texttt{ababa}$, it holds
  $\OccTwo{\Pat}{\Text} = \{15, 10, 8, 6\} =
  \{\SA{\Text}[i] : i \in (6 \dd 10]\}$, so
  $\RangeBegTwo{\Pat}{\Text} = 6$ and
  $\RangeEndTwo{\Pat}{\Text} = 10$.
\end{example}

\begin{definition}[Inverse suffix array]\label{def:isa}
  The \emph{inverse suffix array} of a text $\Text \in \Sigma^{\Textlen}$
  of length $\Textlen \geq 1$
  is an array $\ISA{\Text}[1 \dd \Textlen]$ containing the inverse
  permutation of $\SA{\Text}$ (\cref{def:sa}), i.e., $\ISA{\Text}[j] = i$ holds if and
  only if $\SA{\Text}[i] = j$. Equivalently, $\ISA{\Text}[j]$ stores
  the lexicographic \emph{rank} of $\Text[j \dd \Textlen]$ among the
  suffixes of $\Text$, that is,
  $\ISA{\Text}[j] = 1 + \RangeBegTwo{\Text[j \dd \Textlen]}{\Text}$
  (\cref{def:occ}).
  See \cref{fig:text-order} for an example.
\end{definition}

\begin{definition}[Lexicographical predecessor array $\Phi$]\label{def:phi-array}
  For any string $\Text \in \Sigma^{\Textlen}$ of length
  $\Textlen \geq 1$, we define $\PhiArray{\Text}[1 \dd \Textlen]$ so that
  $\PhiArray{\Text}[\SA{\Text}[1]] = \SA{\Text}[\Textlen]$, and for
  every $j \in [1 \dd \Textlen] \setminus \{\SA{\Text}[1]\}$,
  \[
    \PhiArray{\Text}[j] = \SA{\Text}[\ISA{\Text}[j] - 1]
  \]
  (see \cref{def:sa,def:isa}). Equivalently,
  $\PhiArray{\Text}[1 \dd \Textlen]$ is the unique array that satisfies
  $\PhiArray{\Text}[\SA{\Text}[i]] = \SA{\Text}[i-1]$ for every
  $i \in (1 \dd \Textlen]$, and
  $\SA{\Text}[\SA{\Text}[1]] = \SA{\Text}[\Textlen]$.
  See \cref{fig:text-order} for an example.
\end{definition}

\begin{definition}[Inverse lexicographical predecessor array $\Phi^{-1}$]\label{def:inv-phi-array}
  The \emph{inverse lexicographical predecessor array} (or simply
  \emph{lexicographical successor array}) of a text
  $\Text \in \Sigma^{\Textlen}$ of length $\Textlen \geq 1$ is an array
  $\InvPhiArray{\Text}[1 \dd \Textlen]$ containing the inverse
  permutation of $\PhiArray{\Text}$ (\cref{def:phi-array}),
  i.e., such that $\InvPhiArray{\Text}[j] = i$ holds if and only if
  $\PhiArray{\Text}[i] = j$. Equivalently,
  $\InvPhiArray{\Text}[1 \dd \Textlen]$ is the unique array that
  satisfies the equation
  $\InvPhiArray{\Text}[\SA{\Text}[i]] = \SA{\Text}[i + 1]$ for every
  $i \in [1 \dd \Textlen)$, and
  $\InvPhiArray{\Text}[\SA{\Text}[\Textlen]] = \SA{\Text}[1]$.
  See \cref{fig:text-order} for an example.
\end{definition}

\begin{definition}[Last-to-First mapping LF]\label{def:lf}
  The \emph{Last-to-First (LF) mapping} of a text
  $\Text \in \Sigma^{\Textlen}$ of length $\Textlen \geq 1$
  is an array $\LF{\Text}[1 \dd \Textlen]$
  defined so that, for every $i \in [1 \dd \Textlen]$,
  \[
    \LF{\Text}[i] :=
      \begin{cases}
        \ISA{\Text}[\SA{\Text}[i] - 1] & \text{if }\SA{\Text}[i] \ne 1,\\
        \ISA{\Text}[\Textlen] & \text{otherwise}.
      \end{cases}
  \]
  Equivalently, $\LF{\Text}[1 \dd \Textlen]$ is the unique array that
  satisfies the equation
  $\LF{\Text}[\ISA{\Text}[j]] = \ISA{\Text}[j-1]$ for every
  $j \in (1 \dd \Textlen]$ and
  $\LF{\Text}[\ISA{\Text}[1]] = \ISA{\Text}[\Textlen]$.
  See \cref{fig:lex-order} for an example.
\end{definition}

\begin{definition}[First-to-Last mapping $\ILF{}$]\label{def:ilf}
  The \emph{First-to-Last} \emph{(inverse LF) mapping} of a text
  $\Text \in \Sigma^{\Textlen}$ of length $\Textlen \geq 1$ is an array
  $\ILF{\Text}[1 \dd \Textlen]$ containing the inverse permutation of
  $\LF{\Text}$ (\cref{def:lf}), i.e., $\ILF{\Text}[j] = i$ holds if
  and only if $\LF{\Text}[i] = j$.  Equivalently,
  $\ILF{\Text}[1 \dd \Textlen]$ is the unique array that satisfies
  the equation $\ILF{\Text}[\ISA{\Text}[j]] = \ISA{\Text}[j + 1]$
  for every $j \in [1 \dd \Textlen)$ and
  $\ILF{\Text}[\ISA{\Text}[\Textlen]] = \ISA{\Text}[1]$.
  See \cref{fig:lex-order} for an example.
\end{definition}

\begin{remark}\label{rm:ilf}
  The inverse LF mapping (\cref{def:ilf}) for text $\Text$ is also
  denoted as $\Psi_{\Text}$~\cite{GrossiV00}.
\end{remark}

\subsection{LCP Arrays}\label{sec:prelim-lcp}

\begin{definition}[Longest common prefix (LCP) array]\label{def:lcp-array}
  For any string $\Text \in \Sigma^{\Textlen}$ of length
  $\Textlen \geq 1$, the \emph{longest common prefix (LCP) array}
  $\LCP{\Text}[1 \dd \Textlen]$ is an array such that $\LCP{\Text}[1] = 0$,
  and for every $i \in [2 \dd \Textlen]$,
  \[
    \LCP{\Text}[i] = \LCE{\Text}{\SA{\Text}[i]}{\SA{\Text}[i-1]}
  \]
  (see \cref{sec:prelim-basic} and \cref{def:sa}).
  See \cref{fig:lex-order} for an example.
\end{definition}

\begin{definition}[Permuted longest common prefix (PLCP) array]\label{def:plcp-array}
  For any string $\Text \in \Sigma^{\Textlen}$ of length $\Textlen \geq 1$,
  the \emph{permuted longest common prefix (PLCP) array}
  $\PLCP{\Text}[1 \dd \Textlen]$ is defined so that
  $\PLCP{\Text}[\SA{\Text}[1]] = 0$ and, for every
  $j \in [1 \dd \Textlen] \setminus \{\SA{\Text}[1]\}$,
  \[
    \PLCP{\Text}[j]
      = \LCE{\Text}{j}{\PhiArray{\Text}[j]}
      = \LCE{\Text}{j}{\SA{\Text}[\ISA{\Text}[j]-1]}
  \]
  (see \cref{sec:prelim-basic} and
  \cref{def:phi-array,def:sa,def:isa}).  Equivalently, $\PLCP{\Text}$
  is the unique array satisfying the equation
  $\PLCP{\Text}[\SA{\Text}[i]] = \LCP{\Text}[i]$ for every
  $i \in [1 \dd \Textlen]$ (see \cref{def:lcp-array}).
  See \cref{fig:text-order} for an example.
\end{definition}

\subsection{Repetitiveness Measures}\label{sec:prelim-repetitiveness}

\subsubsection{Lempel--Ziv Factorization}\label{sec:prelim-lz}

\begin{definition}[LZ77-like factorization]\label{def:lz77-like}
  A partition $\Text = f_1 f_2 \dots f_k$ of a text
  $\Text \in \Sigma^{*}$ is called an \emph{LZ77-like factorization}
  of $\Text$, if for every $i \in [1 \dd k]$, it holds either $|f_i| = 1$
  or, letting $j = |f_1 f_2 \dots f_{i-1}|$, there exists
  $j' \in [1 \dd j)$ satisfying $\LCE{\Text}{j}{j'} \geq |f_i|$.
  Each substring $f_i$ in the LZ77-like factorization is called a
  \emph{phrase}.  If $|f_i| = 1$, then $f_i$ is called a \emph{literal}
  phrase. Otherwise, it is called a \emph{repeat} phrase. If $f_i$ is
  a repeat phrase, and $j' \in [1 \dd j)$ (where $j = |f_1 \dots f_{i-1}|$)
  is an index satisfying $\LCE{\Text}{j}{j'} \geq |f_i|$, then the substring
  $\Text[j' \dd j' + |f_i|)$ is called a \emph{source} of phrase $f_i$.
\end{definition}

\begin{remark}\label{rm:lz}
  Observe that if $\Text$ has an LZ77-like factorization consisting of
  $k$ phrases, then $\Text$ can be encoded in $\bigO(k)$ space. In the
  underlying representation, the literal phrase $f_{i}$ is encoded as
  a pair $(\Text[j+1], 0)$, where $j = |f_1 \cdots f_{i-1}|$.  A
  repeat phrase $f_i$ is encoded as $(j',|f_i|)$, where
  $j' \in [1 \dd j)$ (with $j = |f_1 \dots f_{i-1}|$) is any index
  satisfying $\LCE{\Text}{j}{j'} \geq |f_i|$.
\end{remark}

\begin{definition}[Longest previous factor (LPF) array]\label{def:lpf}
  For every text $\Text \in \Sigma^{\Textlen}$ of length $\Textlen \geq 1$, we define the
  \emph{longest previous factor (LPF) array} $\LPF{\Text}[1 \dd \Textlen]$
  such that $\LPF{\Text}[1] = 0$, and for every
  $j \in [2 \dd \Textlen]$, it holds (see \cref{def:occ})
  \[
    \LPF{\Text}[j] = \max\{\ell \in [0 \dd \Textlen - j + 1] :
    \min \OccTwo{\Text[j \dd j+\ell)}{\Text} < j\}.
  \]
\end{definition}

\begin{definition}[Lempel--Ziv (LZ77) factorization~\cite{LZ77}]\label{def:lz}
  The \emph{LZ77 factorization} of $\Text$ is a \emph{greedy} left-to-right
  LZ77-like factorization of $\Text = f_1 f_2 \dots f_k$ defined such
  that, for every $i \in [1 \dd k]$,
  \[
    |f_{i}| = \max(1, \LPF{\Text}[j + 1])
  \]
  (see \cref{def:lpf}),
  where $j = |f_1 \dots f_{i-1}|$.
  We denote the number of phrases in the LZ77 factorization by
  $\LZSize{\Text}$ (i.e.., $k = \LZSize{\Text}$).
\end{definition}

\begin{example}\label{ex:lz77}
  The text $\Text$ in \cref{fig:lex-order} has the LZ77 factorization
  $\Text = \texttt{b}\cdot \texttt{b}\cdot \texttt{a}\cdot \texttt{ba}
    \cdot \texttt{aba}\cdot \texttt{bababa} \cdot \texttt{ababa}$ with
  $\LZSize{\Text} = 7$ phrases, and its example representation is $
  (\texttt{b},0), (1,1), (\texttt{a},0), (2,2), (3,3), (7,6), (10,5)$.
\end{example}

\begin{theorem}[{\cite[Theorem~1]{LZ76}}]\label{th:lz77-size}
  For every LZ77-like factorization $\Text = f_1 f_2 \dots f_k$ of a
  text $\Text$, it holds $\LZSize{\Text} \leq k$.
\end{theorem}

\begin{observation}\label{ob:rl}
  For any text $\Text$ satisfying $\RL{\Text} = k$, it holds
  $\LZSize{\Text} = \bigO(k)$.
\end{observation}
\begin{proof}
  It is easy to write the LZ-like factorization of $\Text$ consisting
  of $2k$ phrases. The claim thus follows by \cref{th:lz77-size}.
\end{proof}

\subsubsection{Grammar Compression}\label{sec:prelim-grammar}

\begin{definition}[Context-free grammar (CFG)]\label{def:cfg}
  A \emph{context-free grammar} (CFG) is a tuple $G = (V, \Sigma, R,
  S)$, where $V$ is a finite set of \emph{nonterminals} (or
  \emph{variables}), $\Sigma$ is a finite set of \emph{terminals}, and
  $R \subseteq V \times (V \cup \Sigma)^*$ is a set of
  \emph{productions} (or \emph{rules}). We assume
  $V \cap \Sigma = \emptyset$ and $S \in V$.
  The nonterminal $S$ is called the \emph{start symbol}.
  Nonterminals in $V \setminus \{S\}$ are called \emph{secondary}.
  If $(N, \gamma) \in R$, then we write $N \rightarrow \gamma$.
  For $u, v \in (V \cup \Sigma)^*$, we write $u \Rightarrow v$
  if there exist $u_1, u_2 \in (V\cup \Sigma)^*$ and a
  rule $N \rightarrow \gamma$ such that $u = u_1 N u_2$ and
  $v = u_1 \gamma u_2$. We say that $u$ \emph{derives} $v$ and write
  $u \Rightarrow^* v$, if there exists a sequence $u_1, \ldots, u_k$,
  $k \geq 1$, such that $u = u_1$, $v = u_k$, and
  $u_i \Rightarrow u_{i+1}$ for $1 \leq i < k$.
  The \emph{language} of grammar $G$ is the set
  $\Lang{G} := \{w \in \Sigma^* \mid S \Rightarrow^* w\}$.
\end{definition}

\begin{definition}[Straight-line grammar (SLG)]\label{def:slg}
  A CFG $G = (V,\Sigma,R,S)$ is a \emph{straight-line grammar (SLG)}
  if it satisfies the following two conditions:
  \begin{enumerate}
  \item there exists an ordering $(N_1, \dots, N_{|V|})$ of all elements
    in $V$, such that, for every $(N,\gamma) \in R$, letting $i \in [1
      \dd |V|]$ be such that $N = N_i$, it holds $\gamma \in (\{N_{i+1},
    \dots, N_{|V|}\} \cup \Sigma)^{*}$, and
  \item for every nonterminal $N \in V$, there exists exactly one
    $\gamma \in (V \cup \Sigma)^{*}$ such that $(N,\gamma) \in R$.
  \end{enumerate}
  The unique string $\gamma \in (V \cup \Sigma)^{*}$ such that
  $(N,\gamma) \in R$ is called the \emph{definition} of nonterminal $N$,
  and is denoted $\Rhs{G}{N}$. Note that in an SLG, for every $\alpha
  \in (V \cup \Sigma)^{*}$, there exists exactly one string $\gamma \in
  \Sigma^{*}$ satisfying $\alpha \Rightarrow^{*} \gamma$. Such $\gamma$
  is called the \emph{expansion} of $\alpha$, and is denoted
  $\Exp{G}{\alpha}$. In particular, in an SLG, we have $|\Lang{G}| = 1$.
  We define the size of an SLG as $|G| =
  \sum_{N \in V}\max(|\Rhs{G}{N}|, 1)$.
\end{definition}

\begin{remark}\label{rm:slg-size}
  Note that every SLG $G$ of size $|G| = g$ can be encoded in
  $\bigO(g)$ space: we choose an ordering of nonterminals and write
  down the definitions of all variables, replacing nonterminals with
  their index in this order.
\end{remark}

\begin{definition}[Straight-line program (SLP)]\label{def:slp}
  An SLG $G = (V, \Sigma, R, S)$ is called a \emph{straight-line
  program (SLP)} if for every $N \in V$, either $\Rhs{G}{N} = AB$
  for some $A,B \in V$, or $\Rhs{G}{N} = c$ for some $c \in \Sigma$.
\end{definition}

\begin{definition}[Parse tree]\label{def:parse-tree}
  Let $G = (V, \Sigma, R, S)$ be an SLG. The \emph{parse tree} of
  $A \in V \cup \Sigma$ is a rooted, ordered tree $\mathcal{T}_{G}(A)$,
  where each node $v$ is associated with a symbol
  $s(v) \in V \cup \Sigma$.
  The root of $\mathcal{T}_{G}(A)$ is a node $\rho$ such
  that $s(\rho) = A$.  If $A \in \Sigma$, then $\rho$ has no
  children. If $A \in V$ and $\Rhs{G}{A} = B_1 \cdots B_k$, then
  $\rho$ has $k$ children, and the subtree rooted at the $i$th child
  is (a copy of) $\mathcal{T}_{G}(B_i)$. The parse tree of $G$ is
  defined as the parse tree $\mathcal{T}_{G}(S)$ of the start symbol $S$.
\end{definition}

\begin{definition}[Height of SLG]\label{def:slg-height}
  Let $G = (V, \Sigma, R, S)$ be an SLG. The \emph{height} of $G$ is
  defined as the height of its parse tree.
\end{definition}

\subsubsection{Run-Length Burrows--Wheeler Transform}\label{sec:prelim-bwt}

\begin{definition}[Burrows--Wheeler transform (BWT)~\cite{bwt}]\label{def:bwt}
  The \emph{Burrows--Wheeler Transform} (BWT) of a text
  $\Text \in \Sigma^{\Textlen}$ of length $\Textlen \geq 1$, denoted $\BWT{\Text}[1 \dd \Textlen]$,
  is a permutation of the symbols in $\Text$ such that, for every
  $i \in [1 \dd \Textlen]$,
  \[
    \BWT{\Text}[i] :=
      \begin{cases}
        \Text[\SA{\Text}[i] - 1] & \text{if }\SA{\Text}[i] > 1,\\
        \Text[\Textlen] & \text{otherwise}.
      \end{cases}
  \]
  Equivalently, $\BWT{\Text}[i] = \Textinf[\SA{\Text}[i] - 1]$.  By
  $\RLBWTSize{\Text} := |\RL{\BWT{\Text}[1 \dd \Textlen]}|$, we denote
  the number of equal-letter runs in the BWT of $\Text$.
  See \cref{fig:lex-order} for an example.
\end{definition}

\begin{example}\label{ex:bwt}
  For the example text $\Text$ in \cref{fig:lex-order}, we have
  $\RLBWTSize{\Text} =
  |\RL{{\tt b}^6{\tt a}^1{\tt b}^2{\tt a}^6{\tt b}^1{\tt a}^3}| = 6$.
\end{example}

\begin{theorem}[{\cite{Gagie2020}}]\label{th:rindex}
  Let $\epsilon \in (0,1)$ be a constant. For every nonempty text
  $\Text \in \Sigma^{\Textlen}$, there exists a data structure of size
  $\bigO(\RLBWTSize{\Text} \log^{1+\epsilon} \Textlen)$ (see
  \cref{def:bwt}) that answers suffix array, inverse suffix array, and
  LCP array queries on $\Text$ (that, given any
  $i \in [1 \dd \Textlen]$, return $\SA{\Text}[i]$, $\ISA{\Text}[i]$,
  and $\LCP{\Text}[i]$; see \cref{def:sa,def:isa,def:lcp-array}) in
  $\bigO\big(\tfrac{\log \Textlen}{\log \log \Textlen}\big)$ time.
\end{theorem}

\begin{remark}\label{rm:rindex}
  Although the above results are not directly stated
  in~\cite{Gagie2020} (the data structures supporting SA, inverse SA,
  and LCP queries described in~\cite[Theorems~5.4, 5.7, and 5.8]{Gagie2020}
  for a text $\Text \in \Sigma^{\Textlen}$ use
  $\bigO(\RLBWTSize{\Text} \log \Textlen)$ space and support the
  queries in $\bigO(\log \Textlen)$ time), with a minor modification,
  the query time can be improved to
  $\bigO\left(\frac{\log \Textlen}{\log \log \Textlen}\right)$ to match
  the lower bound established in this paper
  (at the price of a small space increase).

  Specifically, instead of storing two half-blocks around each BWT
  block boundary, we store $\tau = \Theta(\log^{\epsilon} \Textlen)$
  blocks (for any constant $\epsilon > 0$), and adjust the block
  length to be a power of~$\tau$ (rather than a power of~$2$). This is
  a standard technique, used for instance in~\cite{attractors,blocktree}.

  The resulting data structures support SA, inverse SA, and LCP
  queries in $\bigO(\log_{\tau} \Textlen) =
  \bigO\left(\frac{\log \Textlen}{\log \log \Textlen}\right)$
  time and achieve the space stated in \cref{th:rindex}.
\end{remark}

\begin{theorem}[\cite{Gagie2020}]\label{th:plcp-upper-bound}
  For every nonempty text $\Text \in \Sigma^{\Textlen}$, there exists a data
  structure of size $\bigO(\RLBWTSize{\Text})$ (see \cref{def:bwt})
  that answers PLCP array queries on $\Text$ (that, given any
  $j \in [1 \dd \Textlen]$, return $\PLCP{\Text}[j]$; see
  \cref{def:plcp-array}) in $\bigO(\log \log \Textlen)$ time.
\end{theorem}

\subsubsection{Substring Complexity}\label{sec:prelim-delta}

\begin{definition}[Substring counting function]\label{def:substr-count}
  For a string $\Text\in \Sigma^{\Textlen}$ and $\ell \in \Zp$, we
  denote the number of length-$\ell$ substrings by
  \[
    \SubstrCount{\ell}{\Text} = |\{\Text[i \dd i + \ell) : i \in [1\dd
      \Textlen-\ell+1]\}|.
  \]
  Note that for every $\ell > \Textlen$, $\SubstrCount{\ell}{\Text} = 0$.
\end{definition}

\begin{definition}[Substring complexity~\cite{delta}]\label{def:delta}
  The \emph{substring complexity} of $\Text$ is defined as
  $\SubstringComplexity{\Text} =
  \max_{\ell=1}^{\Textlen}\frac{1}{\ell}\SubstrCount{\ell}{\Text}$.
\end{definition}

\begin{lemma}[{\cite[Lemma~II.4]{delta}}]\label{lm:lz-upper-bound}
  For every text $\Text \in \IntegerAlphabet^{\Textlen}$, it holds
  \[
    \LZSize{\Text} = \bigO\Big(\SubstringComplexity{\Text} \log
    \tfrac{\Textlen \log \AlphabetSize}{\SubstringComplexity{\Text} \log \Textlen}\Big).
  \]
\end{lemma}

\begin{remark}\label{rm:lz-upper-bound}
 A simplified statement of the upper bound from
 \cref{lm:lz-upper-bound} is $\LZSize{\Text} =
 \bigO(\SubstringComplexity{\Text} \log \Textlen)$.
\end{remark}

\begin{theorem}[{\cite[Theorem~III.7]{resolution}}]\label{th:rlbwt-size}
  For every text $\Text \in \Sigma^{\Textlen}$, it holds
  \[
    \RLBWTSize{\Text} = \bigO\left(\big(\SubstringComplexity{\Text} \log
    \SubstringComplexity{\Text}\big) \cdot  \max\Big(1,
    \log \tfrac{\Textlen}{\SubstringComplexity{\Text} \log \SubstringComplexity{\Text}}\Big)\right).
  \]
  The above bound is asymptotically tight for all combinations of
  $\Textlen$, $\RLBWTSize{\Text}$, and $\SubstringComplexity{\Text}$.
\end{theorem}

\begin{remark}\label{rm:rlbwt-size}
 A simplified statement of the upper bound from \cref{th:rlbwt-size}
 is $\RLBWTSize{\Text} = \bigO(\SubstringComplexity{\Text} \log^2
 \Textlen)$.
\end{remark}

By combining~\cite[Lemma~3.7 and Theorem~3.9]{attractors}
and~\cite[Lemma~II.3]{delta}, we obtain the following result.

\begin{theorem}[{\cite{attractors,delta}}]\label{lm:z-and-r-upper-bound}
  For every text $\Text$, it holds
  \begin{enumerate}
  \item\label{lm:z-and-r-upper-bound-it-1}
    $\SubstringComplexity{\Text} = \bigO(\LZSize{\Text})$,
  \item\label{lm:z-and-r-upper-bound-it-2}
    $\SubstringComplexity{\Text} = \bigO(\RLBWTSize{\Text})$.
  \end{enumerate}
\end{theorem}

By combining~\cite{VerbinY13} with \cref{lm:lz-upper-bound}, we obtain
the following result.

\begin{theorem}[{\cite{VerbinY13,delta}}]\label{th:random-access-lower-bound}
  There is no data structure that, for every text $\Text \in
  \BinaryAlphabet^{\Textlen}$, uses
  $\bigO(\SubstringComplexity{\Text} \log^{\bigO(1)} \Textlen)$
  space and answers random access queries on $\Text$
  (that, given any $j \in [1 \dd \Textlen]$, return $\Text[j]$)
  in $o(\tfrac{\log \Textlen}{\log \log \Textlen})$ time.
\end{theorem}

The above lower bound is tight. By combining~\cite{blocktree} with
\cref{lm:lz-upper-bound}, we obtain the following result.

\begin{theorem}[{\cite{blocktree,delta}}]\label{th:random-access-upper-bound}
  Let $\epsilon > 0$ be a positive constant.  There exists a data
  structure that, for every $\Text \in \Sigma^{\Textlen}$, uses
  $\bigO(\SubstringComplexity{\Text} \log^{2+\epsilon} \Textlen)$
  space and answers random access queries on $\Text$ (that, given any
  $j \in [1 \dd \Textlen]$, return $\Text[j]$) in
  $\bigO(\tfrac{\log \Textlen}{\log \log \Textlen})$ time.
\end{theorem}

\subsection{Predecessor Queries}\label{sec:prelim-predecessor}

\begin{definition}[Predecessor]\label{def:predecessor}
  Consider a nonempty set $A \subseteq \Z$. For every $x \in \Z$, we
  define
  \[
    \Predecessor{A}{x} = \max \{y \in A: y < x\} \cup \{-\infty\}.
  \]
\end{definition}

\begin{example}\label{ex:predecessor}
  Let $A = \{2, 5, 7, 8, 10, 12\}$.
  Then, it holds
  $\Predecessor{A}{9} = 8$,
  $\Predecessor{A}{5} = 2$, and
  $\Predecessor{A}{2} = -\infty$.
\end{example}

\begin{theorem}[Y-fast tries~\cite{Willard83}]\label{th:predecessor-yfast-trie}
  Consider a sequence $(a_i)_{i \in [0 \dd m]}$, where $m \geq 1$,
  such that $a_{0} = -\infty$, $a_1 < a_2 < \dots < a_m$, and there
  exists $u \in \Zp$ such that, for every $i \in [1 \dd m]$, it holds
  $a_i \in [0 \dd u]$.  Then, there exists a data structure of size
  $\bigO(m)$ that, given any $x \in \Z$, in $\bigO(\log \log u)$ time
  returns $i \in [0 \dd m]$ satisfying $a_{i} = \Predecessor{A}{x}$
  (\cref{def:predecessor}), where $A = \{a_1, a_2, \dots, a_m\}$.
\end{theorem}

\begin{theorem}[Fusion trees~\cite{FredmanW93}]\label{th:predecessor-fusion-tree}
  Consider a sequence $(a_i)_{i \in [0 \dd m]}$, where $m \geq 1$,
  such that $a_{0} = -\infty$, $a_1 < a_2 < \dots < a_m$, and, for
  every $i \in [1 \dd m]$, it holds $a_i \in [0 \dd u]$, where
  $u = 2^w$. In the word RAM model with $w$-bit word size, there exists a
  data structure of size $\bigO(m)$ that, given any $x \in \Z$, in
  $\bigO(1 + \log_{w} m)$ time returns $i \in [0 \dd m]$ satisfying
  $a_{i} = \Predecessor{A}{x}$ (\cref{def:predecessor}), where
  $A = \{a_1, a_2, \dots, a_m\}$.
\end{theorem}

\begin{theorem}[{\cite{PatrascuT06}}]\label{th:predecessor-lower-bound}
  There is no data structure that, for every nonempty set
  $A \subseteq[1 \dd m^2]$ of size $|A| = m$, uses
  $\bigO(m \log^{\bigO(1)} m)$ space and answers predecessor queries
  on $A$ (that given any $x \in \Z$, return $\Predecessor{A}{x}$;
  see \cref{def:predecessor}) in $o(\log \log m)$ time.
\end{theorem}

\subsection{Colored Predecessor Queries}\label{sec:prelim-colored-predecessor}

\begin{definition}[Colored predecessor]\label{def:colored-predecessor}
  Consider a nonempty set $A \subseteq \Z$. For every $x \in \Z$, we
  define\footnote{This definition of
    the \emph{colored predecessor problem} is equivalent to the original
    definition~\cite{PatrascuT06}, in which the color of each integer is
    an arbitrary bit, and two consecutive integers may have the same
    color. It is easy to see that if two adjacent integers have the same
    color, the larger can be removed without affecting the query
    answer. Thus, we may assume adjacent elements have alternating
    colors, which is equivalent to defining the color of each element as
    the parity of its rank.}
  \[
    \PredecessorColor{A}{x} = |\{y \in A: y < x\}| \bmod 2.
  \]
\end{definition}

\begin{example}\label{ex:colored-predecessor}
  Let $A = \{2, 5, 7, 8, 10, 12\}$.
  Then, it holds
  $\PredecessorColor{A}{9} = 0$ and
  $\PredecessorColor{A}{8} = 1$.
\end{example}

\begin{theorem}[{\cite{PatrascuT06}}]\label{th:colored-predecessor-lower-bound}
  There is no data structure that, for every nonempty set
  $A \subseteq[1 \dd m^2]$ of size $|A| = m$, uses
  $\bigO(m \log^{\bigO(1)} m)$ space and answers colored predecessor
  queries on $A$ (that given any $x \in \Z$, return $\PredecessorColor{A}{x}$;
  see \cref{def:colored-predecessor}) in $o(\log \log m)$ time.
\end{theorem}

\subsection{Range Queries}\label{sec:prelim-range-queries}

\begin{definition}[Range counting and selection queries]\label{def:range-count-and-select}
  Let $A[1 \dd m]$ be an array of $m \geq 0$ nonnegative integers.
  For every $j \in [0 \dd m]$ and $v \geq 0$, we define
  \[
    \TwoSidedRangeCount{A}{j}{v} := |\{i \in (0 \dd j] : A[i] \geq v\}|.
  \]
  For every $v \geq 0$ and every
  $r \in [1 \dd \TwoSidedRangeCount{A}{m}{v}]$,
  by $\RangeSelect{A}{r}{v}$
  we define the $r$th smallest element of
  $\{i \in [1 \dd m] : A[i] \geq v\}$.
\end{definition}

\begin{definition}[Range minimum query (RMQ)]\label{def:rmq}
  Let $A[1 \dd m]$ be an array of $m \geq 0$ integers.  For
  every $b, e \in [0 \dd m]$ such that $b < e$, we define
  (see \cref{rm:argmin})
  \[
    \RMQ{A}{b}{e} := \argmin_{i \in (b \dd e]} A[i].
  \]
\end{definition}

\begin{remark}\label{rm:argmin}
  We assume that $\argmin\, \{f(x) : x \in S\}$ returns the
  \emph{smallest} $y \in S$ such that $f(y) = \min\, \{f(x) : x \in S\}$.
\end{remark}

\begin{example}\label{ex:range-queries}
  Let $A = [5, 1, 2, 8, 4, 7, 6, 2, 9]$. Then,
  $\TwoSidedRangeCount{A}{6}{4} = 4$,
  $\RangeSelect{A}{4}{5} = 7$, and
  $\RMQ{A}{2}{9} = 3$.
\end{example}

\begin{theorem}[{\cite{FischerH11}}]\label{th:rmq}
  For every array $A[1 \dd m]$ of $m$ integers, there exists a data
  structure of $\bigO(m)$ bits that answers range minimum queries on
  $A$ in $\bigO(1)$ time and can be constructed in $\bigO(m)$ time.
\end{theorem}

\begin{theorem}[{\cite{Patrascu07}}]\label{th:range-count-lower-bound}
  There is no data structure that,
  for every array $A[1 \dd n]$ containing a permutation of
  $\{1, \dots, n\}$, uses
  $\bigO(n \log^{\bigO(1)} n)$ space and answers range counting
  queries on $A$ (given any $j \in [0 \dd m]$ and $v \geq 0$, return
  $\TwoSidedRangeCount{A}{j}{v}$; see \cref{def:range-count-and-select})
  in $o(\tfrac{\log n}{\log \log n})$ time.
\end{theorem}

\begin{theorem}[{\cite{JorgensenL11}}]\label{th:range-select-lower-bound}
  There is no data structure that, for every array $A[1 \dd n]$
  containing a permutation of $\{1, \dots, n\}$, uses
  $\bigO(n \log^{\bigO(1)} n)$ space and answers range selection
  queries on $A$ (given any $v \geq 0$ and
  $r \in [1 \dd \TwoSidedRangeCount{A}{m}{v}]$,
  return $\RangeSelect{A}{r}{v}$; see \cref{def:range-count-and-select})
  in $o(\tfrac{\log n}{\log \log n})$ time.
\end{theorem}

\subsection{Model of Computation}

We use the standard word RAM model of computation~\cite{Hagerup98}
with $w$-bit \emph{machine words}, where $w \ge \log \Textlen$, and
all standard bitwise and arithmetic operations take $\bigO(1)$ time.
Unless explicitly stated otherwise, we measure space complexity in
machine words.

In the RAM model, strings are usually represented as arrays, with each
character occupying one memory cell. A single character, however, only
needs $\lceil \log \AlphabetSize \rceil$ bits, which might be much
less than $w$. We can therefore store (the \emph{packed
representation} of) a text $\Text \in \IntegerAlphabet^{\Textlen}$
using $\bigO(\lceil \tfrac{\Textlen \log \AlphabetSize}{w} \rceil)$
words.

\section{Technical Overview}\label{sec:overview}

\subsection{Overview of the \texorpdfstring{\boldmath $\Omega(\tfrac{\log n}{\log \log n})$}{Ω(log n / log log n)} Lower Bounds}\label{sec:overview-slow}

In this section, we sketch the ideas of our lower bound for LCP array
access queries showing that there is no data structure that, for every
$\Text \in \BinaryAlphabet^{\Textlen}$, uses
$\bigO(\SubstringComplexity{\Text} \log^{\bigO(1)} \Textlen)$ space
and answers LCP array queries (that, given any
$i \in [1 \dd \Textlen]$, return $\LCP{\Text}[i]$; see
\cref{def:lcp-array,sec:lcp-problem-def}) in
$o(\tfrac{\log \Textlen}{\log \log \Textlen})$.
This lower bound is asymptotically tight, since LCP queries can be
answered in $\bigO(\RLBWTSize{\Text} \log^{1+\epsilon} \Textlen) =
\bigO(\SubstringComplexity{\Text} \log^{\bigO(1)} \Textlen)$ space and
$\bigO(\tfrac{\log \Textlen}{\log \log \Textlen})$ time (see
\cref{tab:query-bounds}).

It is known that there is no data structure that, for every
array $A[1 \dd m]$ containing a permutation of $\{1, \dots, m\}$,
uses near-linear (i.e., $\bigO(m \log^{\bigO(1)} m)$)
space and answers range counting or selection queries in
$o(\tfrac{\log m}{\log \log m})$ time~\cite{Patrascu07,JorgensenL11};
see \cref{th:range-count-lower-bound,th:range-select-lower-bound}.

The central idea in our reduction is to prove that if,
for every binary text $\Text \in \BinaryAlphabet^{\Textlen}$,
there exists a data structure using
$\bigO(\SubstringComplexity{\Text} \log^{\bigO(1)} \Textlen)$ space
and answering LCP array queries in
$t_{\rm LCP}(|\Text|) = o(\tfrac{\log |\Text|}{\log \log |\Text|})$
time, then we could break the lower bound for answering range
select queries mentioned above; i.e., given any array $A[1 \dd m]$
containing a permutation of $\{1, \dots, m\}$, there exists a data
structure of size $\bigO(m \log^{\bigO(1)} m)$ that answers
range selection queries on $A$ in $o(\tfrac{\log m}{\log \log m})$
time.

For an array $A[1 \dd m]$ containing a permutation of
$\{1, \dots, m\}$, we define the following string:
\[
  \Text_{A} =
    \Big(\textstyle\bigodot_{i=1}^{m} (\zero^{A[i]} \one^{i}) \Big) \cdot
    \zero^{m+1}\one^{m+1}
    \in \BinaryAlphabet^{*}.
\]

\begin{description}[style=sameline,itemsep=1ex,font={\normalfont\itshape}]

\item[Observation: The SA block corresponding to the suffixes of $\Text_{A}$
  prefixed with $\zero^{v}\one$ can be used to answer queries
  $\RangeSelect{A}{r}{v}$.]
  Recall that the query $\RangeSelect{A}{r}{v}$ returns the $r$th smallest
  element of the set $\mathcal{I} = \{i \in [1 \dd m] : A[i] \geq v\}$.
  Observe that, by definition of $\Text_{A}$, it follows that $\zero^{v}\one$
  occurs within the substring $\zero^{A[i]}\one^{i}$ of $\Text_{A}$ only when
  $A[i] \geq v$. This implies that the set $\OccTwo{\zero^{v}\one}{\Text_{A}}$
  captures precisely the set we are interested in (plus one extra occurrence within
  the suffix $\zero^{m+1}\one^{m+1}$ of $\Text_A$). Let $(a_i)_{i \in [1 \dd k]}$
  be the increasing sequence containing all the elements of $\mathcal{I}$,
  and let $a_{k+1} = m+1$. Observe that since
  $\zero^{v}\one^{a_1}\zero \prec
  \zero^{v}\one^{a_2}\zero \prec \dots \prec
  \zero^{v}\one^{a_k}\zero \prec \zero^{v}\one^{m+1}
  = \zero^{v}\one^{a_{k+1}}$, it follows that, for every $i \in [1 \dd k+1]$,
  the $i$th element in the SA block corresponding to
  $\OccTwo{\zero^{v}\one}{\Text_{A}}$ is prefixed with $\zero^{v}\one^{a_i}$
  but not with $\zero^{v}\one^{a_{i}+1}$; i.e., letting
  $b = \RangeBegTwo{\zero^{v}\one}{\Text_{A}}$, the suffix
  $\Text_{A}[\SA{\Text_{A}}[b + i] \dd |\Text_{A}|]$ is prefixed
  with $\zero^{v}\one^{a_i}$ but not $\zero^v\one^{a_i+1}$.
  By definition of the LCP array, this implies that,
  for every $r \in [1 \dd k]$, it holds
  \[
    \LCP{\Text_{A}}[b + r + 1] =
    \lcp{\Text_{A}[\SA{\Text_{A}}[b + r + 1] \dd |\Text_{A}|]}{\Text_{A}[\SA{\Text_{A}}[b + r] \dd |\Text_{A}|]} =
    v + a_{r}.
  \]
  Consequently, we can compute $a_r$ by asking the LCP query with index
  $b + r + 1$ and subtracting $v$ from the resulting value;
  see \cref{lm:reduce-range-select-to-lcp}.
  Since $\RangeSelect{A}{r}{v} = a_r$ holds by definition, we thus obtain a way
  to answer range select queries by LCP array queries on $\Text_{A}$.
\end{description}

To finish our argument, let us now assume that we can answer LCP array
queries on any $\Text \in \BinaryAlphabet^{*}$ in
$t_{\rm LCP}(|\Text|) = o(\tfrac{\log |\Text|}{\log \log |\Text|})$ time and
using $\bigO(\SubstringComplexity{\Text} \log^{c} |\Text|)$ space (for
some positive constant $c$). Using this structure, we design a
structure for answering range selection queries on $A[1 \dd m]$ so
that it consists of the following components:
\begin{enumerate}
\item An array $R[1 \dd m]$ defined so that
  $R[i] = \RangeBegTwo{\zero^i\one}{\Text_{A}}$. It needs $\bigO(m)$ space.
\item The above hypothetical data structure answering LCP queries on
  $\Text_{A}$. To bound the space for this structure, we need to show that
  $\SubstringComplexity{\Text_{A}} = \bigO(m \log^{\bigO(1)} m)$.
  To this end, observe that $\Text_{A}$ can be run-length compressed to size
  $2(m+1)$. This implies that the LZ77 factorization of $\Text_{A}$ has size
  $\bigO(m)$ (\cref{ob:rl}), and hence, by the bound
  $\SubstringComplexity{S} = \bigO(\LZSize{S})$
  (\cref{lm:z-and-r-upper-bound}\eqref{lm:z-and-r-upper-bound-it-1})
  holding for all strings $S$, we obtain that
  $\SubstringComplexity{\Text_{A}} = \bigO(m)$.  Note also that
  $|\Text_{A}| = \bigO(m^2)$.  Thus, the data structure answering LCP
  queries on $\Text_{A}$ needs
  $\bigO(\SubstringComplexity{\Text_{A}} \log^c |\Text_{A}|) =
  \bigO(m \log^c (m^2)) = \bigO(m \log^c m)$ space.
\end{enumerate}
In total, our new data structure for range select queries needs
$\bigO(m \log^c m) = \bigO(m \log^{\bigO(1)} m)$ space.

Let $v \geq 0$ and $r \in [1 \dd \TwoSidedRangeCount{A}{m}{v}]$. To compute
$\RangeSelect{A}{r}{v}$ using the above hypothetical structure, we proceed
as follows (assuming that $v \in [1 \dd m]$; otherwise, we can answer the query
immediately; see \cref{th:lcp-lower-bound}):
\begin{enumerate}
\item First, in $\bigO(1)$ time we fetch the value
  $b = R[v] = \RangeBegTwo{\zero^{v}\one}{\Text_{A}}$.
\item Then, in $\bigO(t_{\rm LCP}(|\Text_{A}|))$ time we compute
  $\ell = \LCP{\Text_{A}}[b + r + 1]$ and return $\ell - v$.
  By the above discussion, $\ell - v = \RangeSelect{A}{r}{v}$.
\end{enumerate}
In total, we spend
$\bigO(t_{\rm LCP}(|\Text_{A}|)) =
o(\tfrac{\log |\Text_{A}|}{\log \log |\Text_{A}|}) =
o(\tfrac{\log (m^2)}{\log \log (m^2)}) =
o(\tfrac{\log m}{\log \log m})$ time.

We have obtained a data structure answering range selection queries
in $o(\tfrac{\log m}{\log \log m})$ time and using $\bigO(m \log^{\bigO(1)} m)$
space. This contradicts the lower bound from~\cite{JorgensenL11}
(see also \cref{th:range-select-lower-bound}). Thus, the data structure
for LCP queries we assumed at the beginning cannot exist.

\subsection{Overview of the \texorpdfstring{\boldmath $\Omega(\log \log n)$}{Ω(log log n)} Lower Bounds}\label{sec:overview-fast}

In this section, we sketch the key ideas of our lower bounds for
faster queries, presented in \cref{sec:fast-queries}.

It is known that there is no data structure that, for every
$A \subseteq [1 \dd m^2]$ of size $|A| = m$, uses near-linear
(i.e., $\bigO(m \log^{\bigO(1)} m)$) space and answers
$\PredecessorColor{A}{x}$ queries in $o(\log \log m)$
time~\cite{PatrascuT06} (see also \cref{th:colored-predecessor-lower-bound}).
Note that this implies the same lower bound for any data structure that
returns the index of the predecessor (see \cref{th:predecessor-lower-bound})
since such a data structure could be used to answer $\PredecessorColor{A}{x}$
queries (it suffices to store the parity of each element's rank in an array
and return it after finding the index of the predecessor).

\paragraph{Overview of the Lower Bound for BWT Queries}

We now give an overview of our lower bound for BWT queries
presented in \cref{sec:bwt}. Specifically, we outline how to show
that there is no data structure that, for every
$\Text \in \BinaryAlphabet^{\Textlen}$,
uses $\bigO(\SubstringComplexity{\Text} \log^{\bigO(1)} \Textlen)$ space and
answers BWT queries (that, given any $i \in [1 \dd \Textlen]$, return
$\BWT{\Text}[i]$; see \cref{sec:bwt-problem-def})
in $o(\log \log \Textlen)$ time.
To this end, we will demonstrate that the existence of such a structure
implies that, for every set $A$ of $m$ integers in the range
$[1 \dd m^2]$, there exists a data structure of size
$\bigO(m \log^{\bigO(1)} m)$ that answers colored predecessor queries
(that, given any $x \in \Z$, return $\PredecessorColor{A}{x}$) in
$o(\log \log m)$ time, contradicting the above lower bound
(\cref{th:colored-predecessor-lower-bound}).

Consider any set $A \subseteq [1 \dd m^2]$ of size $|A| = m$. Denote
$A = \{a_1, \dots, a_m\}$ so that $a_1 < a_2 < \dots < a_m$. We also
define $a_0 = 0$ and $a_{m+1} = m^2$. Finally, let
$(b_i)_{i \in [0 \dd m]}$ be a sequence defined by $b_i = i \bmod 2$.
We then define the following string:
\[
  \Text_{A} =
    \textstyle\bigodot_{i=0}^{m}
      \big(b_i \cdot \ebin{k}{i} \big)^{a_{i+1}-a_{i}}
    \in \BinaryAlphabet^{*},
\]
where $k = 1 + \lfloor \log m \rfloor$, $\bin{k}{x}$ denotes a bitstring
of length exactly $k$ that contains the binary representation of integer
$x$, with leading zeros, and
$\ebin{k}{x} = \one^{k+1} \cdot \zero \cdot \bin{k}{x} \cdot \zero$.

\begin{description}[style=sameline,itemsep=1ex,font={\normalfont\itshape}]

\item[Observation: The BWT of $\Text_{A}$ can be used to answer
  queries $\PredecessorColor{A}{x}$.]  Consider the block
  $\SA{\Text_{A}}(b \dd e]$ in the suffix array corresponding to
  $\OccTwo{\one^{k+1}\zero}{\Text}$.  Note that $e - b = m^2$ since
  the string $\one^{k+1}\zero$ occurs only at positions of the form
  $(2k+4) \cdot t - (2k + 2)$ with $t \in [1 \dd m^2]$.  Moreover,
  observe that, for every $x_1, x_2 \in [0 \dd 2^k)$ with $x_1 < x_2$,
  we have $\ebin{k}{x_1} \prec \ebin{k}{x_2}$. By the definition of
  $\Text_{A}$, this implies that the first $a_1 - a_0$ suffixes in
  $\SA{\Text_{A}}(b \dd e]$ have $\ebin{k}{1}$ as a prefix, the next
  $a_2 - a_1$ suffixes have $\ebin{k}{2}$ as a prefix, and so on.
  Consequently, the first $a_1 - a_0$ symbols in $\BWT{\Text_{A}}(b
  \dd e]$ are $b_1 = \one$, the next $a_2 - a_1$ symbols are
  $b_2 = \zero$, and so on.  More generally, we obtain
  $\BWT{\Text_{A}}(b \dd e] = S$, where
  $S = b_0^{a_1-a_0}b_1^{a_2-a_1} \cdots b_m^{a_{m+1}-a_{m}} \in
  \BinaryAlphabet^{m^2}$. It remains to observe that $S$ is the
  answer string for $\PredecessorColor{A}{x}$ queries, i.e., for
  every $x \in [1 \dd m^2]$, it holds $\PredecessorColor{A}{x} = S[x]$.
  Thus, we obtain $\PredecessorColor{A}{x} = \BWT{\Text_{A}}[b + x]$.
\end{description}

Using the above observation, we develop our lower bound as follows.
Assume that we can answer BWT queries on any
$\Text \in \BinaryAlphabet^{\Textlen}$ in
$t_{\rm BWT}(|\Text|) = o(\log \log |\Text|)$ time and using
$\bigO(\SubstringComplexity{\Text} \log^{c} |\Text|)$ space
(for some positive constant $c$). Using this structure, we obtain a
structure for answering colored predecessor queries on $A$ by storing
the position $b = \RangeBegTwo{\one^{k+1}\zero}{\Text_{A}}$, and
the BWT structure for the string $\Text_{A}$.

To bound the space for the BWT structure for $\Text_{A}$, we need to
upper bound $\SubstringComplexity{\Text_{A}}$. To this end, first note
that the substring $(b_i \cdot \ebin{k}{i})^{a_{i+1}-a_i}$ has an
LZ77-like factorization (see \cref{def:lz77-like}) of size $2k+5$ (the
first $2k+4$ phrases are of length one, and the last phrase encodes
the remaining suffix). This implies that $\Text_{A}$ has an LZ77-like
factorization of size $(m+1)(2k+5)$. Since the LZ77 factorization is
the smallest LZ77-like factorization (\cref{th:lz77-size}), we thus
obtain $\LZSize{\Text_{A}} \leq (m+1)(2k+5) = \bigO(m \log m)$. By
\cref{lm:z-and-r-upper-bound}\eqref{lm:z-and-r-upper-bound-it-1}, we
obtain $\SubstringComplexity{\Text_{A}} = \bigO(m \log m)$. It remains
to observe that $|\Text_{A}| = \bigO(m^2 \log m)$. The data structure
for BWT queries applied on $\Text_{A}$ hence needs
\[
  \bigO(\SubstringComplexity{\Text_{A}} \log^c |\Text_{A}|)
    = \bigO((m \log m) \cdot \log^c (m^2 \log m))
    = \bigO(m \log^{c+1} m)
    = \bigO(m \log^{\bigO(1)} m)
\]
space. Given this data structure, we can compute
$\PredecessorColor{A}{x}$ for any $x \in [1 \dd m^2]$ (for other $x$,
we can return the answer immediately) by issuing a BWT query to
compute $\BWT{\Text_{A}}[b + x]$ in $\bigO(t_{\rm BWT}(|\Text_{A}|)) =
o(\log \log |\Text_{A}|) = o(\log \log (m^2 \log m)) = o(\log \log m)$
time.

We have obtained a data structure answering colored predecessor
queries in $o(\log \log m)$ time and using
$\bigO(m \log^{\bigO(1)} m)$ space. This contradicts the lower bound
from~\cite{PatrascuT06} (see also \cref{th:colored-predecessor-lower-bound}).
Thus, the data structure for BWT queries we assumed at the beginning
cannot exist.

\paragraph{Overview of the Lower Bound for PLCP Queries}

We now give an overview of a slightly different technique for proving
a lower bound for PLCP queries presented in
\cref{sec:plcp}. Specifically, we outline how to show that there is no
data structure that, for every $\Text \in \BinaryAlphabet^{\Textlen}$,
uses $\bigO(\SubstringComplexity{\Text} \log^{\bigO(1)} \Textlen)$
space and answers PLCP array queries (that, given any
$j \in [1 \dd \Textlen]$, return $\PLCP{\Text}[j]$; see
\cref{sec:plcp-problem-def}) in $o(\log \log \Textlen)$ time. To this
end, we will demonstrate that the existence of such a structure
implies that, for every set $A = \{a_1, a_2, \dots, a_m\}$ of $m$
integers (we assume $a_1 < a_2 < \dots < a_m$, and also set
$a_0 = -\infty$), in the range $[1 \dd m^2]$, there exists a data
structure of size $\bigO(m \log^{\bigO(1)} m)$ that, given any $x \in
\Z$, in $o(\log \log m)$ time returns the index $i \in [0 \dd m]$
satisfying $\Predecessor{A}{x} = a_i$. With an additional array whose
$i$th entry contains the value $a_i$, this allows determining
$\Predecessor{A}{x}$ in $o(\log \log m)$ time, contradicting the
predecessor lower bound from~\cite{PatrascuT06}
(\cref{th:predecessor-lower-bound}).

Consider any set $A \subseteq [1 \dd m^2]$ of size $|A| = m$. Denote
$A = \{a_1, \dots, a_m\}$, where $a_1 < a_2 < \dots < a_m$. Let also
$a_0 = -\infty$ and $a_{m+1} = m^2 + 1$. We then define the following
string:
\[
  \Text_{A} =
    \big( \textstyle\bigodot_{i=1}^{m} (\zero^{a_i}\one^{m-i+2}) \big) \cdot
    \big( \zero^{m^2+1} \one \big) \cdot
    \big( \zero^{m^2} \one^{m+2} \big)
    \in \BinaryAlphabet^{*}.
\]

\begin{description}[style=sameline,itemsep=1ex,font={\normalfont\itshape}]

\item[Observation: The PLCP array of $\Text_{A}$ can be used to answer
  $\Predecessor{A}{x}$ queries.]  Assume $x \in [2 \dd m^2]$ and
  consider the problem of computing $p = \min \{i \in [1 \dd m + 1] :
  a_i \geq x\}$; observe that $\Predecessor{A}{x} = a_{p-1}$ and
  $\Predecessor{A}{1}=-\infty$. Denote
  $\mathcal{I} = \{i \in [1 \dd m + 1] : a_i \geq x\}$, and note that,
  since $a_1 < a_2 < \dots < a_m < a_{m+1}$, we have $\mathcal{I} =
  \{p, p + 1, \dots, m+1\}$.  For every $i \in [1 \dd m + 1]$,
  $\zero^{x}\one$ occurs within the substring
  $\zero^{a_i}\one^{m-i+2}$ of $\Text_{A}$ if and only if
  $a_i \geq x$.  The set $\OccTwo{\zero^{x}\one}{\Text_{A}}$ thus
  captures the set $\mathcal{I}$ (plus one extra position
  corresponding to the occurrence of $\zero^{x}\one$ within the
  substring $\zero^{m^2}\one^{m+2}$). More precisely, (1) it holds
  $|\OccTwo{\zero^{x}\one}{\Text_{A}}| = m + 3 - p$, (2) for every
  $i \in \mathcal{I}$, there exists exactly one position in
  $\OccTwo{\zero^{x}\one}{\Text_{A}}$ having
  $\zero^{x}\one^{m-i+2}\zero$ as a prefix, and (3) there is exactly
  one position in $\OccTwo{\zero^{x}\one}{\Text_{A}}$ prefixed by
  $\zero^{x}\one^{m+2}$.  Since it holds
  \[
    \zero^{x}\one^{m+2}\zero \succ
    \zero^{x}\one^{m-p+2}\zero \succ
    \zero^{x}\one^{m-(p+1)+2}\zero \succ \dots \succ
    \zero^{x}\one^{m-(m+1)+2}\zero = \zero^x \one \zero,
  \]
  we conclude that, letting $b =
  \RangeBegTwo{\zero^{x}\one}{\Text_{A}}$, for every
  $i \in [1 \dd m-p+2]$, the suffix
  $\Text_{A}[\SA{\Text_{A}}[b + i] \dd |\Text_{A}|]$ has
  $\zero^{x}\one^{i}\zero$ as a prefix, and
  $\Text_{A}[\SA{\Text_{A}}[b + (m+3-p)] \dd |\Text_{A}|]$ has
  $\zero^{x}\one^{m+2}$ as a prefix. Since $\zero^{x}\one^{m+2}$ has
  only a single occurrence in $\Text_{A}$
  (at position $j = |\Text_A| - m - x -1$), we obtain that
  \[
    \PLCP{\Text_{A}}[j] = \lcp{\zero^{x}\one^{m+2}}{\zero^{x}\one^{m-p+2}\zero} = x+m-p+2.
  \]

  Consequently, letting $y = \PLCP{\Text_{A}}[j]$, it holds
  $p = (x + m + 2) - y$.  Recall that $\Predecessor{A}{x} = a_{p-1}$. Thus,
  letting $i = (x + m + 1) - y =
  (x + m + 1) - \PLCP{\Text_{A}}[|\Text_A| -m - x -1]$,
  we have $\Predecessor{A}{x} = a_i$, i.e., we have reduced the
  computation of the predecessor index to a PLCP query on $\Text_{A}$.
\end{description}

Using the above observation, we develop our lower bound as follows.
Assume that we can answer PLCP queries on any $\Text \in
\BinaryAlphabet^{\Textlen}$ in $t_{\rm PLCP}(|\Text|) =
o(\log \log |\Text|)$ time using
$\bigO(\SubstringComplexity{\Text} \log^{c} |\Text|)$ space
(for some positive constant $c$). Using this structure, we obtain a
structure for answering predecessor queries on $A$ by storing
$|\Text_{A}|$, $m$, and the PLCP structure for the string $\Text_{A}$.

To bound the space for the PLCP structure on $\Text_{A}$, note that
the string $\Text_{A}$ consists of exactly $2(m+1)$ equal-letter runs.
By \cref{ob:rl} and
\cref{lm:z-and-r-upper-bound}\eqref{lm:z-and-r-upper-bound-it-1}, we
obtain $\SubstringComplexity{\Text_{A}} =
\bigO(\LZSize{\Text_{A}}) = \bigO(m)$.
It remains to observe that $|\Text_{A}| = \bigO(m^3)$.
The data structure for PLCP queries on $\Text_{A}$ thus needs
$\bigO(\SubstringComplexity{\Text_{A}} \log^c |\Text_{A}|)
  = \bigO(m \log^c (m^3))
  = \bigO(m \log^{c+1} m)
  = \bigO(m \log^{\bigO(1)} m)$
space.
Given this data structure, we can compute $i \in [0 \dd m]$ satisfying
$\Predecessor{A}{x} = a_i$ for any $x \in [1 \dd m^2]$ (for other $x$,
we can return the answer immediately) by issuing a PLCP query to
compute $y = \PLCP{\Text_{A}}[|\Text_A|-m - x -1]$ in
$\bigO(t_{\rm PLCP}(|\Text_{A}|)) = o(\log \log |\Text_{A}|)
= o(\log \log (m^3))
= o(\log \log m)$ time, and returning $i = (x + m + 1) - y$.

We have thus obtained a data structure answering predecessor queries
in $o(\log \log m)$ time and using $\bigO(m \log^{\bigO(1)} m)$
space. This contradicts the lower bound from~\cite{PatrascuT06} (see
also \cref{th:predecessor-lower-bound}). Thus, the data structure for
PLCP queries we assumed at the beginning cannot exist.

\section{\boldmath Queries
  with \texorpdfstring{$\Theta(\tfrac{\log n}{\log \log n})$}{Θ(log n / log log n)} Complexity}\label{sec:slow-queries}

\subsection{Longest Common Prefix (LCP) Queries}\label{sec:lcp}

\subsubsection{Problem Definition}\label{sec:lcp-problem-def}
\vspace{-1.5ex}

\begin{framed}
  \noindent
  \probname{Indexing for Longest Common Prefix (LCP) Queries}
  \begin{bfdescription}
  \item[Input:]
    A string $\Text \in \Sigma^{\Textlen}$.
  \item[Output:]
    A data structure that, given any index $i \in [1 \dd \Textlen]$,
    returns $\LCP{\Text}[i]$ (\cref{def:lcp-array}).
  \end{bfdescription}
\end{framed}
\vspace{2ex}

\subsubsection{Lower Bound}\label{sec:lcp-lower-bound}

\begin{lemma}\label{lm:reduce-range-select-to-lcp}
  Let $A[1 \dd n]$ be a nonempty array containing a permutation
  of $\{1, 2, \dots, n\}$. Let
  \[
    \Text =
      \Big(\textstyle\bigodot_{i=1}^{n} (\zero^{A[i]} \one^{i}) \Big) \cdot
      \zero^{n+1}\one^{n+1}
      \in \BinaryAlphabet^{*}
  \]
  (brackets added for clarity). Then, for every $v \in [1 \dd n]$
  and every $r \in [1 \dd \TwoSidedRangeCount{A}{n}{v}]$
  (see \cref{def:range-count-and-select}), it holds
  \[
    \RangeSelect{A}{r}{v} = \LCP{\Text}[b + r + 1] - v,
  \]
  where $b = \RangeBegTwo{\zero^{v}\one}{\Text}$ (\cref{def:occ}).
\end{lemma}
\begin{proof}
  Denote $k = \TwoSidedRangeCount{A}{n}{v}$. Let $(a_i)_{i \in [1 \dd k+1]}$
  be a sequence such that, for every $i \in [1 \dd k]$,
  \[
    a_i = \RangeSelect{A}{i}{v},
  \]
  and $a_{k+1} = n+1$.
  By definition, we then have $a_1 < a_2 < \dots < a_k < a_{k+1}$
  and $\RangeSelect{A}{r}{v} = a_r$. Next, let $(s_i)_{i \in [1 \dd k+1]}$
  be a sequence defined such that, for every $i \in [1 \dd k]$,
  \[
    s_i = \Big(\textstyle\sum_{t=1}^{a_i-1} (A[t]+t)\Big) +
          \Big(A[a_i]-v+1\Big),
  \]
  and $s_{k+1} = \big( \sum_{t=1}^{n}(A[t]+t) \big) + (n+2-v)$.
  Observe, that there is exactly $k+1$ occurrences of the substring
  $\zero^v \one$ in $\Text$, and it holds
  $\OccTwo{\zero^v \one}{\Text} = \{s_i\}_{i \in [1 \dd k+1]}$.
  Note also that for every $i \in [1 \dd k]$, the suffix
  $\Text[s_i \dd |\Text|]$ has the string $\zero^v \one^{a_i} \zero$
  as a prefix. Moreover, $\Text[s_{k+1} \dd |\Text|]$
  has $\zero^{v} \one^{n+1}$ as a prefix.
  By $a_1 < a_2 < \dots < a_k < a_{k+1}$, we thus have
  \[
    \Text[s_1 \dd |\Text|] \prec
    \Text[s_2 \dd |\Text|] \prec \dots \prec
    \Text[s_{k} \dd |\Text|] \prec
    \Text[s_{k+1} \dd |\Text|].
  \]
  Consequently, by definition of $b = \RangeBegTwo{\zero^{v}\one}{\Text}$
  and \cref{rm:occ}, for every $i \in [1 \dd k+1]$, it
  holds
  \[
    \SA{\Text}[b+i] = s_{i}.
  \]
  Lastly, note that $\zero^{v} \one^{a_i} \zero$ being a prefix of
  $\Text[s_i \dd |\Text|]$ for $i \in [1 \dd k]$, and $\zero^{v} \one^{n+1}$
  being a prefix of $\Text[s_{k+1} \dd |\Text|]$ imply that, for every
  $i \in [1 \dd k]$, it holds
  $\LCE{\Text}{s_{i}}{s_{i+1}} = v + a_{i}$.
  Putting everything together, we thus obtain
  \begin{align*}
    \LCP{\Text}[b+r+1]
      &= \LCE{\Text}{\SA{\Text}[b+r]}{\SA{\Text}[b+r+1]}\\
      &= \LCE{\Text}{s_{r}}{s_{r+1}}\\
      &= v + a_{r} \\
      &= v + \RangeSelect{A}{r}{v}.
  \end{align*}
  which is equivalent to the claim.
\end{proof}

\begin{theorem}\label{th:lcp-lower-bound}
  There is no data structure that, for every text $\Text \in \BinaryAlphabet^{\Textlen}$,
  uses $\bigO(\SubstringComplexity{\Text} \log^{\bigO(1)}
  \Textlen)$ space and answers LCP array queries on $\Text$ (that,
  given any $i \in [1 \dd \Textlen]$, return $\LCP{\Text}[i]$; see
  \cref{def:lcp-array}) in $o(\tfrac{\log \Textlen}{\log \log \Textlen})$
  time.
\end{theorem}
\begin{proof}

  Suppose that the claim does not hold. Let $D_{\rm LCP}$ denote the
  hypothetical data structure answering LCP queries, and let
  $c = \bigO(1)$ be a positive constant such that, that for
  a text $\Text \in \BinaryAlphabet^{\Textlen}$, $D_{\rm LCP}$ uses
  $\bigO(\SubstringComplexity{\Text} \log^c \Textlen)$ space, and
  answers LCP queries on $\Text$ in $t_{\rm LCP}(|\Text|) =
  o(\tfrac{\log |\Text|}{\log \log |\Text|})$ time.  We will prove
  that this implies that there exists a data structure answering range
  selection queries that contradicts
  \cref{th:range-select-lower-bound}.

  Consider any array $A[1 \dd n]$ containing a permutation of
  $\{1, 2, \dots, n\}$, where $n \geq 1$.
  Let
  $\Text_{A} = (\bigodot_{i=1}^{n}(\zero^{A[i]}\one^{i})) \cdot
  (\zero^{n+1}\one^{n+1}) \in
  \BinaryAlphabet^{*}$ be a text defined as in
  \cref{lm:reduce-range-select-to-lcp}. Let $R[1 \dd n]$ be array
  defined so that, for every $i \in [1 \dd n]$, $R[i] =
  \RangeBegTwo{\zero^{i}\one}{\Text_{A}}$ (\cref{def:occ}).

  Let $D_{\rm select}$ denote the data structure consisting of the
  following components:
  \begin{enumerate}
  \item The array $R[1 \dd n]$ stored in plain form. It uses
    $\bigO(n)$ space.
  \item The data structure $D_{\rm LCP}$ for the string $\Text_{A}$.
    Note that
    \[
      |\Text_{A}|
        = (\textstyle\sum_{i=1}^{n}(A[i]+i)) + 2(n+1)
        = (n+2)(n+1) \in \bigO(n^2).
    \]
    Moreover, note that $|\RL{\Text_{A}}| = 2(n+1)$. By applying
    \cref{ob:rl} and
    \cref{lm:z-and-r-upper-bound}\eqref{lm:z-and-r-upper-bound-it-1},
    we thus obtain $\SubstringComplexity{\Text_{A}} = \bigO(n)$.
    Consequently, $D_{\rm LCP}$ for $\Text_{A}$ needs
    $\bigO(\SubstringComplexity{\Text_{A}} \log^c |\Text_{A}|) =
    \bigO(n \log^c n)$ space.
  \end{enumerate}
  In total, $D_{\rm select}$ needs $\bigO(n \log^c n)$ space.

  Given any $v \geq 0$ and $r \in [1 \dd \TwoSidedRangeCount{A}{n}{v}]$,
  we compute $\RangeSelect{A}{r}{v}$ as follows:
  \begin{enumerate}
  \item Note that since $A[1 \dd n]$ contains a permutation
    of $\{1, \dots, n\}$, the assumption
    $r \in [1 \dd \TwoSidedRangeCount{A}{n}{v}]$
    implies that $v \leq n$ (otherwise,
    $\TwoSidedRangeCount{A}{n}{v} = 0$). Additionally, note that if
    $v = 0$, then we can set $v = 1$ without changing the answer.
    Let us thus assume that $v \in [1 \dd n]$. First,
    using array $R$, in $\bigO(1)$ time we compute
    $b = \RangeBegTwo{\zero^{v}\one}{\Text_{A}} = R[v]$.
  \item Using $D_{\rm LCP}$, in $\bigO(t_{\rm LCP}(|\Text_{A}|))$
    time we compute $\ell = \LCP{\Text_{A}}[b + r + 1]$.
    By \cref{lm:reduce-range-select-to-lcp}, it holds
    $\RangeSelect{A}{r}{v} = \ell - v$. Thus, we return
    $\ell - v$ as the answer.
  \end{enumerate}
  In total, the query time is
  \[
    \bigO(t_{\rm LCP}(|\Text_{A}|))
     = o\Big(\tfrac{\log |\Text_{A}|}{\log \log |\Text_{A}|}\Big)
     = o\Big(\tfrac{\log (n^2)}{\log \log (n^2)}\Big)
     = o\Big(\tfrac{\log n}{\log \log n}\Big).
  \]

  We have thus proved that there exists a data structure that,
  for every array $A[1 \dd n]$ containing a permutation of $\{1, \dots, n\}$,
  uses $\bigO(n \log^c n) = \bigO(n \log^{\bigO(1)} n)$ space, and answers
  range selection queries on $A$ in $o(\tfrac{\log n}{\log \log n})$ time.
  This contradicts \cref{th:range-select-lower-bound}.
\end{proof}

\subsection{Suffix Array (SA) Queries}\label{sec:sa}

\subsubsection{Problem Definition}\label{sec:sa-problem-def}
\vspace{-1.5ex}

\begin{framed}
  \noindent
  \probname{Indexing for Suffix Array (SA) Queries}
  \begin{bfdescription}
  \item[Input:]
    A string $\Text \in \Sigma^{\Textlen}$.
  \item[Output:]
    A data structure that, given any index $i \in [1 \dd \Textlen]$,
    returns $\SA{\Text}[i]$ (\cref{def:sa}).
  \end{bfdescription}
\end{framed}
\vspace{2ex}

\subsubsection{Lower Bound}\label{sec:sa-lower-bound}

\begin{theorem}\label{th:sa-lower-bound}
  There is no data structure that, for every text $\Text \in \BinaryAlphabet^{\Textlen}$,
  uses $\bigO(\SubstringComplexity{\Text} \log^{\bigO(1)}
  \Textlen)$ space and answers suffix array queries on $\Text$ (that,
  given any $i \in [1 \dd \Textlen]$, return $\SA{\Text}[i]$; see
  \cref{def:sa}) in $o(\tfrac{\log \Textlen}{\log \log \Textlen})$
  time.
\end{theorem}
\begin{proof}

  Suppose that the claim does not hold. Let $D_{\rm SA}$ denote the
  hypothetical data structure answering suffix array queries, and let
  $c = \bigO(1)$ be a positive constant such that, that for
  a text $\Text \in \BinaryAlphabet^{\Textlen}$, $D_{\rm SA}$ uses
  $\bigO(\SubstringComplexity{\Text} \log^c \Textlen)$ space, and
  answers suffix array queries on $\Text$ in $t_{\rm SA}(|\Text|) =
  o(\tfrac{\log |\Text|}{\log \log |\Text|})$ time. We will prove
  that this implies that there exists a data structure answering LCP
  array queries that contradicts \cref{th:lcp-lower-bound}.

  Let $\Text \in \BinaryAlphabet^{\Textlen}$ be a nonempty text.

  Let $D_{\rm LCP}$ denote the data structure consisting of the
  following components:
  \begin{enumerate}
  \item The data structure from \cref{th:plcp-upper-bound} for
    text $\Text$. It uses $\bigO(\RLBWTSize{\Text}) =
    \bigO(\SubstringComplexity{\Text} \log^2 \Textlen)$
    space (see \cref{th:rlbwt-size,rm:rlbwt-size}).
  \item The data structure $D_{\rm SA}$ for the string $\Text$. It
    uses $\bigO(\SubstringComplexity{\Text} \log^c \Textlen)$ space.
  \end{enumerate}
  In total, $D_{\rm LCP}$ needs $\bigO(\SubstringComplexity{\Text} \log^c \Textlen + 
  \SubstringComplexity{\Text} \log^2 \Textlen)$ space.

  Given any $i \in [1 \dd \Textlen]$, we compute $\LCP{\Text}[i]$
  (\cref{def:lcp-array}) as follows:
  \begin{enumerate}
  \item Using $D_{\rm SA}$, in $\bigO(t_{\rm SA}(\Textlen))$ time
    we compute $j = \SA{\Text}[i]$.
  \item Using the structure from \cref{th:plcp-upper-bound}, in
    $\bigO(\log \log \Textlen)$ time we compute and return
    $x = \PLCP{\Text}[j]$. Note that by \cref{def:plcp-array},
    it holds $x = \PLCP{\Text}[j] = \PLCP{\Text}[\SA{\Text}[i]] = \LCP{\Text}[i]$.
  \end{enumerate}
  In total, the query time is
  \[
   \bigO(\log \log n + t_{\rm SA}(\Textlen))
     = \bigO(\log \log n) + o\Big(\tfrac{\log \Textlen}{\log \log \Textlen}\Big)
     = o\Big(\tfrac{\log n}{\log \log n}\Big).
  \]

  We have thus proved that there exists a data structure that,
  for every text $\Text \in \BinaryAlphabet^{\Textlen}$
  uses $\bigO(\SubstringComplexity{\Text} \log^c \Textlen + \SubstringComplexity{\Text} \log^2 \Textlen)
  = \bigO(\SubstringComplexity{\Text} \log^{\bigO(1)} \Textlen)$ space, and
  answers LCP array queries on $\Text$ in $o(\tfrac{\log n}{\log \log n})$ time.
  This contradicts \cref{th:lcp-lower-bound}.
\end{proof}

\subsection{\boldmath Inverse Suffix Array (\texorpdfstring{$\text{SA}^{-1}$}{SA⁻¹}) Queries}\label{sec:isa}

\subsubsection{Problem Definition}\label{sec:isa-problem-def}
\vspace{-1.5ex}

\begin{framed}
  \noindent
  \probname{Indexing for Inverse Suffix Array ($\text{SA}^{-1}$) Queries}
  \begin{bfdescription}
  \item[Input:]
    A string $\Text \in \Sigma^{\Textlen}$.
  \item[Output:]
    A data structure that, given any index $j \in [1 \dd \Textlen]$,
    returns $\ISA{\Text}[j]$ (\cref{def:isa}).
  \end{bfdescription}
\end{framed}
\vspace{2ex}

\subsubsection{Lower Bound}\label{sec:isa-lower-bound}

\begin{observation}\label{ob:random-access-via-isa}
  Let $\Text \in \BinaryAlphabet^{\Textlen}$ be a nonempty text. Denote $n_0 = |\{j \in [1 \dd \Textlen] : \Text[j] = \zero\}|$.
  For every $j \in [1 \dd \Textlen]$, $\Text[j] = \zero$ holds if and only if $\ISA{\Text}[j] \leq n_0$ (\cref{def:isa}).
\end{observation}

\begin{theorem}\label{th:isa-lower-bound}
  There is no data structure that, for every text $\Text \in \BinaryAlphabet^{\Textlen}$,
  uses $\bigO(\SubstringComplexity{\Text} \log^{\bigO(1)} \Textlen)$ space and answers inverse SA queries on $\Text$
  (that is, given any $i \in [1 \dd \Textlen]$, returns $\ISA{\Text}[i]$; see \cref{def:isa}) in 
  $o(\tfrac{\log \Textlen}{\log \log \Textlen})$ time.
\end{theorem}
\begin{proof}

  Suppose the claim does not hold. Let $D_{\rm ISA}$ denote the
  hypothetical data structure answering inverse SA queries, and let
  $c = \bigO(1)$ be a positive constant such that for
  a text $\Text \in \BinaryAlphabet^{\Textlen}$, $D_{\rm ISA}$ uses
  $\bigO(\SubstringComplexity{\Text} \log^c \Textlen)$ space, and
  answers inverse SA queries on $\Text$ in time 
  $t_{\rm ISA}(|\Text|) = o(\tfrac{\log |\Text|}{\log \log |\Text|})$. 
  We will show that this implies the existence of a data structure
  answering random access queries, contradicting
  \cref{th:random-access-lower-bound}.

  Let $\Text \in \BinaryAlphabet^{\Textlen}$, where $\Textlen \geq 1$,
  and let $n_0 = |\{j \in [1 \dd \Textlen] : \Text[j] = \zero\}|$.

  Let $D_{\rm access}$ denote the data structure consisting of the
  following components:
  \begin{enumerate}
  \item The integer $n_0$, stored in $\bigO(1)$ space.
  \item The data structure $D_{\rm ISA}$ for the string $\Text$,
    using $\bigO(\SubstringComplexity{\Text} \log^c \Textlen)$ space.
  \end{enumerate}
  In total, $D_{\rm access}$ uses
  $\bigO(\SubstringComplexity{\Text} \log^c \Textlen)$ space.

  Given any $j \in [1 \dd \Textlen]$, we compute $\Text[j]$ as follows:
  \begin{enumerate}
  \item Use $D_{\rm ISA}$ to compute $i = \ISA{\Text}[j]$ in
    $\bigO(t_{\rm ISA}(\Textlen))$ time.
  \item Return $\Text[j] = \zero$ if $i \leq n_0$ and $\Text[j] = \one$ otherwise.
    This is correct by \cref{ob:random-access-via-isa}.
  \end{enumerate}
  The total query time is $\bigO(t_{\rm ISA}(\Textlen)) =
  o(\tfrac{\log \Textlen}{\log \log \Textlen})$.

  We have therefore constructed a data structure that, for every 
  $\Text \in \BinaryAlphabet^{\Textlen}$, uses 
  $\bigO(\SubstringComplexity{\Text} \log^c \Textlen) =
  \bigO(\SubstringComplexity{\Text} \log^{\bigO(1)} \Textlen)$ space and answers
  random access queries in $o(\tfrac{\log \Textlen}{\log \log \Textlen})$ time.
  This contradicts \cref{th:random-access-lower-bound}.
\end{proof}

\subsection{Longest Common Extension (LCE) Queries}\label{sec:lce}

\subsubsection{Problem Definition}\label{sec:lce-problem-def}
\vspace{-1.5ex}

\begin{framed}
  \noindent
  \probname{Indexing for Longest Common Extension (LCE) Queries}
  \begin{bfdescription}
  \item[Input:]
    A string $\Text \in \Sigma^{\Textlen}$.
  \item[Output:]
    A data structure that, given any $i,j \in [1 \dd \Textlen]$,
    returns $\LCE{\Text}{i}{j}$ (\cref{sec:prelim-basic}).
  \end{bfdescription}
\end{framed}
\vspace{2ex}

\subsubsection{Lower Bound}\label{sec:lce-lower-bound}

\begin{lemma}\label{lm:delta-after-appending}
  Let $\Sigma$ be a nonempty set. For every nonempty text
  $\Text \in \Sigma^{+}$ and every $c \in \Sigma$, it holds
  $\SubstringComplexity{\Text'} \leq \SubstringComplexity{\Text} + 1$
  (see \cref{def:delta}),
  where $\Text' = \Text \cdot c$.
\end{lemma}
\begin{proof}
  Appending a symbol to $\Text$ creates at most one new 
  substring of any given length, i.e., for every $\ell \in \Zp$, it holds
  $\SubstrCount{\ell}{\Text'} \leq \SubstrCount{\ell}{\Text} + 1$.
  This implies that
  \begin{align*}
    \SubstringComplexity{\Text'}
      &= \textstyle\max_{\ell=1}^{|\Text|+1} \tfrac{1}{\ell} \SubstrCount{\ell}{\Text'}\\
      &\leq \textstyle\max_{\ell=1}^{|\Text|+1} \tfrac{1}{\ell} (\SubstrCount{\ell}{\Text} + 1)\\
      &\leq \textstyle\max_{\ell=1}^{|\Text|+1} (\tfrac{1}{\ell} \SubstrCount{\ell}{\Text} + 1)\\
      &= \textstyle 1 + \max_{\ell=1}^{|\Text|+1} \tfrac{1}{\ell} \SubstrCount{\ell}{\Text}\\
      &= \textstyle 1 + \max_{\ell=1}^{|\Text|} \tfrac{1}{\ell} \SubstrCount{\ell}{\Text}\\
      &= 1 + \SubstringComplexity{\Text}. \qedhere
  \end{align*}
\end{proof}

\begin{theorem}\label{th:lce-lower-bound}
  There is no data structure that, for every text $\Text \in \BinaryAlphabet^{\Textlen}$,
  uses $\bigO(\SubstringComplexity{\Text} \log^{\bigO(1)}
  \Textlen)$ space and answers LCE queries on $\Text$ (that,
  given any $i,j \in [1 \dd \Textlen]$, return $\LCE{\Text}{i}{j}$; see
  \cref{sec:prelim-basic}) in $o(\tfrac{\log \Textlen}{\log \log \Textlen})$
  time.
\end{theorem}
\begin{proof}

  Suppose that the claim does not hold. Let $D_{\rm LCE}$ denote the
  hypothetical data structure answering LCE queries, and let
  $c = \bigO(1)$ be a positive constant such that, that for
  a text $\Text \in \BinaryAlphabet^{\Textlen}$, $D_{\rm LCE}$ uses
  $\bigO(\SubstringComplexity{\Text} \log^c \Textlen)$ space, and
  answers LCE queries on $\Text$ in $t_{\rm LCE}(|\Text|) =
  o(\tfrac{\log |\Text|}{\log \log |\Text|})$ time.  We will prove
  that this implies that there exists a data structure answering random
  access queries to $\Text$ that contradicts
  \cref{th:random-access-lower-bound}.

  Let $\Text \in \BinaryAlphabet^{\Textlen}$ be a nonempty text.
  Denote $\Text' = \zero \cdot \Text \in \BinaryAlphabet^{\Textlen + 1}$.

  Let $D_{\rm access}$ denote the data structure consisting of a
  single component: the data structure $D_{\rm LCE}$ for the string
  $\Text'$. Note that $|\Text'| = \Textlen + 1$. Next, we give
  an upper bound for $\SubstringComplexity{\Text'}$. To this end,
  note that, by \cref{def:delta},
  for every string $S$, it holds $\SubstringComplexity{S} = \SubstringComplexity{\revstr{S}}$
  (where $\revstr{S}$ denotes the reverse of $S$; see \cref{sec:prelim-basic}).
  By \cref{lm:delta-after-appending}, we thus have
  \[
    \SubstringComplexity{\Text'}
      = \SubstringComplexity{\revstr{\Text'}}
      = \SubstringComplexity{\revstr{\Text} \cdot \zero}
      = 1 + \SubstringComplexity{\revstr{\Text}}
      = 1 + \SubstringComplexity{\Text}.
  \]
  Thus, $D_{\rm LCE}$ for $\Text'$ (and hence also $D_{\rm access}$) needs
  $\bigO(\SubstringComplexity{\Text'} \log^{c} |\Text'|)
  = \bigO(\SubstringComplexity{\Text} \log^{c} \Textlen)$ space.

  Given any $j \in [1 \dd \Textlen]$, we compute $\Text[j]$ as follows:
  \begin{enumerate}
  \item Using $D_{\rm LCE}$, in $\bigO(t_{\rm LCE}(|\Text'|))$
    time we compute $\ell = \LCE{\Text'}{1}{j+1}$.
  \item If $\ell \geq 1$, we return $\Text[j] = \zero$. Otherwise,
    we return that $\Text[j] = \one$.
  \end{enumerate}
  In total, the query time is
  \[
    \bigO(t_{\rm LCE}(|\Text'|))
     = o\Big(\tfrac{\log |\Text'|}{\log \log |\Text'|}\Big)
     = o\Big(\tfrac{\log (\Textlen+1)}{\log \log (\Textlen+1)}\Big)
     = o\Big(\tfrac{\log \Textlen}{\log \log \Textlen}\Big).
  \]

  We have thus proved that there exists a data structure that,
  for every $\Text \in \BinaryAlphabet^{\Textlen}$
  uses $\bigO(\SubstringComplexity{\Text} \log^c \Textlen) =
  \bigO(\SubstringComplexity{\Text} \log^{\bigO(1)} \Textlen)$ space, and answers
  random access queries on $\Text$ in $o(\tfrac{\log \Textlen}{\log \log \Textlen})$ time.
  This contradicts \cref{th:random-access-lower-bound}.
\end{proof}

\subsubsection{Upper Bound}\label{sec:lce-upper-bound}

\begin{theorem}\label{th:lce-upper-bound-r}
  Let $\epsilon \in (0,1)$ be a constant.  For every nonempty text
  $\Text \in \Sigma^{\Textlen}$, there exists a data structure of size
  $\bigO(\RLBWTSize{\Text} \log^{2+\epsilon} \Textlen)$
  (see \cref{def:bwt}) that answers LCE queries (that, given any
  $i,j \in [1 \dd \Textlen]$, return $\LCE{\Text}{i}{j}$;
  see \cref{sec:prelim-basic}) in
  $\bigO\big(\tfrac{\log \Textlen}{\log \log \Textlen}\big)$ time.
\end{theorem}
\begin{proof}

  The data structure consists of the following components:
  \begin{enumerate}
  \item The data structure from \cref{th:lcp-rmq-upper-bound} for text $\Text$.
    It needs $\bigO(\RLBWTSize{\Text} \log^{2+\epsilon} \Textlen)$ space.
  \item The data structure from \cref{th:rindex} for text $\Text$.
    It needs $\bigO(\RLBWTSize{\Text} \log^{1+\epsilon} \Textlen)$ space.
  \end{enumerate}
  In total, the data structure needs
  $\bigO(\RLBWTSize{\Text} \log^{2+\epsilon} \Textlen)$ space.

  \DSQueries
  Let $i,j \in [1 \dd \Textlen]$. To compute $\LCE{\Text}{i}{j}$, we proceed
  as follows:
  \begin{enumerate}
  \item If $i = j$, then we return that
    $\LCE{\Text}{i}{j} = \Textlen - i + 1$. Let us thus assume that $i \neq j$.
  \item Using \cref{th:rindex}, in $\bigO(\tfrac{\log \Textlen}{\log \log \Textlen})$ time
    we compute $b := \ISA{\Text}[i]$ and $e = \ISA{\Text}[j]$. If $b > e$, then we
    swap $b$ and $e$. We then set $b := b - 1$.
  \item Using \cref{th:lcp-rmq-upper-bound}, in $\bigO(\tfrac{\log \Textlen}{\log \log \Textlen})$ time
    we compute $i_{\rm min} = \argmin_{t \in (b \dd e]} \LCP{\Text}[t]$.
  \item Using \cref{th:rindex}, in $\bigO(\tfrac{\log \Textlen}{\log \log \Textlen})$ time
    we compute and return $\ell = \LCP{\Text}[i_{\rm min}]$. By definition of suffix and LCP array,
    it holds $\ell = \LCE{\Text}{i}{j}$.
  \end{enumerate}
  In total, the query algorithm takes $\bigO(\tfrac{\log \Textlen}{\log \log \Textlen})$ time.
\end{proof}

\begin{theorem}\label{th:lce-upper-bound-delta}
  Let $\epsilon \in (0,1)$ be a constant.  For every nonempty text
  $\Text \in \Sigma^{\Textlen}$, there exists a data structure of size
  $\bigO(\SubstringComplexity{\Text} \log^{4+\epsilon} \Textlen)$
  (see \cref{def:delta}) that answers LCE queries (that, given any
  $i,j \in [1 \dd \Textlen]$, return $\LCE{\Text}{i}{j}$;
  see \cref{sec:prelim-basic}) in
  $\bigO\big(\tfrac{\log \Textlen}{\log \log \Textlen}\big)$ time.
\end{theorem}
\begin{proof}
  The result follows by combining \cref{th:lce-upper-bound-r} and
  \cref{th:rlbwt-size} (see also \cref{rm:rlbwt-size}).
\end{proof}

\section{\boldmath Queries with \texorpdfstring{$\Theta(\log \log n)$}{Θ(log log n)} Complexity}\label{sec:fast-queries}

\subsection{Burrows--Wheeler Transform (BWT) Queries}\label{sec:bwt}

\subsubsection{Problem Definition}\label{sec:bwt-problem-def}
\vspace{-1.5ex}

\begin{framed}
  \noindent
  \probname{Indexing for Burrows--Wheeler Transform (BWT) Queries}
  \begin{bfdescription}
  \item[Input:]
    A string $\Text \in \Sigma^{\Textlen}$.
  \item[Output:]
    A data structure that, given any index $i \in [1 \dd \Textlen]$,
    returns $\BWT{\Text}[i]$ (\cref{def:bwt}).
  \end{bfdescription}
\end{framed}
\vspace{2ex}

\subsubsection{Lower Bound}\label{sec:bwt-lower-bound}

\begin{definition}[Binary mapping]\label{def:bin}
  For any $k \in \Zp$ and $x \in [0 \dd 2^k)$, by $\bin{k}{x} \in
    \BinaryAlphabet^{k}$ we denote the length-$k$ string containing
    the binary representation of $x$ (with leading zeros).
\end{definition}

\begin{definition}[Extended binary mapping]\label{def:ebin}
  Let $k \in \Zp$. For every $x \in [0 \dd 2^k)$, define
  \[
    \ebin{k}{a} := \one^{k+1} \cdot \zero \cdot \bin{k}{x} \cdot \zero,
  \]
  where $\bin{k}{x}$ is as in \cref{def:bin}.
\end{definition}

\begin{lemma}\label{lm:reduce-colored-predecessor-to-bwt}
  Let $A \subseteq [1 \dd m]$ be a nonempty set of size $|A| = m$.
  Denote $A = \{a_1, a_2, \dots, a_m\}$, where $a_1 < a_2 < \dots <
  a_m$, and let $a_0 = 0$ and $a_{m+1} = m^2$. Let $(b_i)_{i=0}^{m}$
  be a sequence defined by $b_i = i \bmod 2$. Let (see \cref{def:ebin})
  \[
    \Text = \bigodot_{i=0}^{m}
      \Big(b_i \cdot \ebin{k}{i} \Big)^{a_{i+1}-a_{i}}
      \in \BinaryAlphabet^{*}
  \]
  (brackets added for clarity), where $k = 1 + \lfloor \log m
  \rfloor$.  Then, for every $x \in [1 \dd m^2]$, it holds (see
  \cref{def:colored-predecessor,def:bwt})
  \[
    \PredecessorColor{A}{x} = \BWT{\Text}[b + x],
  \]
  where $b = \RangeBegTwo{\one^{k+1}\zero}{\Text}$ (see
  \cref{def:occ}).
\end{lemma}
\begin{proof}

  Denote
  \[
    S =
      b_0^{a_1-a_0} \cdot
      b_1^{a_2-a_1} \cdot
      b_2^{a_3-a_2} \cdot \ldots \cdot
      b_{m-1}^{a_m-a_{m-1}} \cdot
      b_m^{a_{m+1}-a_{m}}
      \in \BinaryAlphabet^{m^2}.
  \]

  The proof consists of two steps:
  \begin{enumerate}

  \item First, we prove that it holds $\BWT{\Text}(b \dd b + m^2] =
    S$. Let $\delta = 2k + 4$.  For every $i \in [0 \dd m]$, we define
    \[
      P_i = \{\delta \cdot t - (2k+2) : t \in (a_{i} \dd a_{i+1}]\}.
    \]
    Let also $P = \bigcup_{i=0}^{m} P_i = \{\delta \cdot t - (2k+2) :
    t \in [1 \dd m^2]\}$.  Observe that, it holds
    \[
      \OccTwo{\one^{k+1}\zero}{\Text} = P.
    \]
    On the other hand, note that since for every $i \in [0 \dd m]$ and
    every $x \in P_i$, the suffix $\Text[x \dd |\Text|]$ has the
    string $\one^{k+1} \cdot \zero \cdot \bin{k}{i}$ as a prefix.
    This implies that if $i_1, i_2 \in [0 \dd m]$ are such that $i_1 <
    i_2$, then for every $x_1 \in P_{i_1}$ and $x_2 \in P_{i_2}$, it holds
    $\Text[x_1 \dd |\Text|] \prec \Text[x_2 \dd |\Text|]$.  Recall now
    that, by $b = \RangeBegTwo{\one^{k+1}\zero}{\Text}$ and
    \cref{def:occ}, it holds $\{\SA{\Text}[b + t] : t \in [1 \dd
    m^2]\} = P$. Combining this with the above observation about
    lexicographical order of suffixes starting at positions in $P_{i_1}$
    and $P_{i_2}$, we thus obtain that, for every $i \in [0 \dd m]$, it
    holds
    \[
      \{\SA{\Text}[b + t] : t \in (a_i \dd a_{i+1}]\} = P_i.
    \]
    Since for every $i \in [0 \dd m]$ and $x \in P_i$, we have
    $\Text[x - 1] = b_i$, it follows by \cref{def:bwt} that, for every
    $i \in [0 \dd m]$,
    \[
      \BWT{\Text}(b + a_i \dd b + a_{i+1}] = b_i^{a_{i+1}-a_{i}}.
    \]
    Hence, we obtain that it holds $\BWT{\Text}(b \dd b + m^2] =
    b_0^{a_1-a_0} \cdot b_1^{a_2-a_1} \cdot \ldots \cdot
    b_m^{a_{m+1}-a_{m}} = S$.

  \item In the second step, we observe that, by
    \cref{def:colored-predecessor}, for every $x \in [1 \dd m^2]$, it
    holds
    \[
      \PredecessorColor{A}{x} = S[x].
    \]
    Combining this observation with the above step, we thus obtain the
    claim.
    \qedhere
  \end{enumerate}
\end{proof}

\begin{theorem}\label{th:bwt-lower-bound}
  There is no data structure that, for every text $\Text \in \BinaryAlphabet^{\Textlen}$,
  uses $\bigO(\SubstringComplexity{\Text} \log^{\bigO(1)}
  \Textlen)$ space and answers BWT queries on $\Text$ (that,
  given any $i \in [1 \dd \Textlen]$, return $\BWT{\Text}[i]$; see
  \cref{def:bwt}) in $o(\log \log \Textlen)$ time.
\end{theorem}
\begin{proof}

  Suppose that the claim does not hold. Let $D_{\rm BWT}$ denote the
  hypothetical data structure answering BWT queries, and let $c
  = \bigO(1)$ be a positive constant such that, that for
  a text $\Text \in \BinaryAlphabet^{\Textlen}$, $D_{\rm BWT}$ uses
  $\bigO(\SubstringComplexity{\Text} \log^c \Textlen)$ space, and
  answers BWT queries on $\Text$ in $t_{\rm BWT}(|\Text|) =
  o(\log \log |\Text|)$ time.  We will prove that this implies that
  there exists a data structure answering colored predecessor queries,
  that contradicts \cref{th:colored-predecessor-lower-bound}.

  Let $m \geq 1$ and let $A \subseteq [1 \dd m^2]$ be a nonempty set
  of size $m$.  Denote $A = \{a_1, a_2, \dots, a_m\}$, where $a_1 <
  a_2 < \dots < a_m$, and let $a_0 = 0$ and $a_{m+1} = m^2$.  Let
  $(b_i)_{i=0}^{m}$ be a sequence defined by $b_i = i \bmod 2$.  Let
  $k = 1 + \lfloor \log m \rfloor$.  Let $\Text_{A} =
  \bigodot_{i=0}^{m}(b_i \cdot \ebin{k}{i})^{a_{i+1}-a_{i}} \in
  \BinaryAlphabet^{*}$ (see
  \cref{def:ebin}) be a text defined as in
  \cref{lm:reduce-colored-predecessor-to-bwt}. Finally, let $b =
  \RangeBegTwo{\one^{k+1}\zero}{\Text}$.

  Let $D_{\rm pred}$ denote the data structure consisting of the
  following components:
  \begin{enumerate}
  \item The integer $b$ stored in $\bigO(1)$ space.
  \item The data structure $D_{\rm BWT}$ for the string $\Text_{A}$.
    Note that
    \[
      |\Text_{A}|
        = \textstyle\sum_{i=0}^{m} (2k+4) \cdot (a_{i+1}-a_{i})
        = (2k+4) \cdot \textstyle\sum_{i=0}^{m} (a_{i+1}-a_{i})
        = (2k+4) \cdot m^2 \in \bigO(m^2 \log m).
    \]
    Next, observe that for every $i \in [0 \dd m]$, the substring
    $(b_i \cdot \ebin{k}{i})^{a_{i+1}-a_{i}}$ of $\Text_{A}$ has an LZ77-like
    factorization consisting of $2k+5$ phrases: the first $2k+4$
    phrases are of length one, and the remaining suffix is encoded
    with a single phrase. Thus, $\Text_{A}$ has an LZ77-like
    factorization of size $(m + 1)\cdot (2k + 5)$.  By
    \cref{th:lz77-size}, we thus obtain $\LZSize{\Text_{A}} \leq (m +
    1) \cdot (2k + 5) = \bigO(m \log m)$.  By
    \cref{lm:z-and-r-upper-bound}\eqref{lm:z-and-r-upper-bound-it-1},
    this in turn implies that $\SubstringComplexity{\Text_{A}} =
    \bigO(\LZSize{\Text_{A}}) = \bigO(m \log m)$.  Consequently,
    $D_{\rm BWT}$ for $\Text_{A}$ needs
    $\bigO(\SubstringComplexity{\Text_{A}} \log^c |\Text_{A}|) =
    \bigO(m \log^{c+1} m)$ space.
  \end{enumerate}
  In total, $D_{\rm pred}$ needs $\bigO(m \log^{c+1} m)$ space.

  Given any $x \in \Z$, we compute $\PredecessorColor{A}{x}$
  (\cref{def:colored-predecessor}) as follows:
  \begin{enumerate}
  \item If $x < 1$ (resp.\ $x > m^2$), we return
    $\PredecessorColor{A}{x} = 0$ (resp.\ $\PredecessorColor{A}{x} = m
    \bmod 2$) in $\bigO(1)$ time, and conclude the query
    algorithm. Let us thus assume that $x \in [1 \dd m^2]$.
  \item Using $D_{\rm BWT}$, in $\bigO(t_{\rm BWT}(|\Text_{A}|))$ time
    we compute and return $c = \BWT{\Text_{A}}[b + x]$ as the
    answer. By \cref{lm:reduce-colored-predecessor-to-bwt}, it holds
    $c = \PredecessorColor{A}{x}$.
  \end{enumerate}
  In total, the query takes $\bigO(t_{\rm BWT}(|\Text_{A}|)) = o(\log
  \log |\Text_{A}|) = o(\log \log (m^2 \log m)) = o(\log \log m)$
  time.

  We have thus proved that there exists a data structure that, for
  every set $A \subseteq [1 \dd m^2]$ of size $|A| = m$, uses $\bigO(m
  \log^{c+1} m) = \bigO(m \log^{\bigO(1)} m)$ space, and answers
  colored predecessor queries on $A$ in $o(\log \log m)$ time. This
  contradicts \cref{th:colored-predecessor-lower-bound}.
\end{proof}

\subsection{Permuted Longest Common Prefix (PLCP) Queries}\label{sec:plcp}

\subsubsection{Problem Definition}\label{sec:plcp-problem-def}
\vspace{-1.5ex}

\begin{framed}
  \noindent
  \probname{Indexing for Permuted Longest Common Prefix (PLCP) Queries}
  \begin{bfdescription}
  \item[Input:]
    A string $\Text \in \Sigma^{\Textlen}$.
  \item[Output:]
    A data structure that, given any index $j \in [1 \dd \Textlen]$,
    returns $\PLCP{\Text}[j]$ (\cref{def:plcp-array}).
  \end{bfdescription}
\end{framed}
\vspace{2ex}

\subsubsection{Lower Bound}\label{sec:plcp-lower-bound}

\begin{lemma}\label{lm:reduce-predecessor-to-plcp}
  Let $A \subseteq [1 \dd m^2]$ be a nonempty set of size $|A| = m$.
  Denote $A = \{a_1, a_2, \dots, a_m\}$, where $a_1 < a_2 < \dots <
  a_m$, and let $a_{0} = -\infty$.  Let
  \[
    \Text =
      \Big( \textstyle\bigodot_{i=1}^{m} (\zero^{a_i}\one^{m-i+2}) \Big) \cdot
      \Big( \zero^{m^2+1}\one \Big) \cdot
      \Big( \zero^{m^2}\one^{m+2} \Big) \in
      \BinaryAlphabet^{*}
  \]
  (brackets added for clarity). Then, for every $x \in [1 \dd m^2]$,
  it holds (see \cref{def:predecessor})
  \[
    \Predecessor{A}{x} = a_i,
  \]
  where $\Delta = |\Text| - (m^2 + m + 2)$
  and $i = (x + m + 1) - \PLCP{\Text}[\Delta + m^2 - x + 1]$
  (\cref{def:plcp-array}).
\end{lemma}
\begin{proof}
  Denote $j = \Delta + m^2 - x + 1$ and let $a_{m+1} = m^2 + 1$. Let
  \[
    p = \min\{t \in [1 \dd m+1] : a_t \geq x\}.
  \]
  Note that $p$ is well-defined, since $a_{m+1} = m^2 + 1$.
  For every $t \in [p \dd m+1]$, let
  $s_t = (\sum_{u=1}^{t-1}(a_u+(m-u+2))) + (a_{t}-x+1)$.
  Observe that, by definition of $\Text$, it holds
  $|\OccTwo{\zero^{x}\one}{\Text}| = n - p + 3$ and, moreover,
  \[
    \OccTwo{\zero^{x}\one}{\Text}
      = \{s_t\}_{t \in [p \dd n+1]} \cup \{j\}.
  \]
  Next, observe that:
  \begin{itemize}
  \item $\Text[j \dd |\Text|]$ has $\zero^{x}\one^{m+2}$ as a prefix
    and
  \item for every $t \in [p \dd m+1]$, $\Text[s_t \dd |\Text|]$ has
    $\zero^{x}\one^{m-t+2}\zero$ as a prefix.
  \end{itemize}
  This implies that it holds
  \[
    \Text[s_{m+1} \dd |\Text|] \prec
    \Text[s_{m} \dd |\Text|] \prec \dots \prec
    \Text[s_{p} \dd |\Text|] \prec
    \Text[j \dd |\Text|].
  \]
  Consequently, it holds $\PhiArray{\Text}[j] = s_{p}$, and hence
  $\PLCP{\Text}[j] = \LCE{\Text}{j}{\PhiArray{\Text}[j]} =
  \LCE{\Text}{j}{s_{p}} = (x + m + 2) - p$. Combining with
  $\Predecessor{A}{x} = a_{p-1}$ (following from the definition of
  $p$), we thus obtain
  \begin{align*}
    i
      &= (x + m + 1) - \PLCP{\Text}[\Delta + m^2 - x + 1]\\
      &= (x + m + 1) - \PLCP{\Text}[j]\\
      &= (x + m + 1) - ((x + m + 2) - p)\\
      &= p-1.
  \end{align*}
  Thus, indeed we have $\Predecessor{A}{x} = a_{p-1} = a_{i}$.
\end{proof}

\begin{theorem}\label{th:plcp-lower-bound}
  There is no data structure that, for every text $\Text \in \BinaryAlphabet^{\Textlen}$,
  uses $\bigO(\SubstringComplexity{\Text} \log^{\bigO(1)}
  \Textlen)$ space and answers PLCP queries on $\Text$ (that, given
  any $j \in [1 \dd \Textlen]$, return $\PLCP{\Text}[j]$; see
  \cref{def:plcp-array}) in $o(\log \log \Textlen)$ time.
\end{theorem}
\begin{proof}

  Suppose that the claim does not hold. Let $D_{\rm PLCP}$ denote the
  hypothetical data structure answering PLCP queries, and let $c =
  \bigO(1)$ be a positive constant such that, that for
  a text $\Text \in \BinaryAlphabet^{\Textlen}$, $D_{\rm PLCP}$
  uses $\bigO(\SubstringComplexity{\Text}
  \log^c \Textlen)$ space, and answers PLCP queries on $\Text$ in
  $t_{\rm PLCP}(|\Text|) = o(\log \log |\Text|)$ time.  We will prove
  that this implies that there exists a data structure answering
  predecessor queries,
  that contradicts \cref{th:predecessor-lower-bound}.

  Let $m \geq 1$ and let $A \subseteq [1 \dd m^2]$ be a nonempty set
  of size $m$.  Denote $A = \{a_1, a_2, \dots, a_m\}$, where $a_1 <
  a_2 < \dots < a_m$.  Let $\Text_{A} =
  (\bigodot_{i=1}^{m}(\zero^{a_i}\one^{m-i+2})) \cdot
  (\zero^{m^2+1}\one) \cdot (\zero^{m^2}\one^{m+2}) \in
  \BinaryAlphabet^{*}$ be a text defined as in
  \cref{lm:reduce-predecessor-to-plcp}.  Denote $\Delta = |\Text_{A}|
  - (m^2 + m + 2)$. Let $A_{\rm val}[0 \dd m]$ be an array such that
  $A_{\rm val}[0] = -\infty$ and, for every $i \in [1 \dd m]$, $A_{\rm
  val}[i] = a_i$.

  Let $D_{\rm pred}$ denote the data structure consisting of the
  following components:
  \begin{enumerate}
  \item The array $A_{\rm val}[0 \dd m]$ stored in plain form. It
    needs $\bigO(m)$ space.
  \item The integer $\Delta$ stored in $\bigO(1)$ space.
  \item The data structure $D_{\rm PLCP}$ for the string $\Text_{A}$.
    Note that
    \[
      |\Text_{A}|
        = \textstyle\sum_{i=1}^{m}a_i +
          \textstyle\sum_{i=2}^{m+1}i +
          2m^2 + m + 4
        \leq m^3 + \tfrac{5}{2}m^2 + \tfrac{3}{2}m + 4
        \in \bigO(m^3).
    \]
    Moreover, note that $|\RL{\Text_{A}}| = 2(m+2)$. By applying
    \cref{ob:rl} and
    \cref{lm:z-and-r-upper-bound}\eqref{lm:z-and-r-upper-bound-it-1},
    we thus obtain $\SubstringComplexity{\Text_{A}} = \bigO(m)$.
    Consequently, $D_{\rm PLCP}$ for $\Text_{A}$ needs
    $\bigO(\SubstringComplexity{\Text_{A}} \log^c |\Text_{A}|) =
    \bigO(m \log^c m)$ space.
  \end{enumerate}
  In total, $D_{\rm pred}$ needs $\bigO(m \log^c m)$ space.

  Given any $x \in \Z$, we compute $\Predecessor{A}{x}$ as follows:
  \begin{enumerate}
  \item If $x < 1$ (resp.\ $x > m^2$), we return $\Predecessor{A}{x} =
    -\infty$ (resp.\ $\Predecessor{A}{x} = A_{\rm val}[m]$) in
    $\bigO(1)$ time, and conclude the query algorithm. Let us thus
    assume that $x \in [1 \dd m^2]$.
  \item Using $D_{\rm PLCP}$, in $\bigO(t_{\rm PLCP}(|\Text_{A}|))$
    time we compute $\ell = \PLCP{\Text_{A}}[\Delta + m^2 - x + 1]$.
    We then let $i = m - (\ell - x) + 1$, and return $p = A_{\rm
    val}[i]$ as the answer.  By
    \cref{lm:reduce-predecessor-to-plcp}, it holds $p =
    \Predecessor{A}{x}$.
  \end{enumerate}
  In total, the query takes $\bigO(t_{\rm PLCP}(|\Text_{A}|)) = o(\log
  \log |\Text_{A}|) = o(\log \log (m^3)) = o(\log \log m)$ time.

  We have thus proved that there exists a data structure that,
  for every set $A \subseteq [1 \dd m^2]$ of size $|A| = m$, uses
  $\bigO(m \log^c m) = \bigO(m \log^{\bigO(1)} m)$ space, and answers
  predecessor queries on $A$ in $o(\log \log m)$ time. This
  contradicts \cref{th:predecessor-lower-bound}.
\end{proof}

\subsection{Last-to-First (LF) Mapping Queries}\label{sec:lf}

\subsubsection{Problem Definition}\label{sec:lf-problem-def}
\vspace{-1.5ex}

\begin{framed}
  \noindent
  \probname{Indexing for Last-to-First (LF) Mapping Queries}
  \begin{bfdescription}
  \item[Input:]
    A string $\Text \in \Sigma^{\Textlen}$.
  \item[Output:]
    A data structure that, given any index $i \in [1 \dd \Textlen]$,
    returns $\LF{\Text}[i]$ (\cref{def:lf}).
  \end{bfdescription}
\end{framed}
\vspace{2ex}

\subsubsection{Lower Bound}\label{sec:lf-lower-bound}

\begin{observation}\label{obs:lf}
  Let $\Text \in \BinaryAlphabet^{\Textlen}$, where $\Textlen \geq 1$.
  Denote $n_0 = |\{j \in [1 \dd \Textlen] : \Text[j] = \zero\}|$.  For
  every $i \in [1 \dd \Textlen]$, $\BWT{\Text}[i] = \zero$
  (\cref{def:bwt}) holds if and only if $\LF{\Text}[i] \leq n_0$
  (\cref{def:lf}).
\end{observation}
\begin{proof}
  The proof consists of two steps.
  \begin{enumerate}
  \item First, we observe that $\BWT{\Text}[i] =
    \Text[\SA{\Text}[\LF{\Text}[i]]]$.  To see this, consider two
    cases:
    \begin{itemize}
    \item If $\SA{\Text}[i] = 1$, then by \cref{def:bwt,def:lf}, we
      have $\BWT{\Text}[i] = \Text[\Textlen]$ and $\LF{\Text}[i] =
      \ISA{\Text}[\Textlen]$.  Putting the two together, we obtain
      $\BWT{\Text}[i] = \Text[\Textlen] =
      \Text[\SA{\Text}[\LF{\Text}[i]]]$.
    \item If $\SA{\Text}[i] > 1$, then by \cref{def:bwt,def:lf}, it holds
      $\BWT{\Text}[i] = \Text[\SA{\Text}[i] - 1]$ and
      $\SA{\Text}[\LF{\Text}[i]] = \SA{\Text}[i] - 1$.  Thus,
      $\BWT{\Text}[i] = \Text[\SA{\Text}[i] - 1] =
      \Text[\SA{\Text}[\LF{\Text}[i]]]$.
    \end{itemize}
  \item By definition of the suffix array, for every $t \in [1 \dd
    \Textlen]$, $\Text[\SA{\Text}[t]] = \zero$ holds if and only if $t
    \leq n_0$.  Letting $t = \LF{\Text}[i]$, we thus obtain the claim.
    \qedhere
  \end{enumerate}
\end{proof}

\begin{theorem}\label{th:lf-lower-bound}
  There is no data structure that, for every text $\Text \in
  \BinaryAlphabet^{\Textlen}$, uses $\bigO(\SubstringComplexity{\Text}
  \log^{\bigO(1)} \Textlen)$ space and answers LF queries on $\Text$
  (that, given any $i \in [1 \dd \Textlen]$, return $\LF{\Text}[i]$;
  see \cref{def:lf}) in $o(\log \log \Textlen)$ time.
\end{theorem}
\begin{proof}

  Suppose that the claim does not hold. Let $D_{\rm LF}$ denote the
  hypothetical data structure answering LF queries, and let $c =
  \bigO(1)$ be a positive constant such that, that for a text $\Text \in
  \BinaryAlphabet^{\Textlen}$, $D_{\rm LF}$ uses
  $\bigO(\SubstringComplexity{\Text} \log^c \Textlen)$ space, and
  answers LF queries on $\Text$ in $t_{\rm LF}(|\Text|) = o(\log \log
  |\Text|)$ time.  We will prove that this implies that there exists a
  data structure answering BWT queries contradicting
  \cref{th:bwt-lower-bound}.

  Let $\Text \in \BinaryAlphabet^{\Textlen}$, where $\Textlen \geq 1$.
  Let $n_0 = |\{j \in [1 \dd \Textlen] : \Text[j] = \zero\}|$.

  Let $D_{\rm BWT}$ denote the data structure consisting of the
  following components:
  \begin{enumerate}
  \item The integer $n_0$ stored in $\bigO(1)$ space.
  \item The data structure $D_{\rm LF}$ for the string $\Text$.
    If uses
    $\bigO(\SubstringComplexity{\Text} \log^c \Textlen)$.
    space.
  \end{enumerate}
  In total, $D_{\rm BWT}$ needs
  $\bigO(\SubstringComplexity{\Text} \log^c \Textlen)$ space.

  Given any $i \in [1 \dd \Textlen]$, we compute $\BWT{\Text}[i]$
  (\cref{def:bwt}) as follows:
  \begin{enumerate}
  \item First, in $\bigO(t_{\rm LF}(\Textlen))$ time we compute $i' =
    \LF{\Text}[i]$.
  \item We return that $\BWT{\Text}[i] = \zero$
    (resp.\ $\BWT{\Text}[i] = \one$) if $i' \leq n_0$ (resp.\ $i' >
    n_0$).  This is correct by \cref{obs:lf}.
  \end{enumerate}
  In total, the query takes $\bigO(t_{\rm LF}(\Textlen)) = o(\log \log
  \Textlen)$ time.

  We have thus proved that there exists a data structure that, for
  every $\Text \in \BinaryAlphabet^{\Textlen}$, uses
  $\bigO(\SubstringComplexity{\Text} \log^c \Textlen) =
  \bigO(\SubstringComplexity{\Text} \log^{\bigO(1)} \Textlen)$ space, and
  answers BWT queries on the text $\Text$ in $o(\log \log
  \Textlen)$ time. This contradicts \cref{th:bwt-lower-bound}.
\end{proof}

\subsection{\boldmath Inverse Last-to-First (\texorpdfstring{$\text{LF}^{-1}$}{LF⁻¹}) Mapping Queries}\label{sec:ilf}

\subsubsection{Problem Definition}\label{sec:ilf-problem-def}
\vspace{-1.5ex}

\begin{framed}
  \noindent
  \probname{Indexing for Inverse Last-to-First ($\mathrm{LF}^{-1}$) Mapping Queries}
  \begin{bfdescription}
  \item[Input:]
    A string $\Text \in \Sigma^{\Textlen}$.
  \item[Output:]
    A data structure that, given any index $i \in [1 \dd \Textlen]$,
    returns $\ILF{\Text}[i]$ (\cref{def:ilf}).
  \end{bfdescription}
\end{framed}
\vspace{2ex}

\subsubsection{Lower Bound}\label{sec:ilf-lower-bound}

\begin{lemma}\label{lm:reduce-predecessor-to-ilf}
  Let $A \subseteq [1 \dd m^2]$ be a nonempty set of size $|A| = m$.
  Denote $A = \{a_1, a_2, \dots, a_m\}$, where $a_1 < a_2 < \dots <
  a_m$, and let $a_0 = 0$ and $a_{m+1} = m^2$. Let $B = A \cup \{a_0\}$.
  Let (see \cref{def:ebin})
  \[
    \Text = \textstyle\bigodot_{i=0}^{m}
      \Big(\one \cdot \ebin{k}{i}\Big)^{a_{i+1}-a_{i}} \cdot
      \Big(\ebin{k}{i} \Big)^{m^2 - (a_{i+1}-a_{i})}
      \in \BinaryAlphabet^{*}
  \]
  (brackets added for clarity), where $k = 1 + \lfloor \log m
  \rfloor$.  Then, for every $x \in [1 \dd m^2]$, it holds (see
  \cref{def:predecessor})
  \[
    \Predecessor{B}{x} = a_i,
  \]
  where $\alpha = \RangeBegTwo{\one^{k+2}\zero}{\Text}$ (\cref{def:occ}),
  $\beta = \RangeBegTwo{\one^{k+1}\zero}{\Text}$, and (see \cref{def:ilf})
  \[
    i = \left\lceil\frac{\ILF{\Text}[\alpha + x] - \beta}{m^2} \right\rceil - 1.
  \]
\end{lemma}
\begin{proof}
  Denote $i_{\rm arg} = \alpha + x$. Let
  \[
    p = \max\{t \in [0 \dd m] : a_t < x\}.
  \]
  Note that since $x \in [1 \dd m^2]$ and $a_0 = 0$, it follows that
  $p$ is well defined and it holds
  \[
    \Predecessor{B}{x} = \Predecessor{A \cup \{a_0\}}{x} = a_p.
  \]
  For every $t \in [0 \dd m+1]$, denote
  $q_t = \beta + t \cdot m^2$
  and
  $s_t = \alpha + a_t$.
  The proof proceeds in four steps:
  \begin{enumerate}

  \item First, we prove that, for $t \in [0 \dd m]$, it holds
    $\RangeBegTwo{\ebin{k}{t}}{\Text} = q_t$ and $\RangeEndTwo{\ebin{k}{t}}{\Text} = q_{t+1}$.
    Observe that every occurrence of $\one^{k+1}\zero$ in $\Text$ coincides with an
    occurrence of $\ebin{k}{t}$ for some $t \in [0 \dd m]$. Formally, $\OccTwo{\one^{k+1}\zero}{\Text}$ is a disjoint union
    \[
      \OccTwo{\one^{k+1}\zero}{\Text} = \textstyle\bigcup_{t=0}^{m} \OccTwo{\ebin{k}{t}}{\Text}.
    \]
    Next, note that for every $t \in [0 \dd m]$, it holds $|\OccTwo{\ebin{k}{t}}{\Text}| = m^2$. Finally,
    observe that by \cref{def:ebin,def:bin}, it follows that for every $t_1, t_2 \in [0 \dd m]$, $t_1 < t_2$ implies
    $\ebin{k}{t_1} \prec \ebin{k}{t_2}$. Since for all $t \in [0 \dd m]$, $\ebin{k}{t}$ is of the same length, it follows
    that for every $t_1, t_2 \in [0 \dd m]$ such that $t_1 < t_2$, it holds $\Text[j_1 \dd |\Text|] \prec \Text[j_2 \dd |\Text|]$
    for all $j_1 \in \OccTwo{\ebin{k}{t_1}}{\Text}$ and $j_2 \in \OccTwo{\ebin{k}{t_2}}{\Text}$.
    Consequently, the SA-interval $\SA{\Text}(\beta \dd \beta + m^2]$ contains
    $\OccTwo{\ebin{k}{0}}{\Text}$, $\SA{\Text}(\beta + m^2 \dd \beta + 2m^2]$ contains $\OccTwo{\ebin{k}{1}}{\Text}$, etc.
    By definition of the sequence $(q_t)_{t \in [0 \dd m+1]}$, we thus obtain that, for every $t \in [0 \dd m]$, it holds
    $\RangeBegTwo{\ebin{k}{t}}{\Text} = \beta + t \cdot m^2 = q_t$ and
    $\RangeEndTwo{\ebin{k}{t}}{\Text} = \beta + (t+1) \cdot m^2 = q_{t+1}$.

  \item In the second step, we prove that, for every $t \in [0 \dd m]$, it holds
    $\RangeEndTwo{\one \cdot \ebin{k}{t}}{\Text} = s_t$ and
    $\RangeEndTwo{\one \cdot \ebin{k}{t}}{\Text} = s_{t+1}$.
    Observe that every occurrence of $\one^{k+2}\zero$ in $\Text$ coincides with an
    occurrence of $\one \cdot \ebin{k}{t}$ for some $t \in [0 \dd m]$.
    Formally, $\OccTwo{\one^{k+2}\zero}{\Text}$ is a disjoint union
    \[
      \OccTwo{\one^{k+2}\zero}{\Text} = \textstyle\bigcup_{t=0}^{m} \OccTwo{\one \cdot \ebin{k}{t}}{\Text}.
    \]
    Next, note that for every $t \in [0 \dd m]$, it holds
    $|\OccTwo{\one \cdot \ebin{k}{t}}{\Text}| = a_{t+1}-a_{t}$. Finally,
    note that by the same argument as above,
    for every $t_1, t_2 \in [0 \dd m]$ such that $t_1 < t_2$, it holds $\Text[j_1 \dd |\Text|] \prec \Text[j_2 \dd |\Text|]$
    for all $j_1 \in \OccTwo{\one \cdot \ebin{k}{t_1}}{\Text}$ and $j_2 \in \OccTwo{\one \cdot \ebin{k}{t_2}}{\Text}$.
    Consequently, the SA-interval
    $\SA{\Text}(\alpha + a_0 \dd \alpha + a_1]$
    contains $\OccTwo{\one \cdot \ebin{k}{0}}{\Text}$,
    $\SA{\Text}(\alpha + a_1 \dd \alpha + a_2]$
    contains $\OccTwo{\one \cdot \ebin{k}{1}}{\Text}$, etc.
    By definition of the sequence $(s_t)_{t \in [0 \dd m+1]}$,
    we thus obtain that, for every $t \in [0 \dd m]$, it holds
    $\RangeBegTwo{\one \cdot \ebin{k}{t}}{\Text} = \alpha + a_t = s_t$
    $\RangeEndTwo{\one \cdot \ebin{k}{t}}{\Text} = \alpha + a_{t+1} = s_{t+1}$.

  \item In the third step, we prove that the suffix $\Text[\SA{\Text}[i_{\rm arg}] + 1 \dd |\Text|]$
    has the string $\ebin{k}{p}$ as a prefix. To this end, observe that, by definition of $p$, it holds
    $a_{p} < x \leq a_{p+1}$. Consequently, it holds
    \[
      s_{p} = \alpha + a_{p} < \alpha + x \leq \alpha + a_{p+1} = s_{p+1}.
    \]
    By the above, it holds $\RangeBegTwo{\one \cdot \ebin{k}{p}}{\Text} = s_{p}$ and
    $\RangeEndTwo{\one \cdot \ebin{k}{p}}{\Text} = s_{p+1}$. This implies that
    $\SA{\Text}[\alpha + x] \in \OccTwo{\one \cdot \ebin{k}{p}}{\Text}$, or equivalently,
    $\SA{\Text}[i_{\rm arg}] \in \OccTwo{\one \cdot \ebin{k}{p}}{\Text}$. Therefore,
    $\Text[\SA{\Text}[i_{\rm arg}] + 1 \dd |\Text|]$ has $\ebin{k}{p}$ as a prefix.

  \item We are now ready to finish the proof. By \cref{def:ilf}, it holds
    $\SA{\Text}[\ILF{\Text}[i_{\rm arg}]] = \SA{\Text}[i_{\rm arg}] + 1$ (note that here
    we utilize the fact that since $\Text[\SA{\Text}[i_{\rm arg}] + 1 \dd |\Text|]$ has $\ebin{k}{p}$ as a prefix,
    we must have $\SA{\Text}[i_{\rm arg}] < |\Text|$).
    We thus also have $\Text[\SA{\Text}[\ILF{\Text}[i_{\rm arg}]] \dd |\Text|] =
    \Text[\SA{\Text}[i_{\rm arg}] + 1 \dd |\Text|]$. Since above we proved
    that $\Text[\SA{\Text}[i_{\rm arg}] + 1 \dd |\Text|]$ has $\ebin{k}{p}$ as a prefix, we thus
    obtain that $\Text[\SA{\Text}[\ILF{\Text}[i_{\rm arg}]] \dd |\Text|]$ has $\ebin{k}{p}$ as a prefix as well.
    By the above characterization of values $\RangeBegTwo{\ebin{k}{p}}{\Text}$ and $\RangeEndTwo{\ebin{k}{p}}{\Text}$,
    it therefore follows that the index
    $\ILF{\Text}[i_{\rm arg}]$ satisfies $q_p < \ILF{\Text}[i_{\rm arg}] \leq q_{p+1}$. Plugging in the
    definition of $q_p$ and $q_{p+1}$, we thus obtain that it holds
    \[
      \beta + p \cdot m^2 < \ILF{\Text}[i_{\rm arg}] \leq \beta + (p+1) \cdot m^2.
    \]
    This is equivalent to $p \cdot m^2 < \ILF{\Text}[i_{\rm arg}] - \beta \leq (p+1) \cdot m^2$,
    which implies that
    \[
      \left\lceil \frac{\ILF{\Text}[i_{\rm arg}] - \beta}{m^2} \right\rceil = p+1.
    \]
    Thus, by definition of $i_{\rm arg}$ and $i$ we obtain
    \[
      i = \left\lceil \frac{\ILF{\Text}[\alpha + x] - \beta}{m^2} \right\rceil - 1
        = \left\lceil \frac{\ILF{\Text}[i_{\rm arg}] - \beta}{m^2} \right\rceil - 1
        = p.
     \]
    Since above we observed that $\Predecessor{B}{x} = a_p$, we thus obtain
    that $\Predecessor{B}{x} = a_p = a_i$, i.e., the claim.
    \qedhere
  \end{enumerate}
\end{proof}

\begin{theorem}\label{th:ilf-lower-bound}
  There is no data structure that, for every text $\Text \in \BinaryAlphabet^{\Textlen}$,
  uses $\bigO(\SubstringComplexity{\Text} \log^{\bigO(1)}
  \Textlen)$ space and answers inverse LF queries on $\Text$ (that, given
  any $j \in [1 \dd \Textlen]$, return $\ILF{\Text}[j]$; see
  \cref{def:ilf}) in $o(\log \log \Textlen)$ time.
\end{theorem}
\begin{proof}

  Suppose that the claim does not hold. Let $D_{\rm ILF}$ denote the
  hypothetical data structure answering inverse LF queries, and let $c =
  \bigO(1)$ be a positive constant such that, that for
  a text $\Text \in \BinaryAlphabet^{\Textlen}$, $D_{\rm ILF}$
  uses $\bigO(\SubstringComplexity{\Text}
  \log^c \Textlen)$ space, and answers inverse LF queries on $\Text$ in
  $t_{\rm ILF}(|\Text|) = o(\log \log |\Text|)$ time.  We will prove
  that this implies that there exists a data structure answering
  predecessor queries,
  that contradicts \cref{th:predecessor-lower-bound}.

  Let $m \geq 1$ and let $A \subseteq [1 \dd m^2]$ be a nonempty set
  of size $m$. Let $k = 1 + \lfloor \log m \rfloor$.
  Denote $A = \{a_1, a_2, \dots, a_m\}$, where $a_1 <
  a_2 < \dots < a_m$. Let also $a_0 = 0$ and $a_{m+1} = m^2$.
  Let $\Text_{A} =
  \bigodot_{i=0}^{m}
  (\one \cdot \ebin{k}{i})^{a_{i+1}-a_{i}} \cdot
  (\ebin{k}{i})^{m^2 - (a_{i+1}-a_{i})} \in \BinaryAlphabet^{*}$
  be a text defined as in \cref{lm:reduce-predecessor-to-ilf}.
  Denote $\alpha = \RangeBegTwo{\one^{k+2} \zero}{\Text}$
  and $\beta = \RangeBegTwo{\one^{k+1} \zero}{\Text}$.
  Let $A_{\rm val}[1 \dd m]$ be an array such that,
  for every $i \in [1 \dd m]$, $A_{\rm val}[i] = a_i$.

  Let $D_{\rm pred}$ denote the data structure consisting of the
  following components:
  \begin{enumerate}
  \item The array $A_{\rm val}[1 \dd m]$ stored in plain form. It
    needs $\bigO(m)$ space.
  \item The integers $\alpha$ and $\beta$ stored in $\bigO(1)$ space.
  \item The data structure $D_{\rm ILF}$ for the string $\Text_{A}$.
    Note that
    \begin{align*}
      |\Text_{A}|
        &= \textstyle\sum_{i=0}^{m} (2k + 4) \cdot (a_{i+1} - a_{i}) + (2k + 3) \cdot (m^2 - (a_{i+1} - a_{i}))\\
        &= \textstyle\sum_{i=0}^{m} (a_{i+1} - a_{i}) + \textstyle\sum_{i=0}^{m} (2k+3) \cdot m^2\\
        &= m^2 + (2k + 3) (m+1) m^2 \in \bigO(m^3 \log m).
    \end{align*}
    Next, observe that for every $i \in [0 \dd m]$, the substring
    $(\one \cdot \ebin{k}{i})^{a_{i+1}-a_{i}} \cdot
    (\ebin{k}{i})^{m^2 - (a_{i+1}-a_{i})}$ of $\Text_{A}$ has
    an LZ77-like factorization consisting of at most $4k + 9$ phrases:
    the first $2k+4$ phrases of length one, then at most a single
    phrase corresponding to the substring $(\one \cdot \ebin{k}{i})^{(a_{i+1}-a_{i})-1}$,
    then $2k+3$ phrases of length one, and finally, at most a single
    phrase corresponding to the substring $(\ebin{k}{i})^{(m^2 - (a_{i+1}-a_{i})) - 1}$.
    Thus, $\Text_{A}$ has an LZ77-like factorization of size at most
    $(m+1) \cdot (4k+9)$. By \cref{th:lz77-size}, we thus obtain
    $\LZSize{\Text_{A}} \leq (m+1) \cdot (4k+9) = \bigO(m \log m)$.
    By \cref{lm:z-and-r-upper-bound}\eqref{lm:z-and-r-upper-bound-it-1},
    this in turn implies that
    $\SubstringComplexity{\Text_{A}} = \bigO(\LZSize{\Text_{A}}) = \bigO(m \log m)$.
    Consequently, $D_{\rm ILF}$ for $\Text_{A}$ needs
    $\bigO(\SubstringComplexity{\Text_{A}} \log^c |\Text_{A}|)
    = \bigO(m \log^{c+1} m)$ space.
  \end{enumerate}
  In total, $D_{\rm pred}$ needs $\bigO(m \log^{c+1} m)$ space.

  Given any $x \in \Z$, we compute $\Predecessor{A}{x}$ as follows:
  \begin{enumerate}
  \item If $x < 1$ (resp.\ $x > m^2$), we return $\Predecessor{A}{x} =
    -\infty$ (resp.\ $\Predecessor{A}{x} = A_{\rm val}[m]$) in
    $\bigO(1)$ time, and conclude the query algorithm. Let us thus
    assume that $x \in [1 \dd m^2]$.
  \item Using $D_{\rm ILF}$, in $\bigO(t_{\rm ILF}(|\Text_{A}|))$
    time we compute $i = \lceil (\ILF{\Text_{A}}[\alpha + x] - \beta) / m^2 \rceil - 1$.
    By \cref{lm:reduce-predecessor-to-ilf}, it holds $\Predecessor{A \cup \{a_0\}}{x} = a_i$.
    If $i = 0$, then we return $\Predecessor{A}{x} = -\infty$. Otherwise, we
    return $\Predecessor{A}{x} = a_i = A_{\rm val}[i]$.
  \end{enumerate}
  In total, the query takes $\bigO(t_{\rm ILF}(|\Text_{A}|)) = o(\log
  \log |\Text_{A}|) = o(\log \log (m^3 \log m)) = o(\log \log m)$ time.

  We have thus proved that there exists a data structure that,
  for every set $A \subseteq [1 \dd m^2]$ of size $|A| = m$, uses
  $\bigO(m \log^{c+1} m) = \bigO(m \log^{\bigO(1)} m)$ space, and answers
  predecessor queries on $A$ in $o(\log \log m)$ time. This
  contradicts \cref{th:predecessor-lower-bound}.
\end{proof}

\subsubsection{Upper Bound}\label{sec:ilf-upper-bound}

\begin{lemma}[{\cite[Lemma~5.2]{Gagie2020}}]\label{lm:lf-within-run}
  Let $\Text \in \Sigma^{\Textlen}$ be a nonempty text such that
  $\Text[\Textlen]$ does not occur in $\Text[1 \dd \Textlen)$.  Then,
  for every $i \in [2 \dd \Textlen]$ satisfying $\BWT{\Text}[i-1]
  = \BWT{\Text}[i]$ (\cref{def:bwt}), it holds $\LF{\Text}[i]
  = \LF{\Text}[i-1] + 1$ (\cref{def:lf}).
\end{lemma}

\begin{lemma}\label{lm:ilf-upper-bound}
  Let $\Text \in \Sigma^{\Textlen}$ be a nonempty text such that
  $\Text[\Textlen]$ does not occur in $\Text[1 \dd \Textlen)$.  Denote
  (see \cref{def:bwt})
  \[
    \mathcal{I} =
      \{i \in [2 \dd \Textlen] : \BWT{\Text}[i-1] \neq \BWT{\Text}[i]\}
      \cup \{1\}.
  \]
  Then, for every $j \in [1 \dd \Textlen]$, it holds (see \cref{def:ilf})
  \[
    \ILF{\Text}[j] = \ILF{\Text}[j_{\rm prev}] + (j-j_{\rm prev}),
  \]
  where $\mathcal{J} = \{\LF{\Text}[i] : i \in \mathcal{I}\}$
  (\cref{def:lf}) and $j_{\rm prev} = \max\{t \in \mathcal{J} : t \leq
  j\}$.
\end{lemma}
\begin{proof}

  We begin by showing that $j_{\rm prev}$ is well-defined. To this end,
  we will show that $1 \in \mathcal{J}$.  Let $i_1
  = \ILF{\Text}[1]$. We claim that $i_1 \in \mathcal{I}$. Suppose that
  this does not hold.  By definition of $\mathcal{I}$, we then have
  $i_1 > 1$ and $\BWT{\Text}[i_1-1] = \BWT{\Text}[i_1]$.
  By \cref{lm:lf-within-run}, we then have $\LF{\Text}[i_1-1]
  = \LF{\Text}[i_1]-1 = 0$, which is not possible (see \cref{def:lf}).
  Thus, we must have $i_1 \in \mathcal{I}$, and hence $\LF{\Text}[i_1]
  = 1 \in \mathcal{J}$.

  We now prove the claim.  If $j = j_{\rm prev}$, then we trivially
  obtain that $\ILF{\Text}[j] = \ILF{\Text}[j_{\rm prev}]$.  Let us
  thus assume that $j_{\rm prev} < j$. Denote $i = \ILF{\Text}[j]$ and
  $i_{\rm prev} = \ILF{\Text}[j_{\rm prev}]$. Let also $i'
  = \max\{t \in \mathcal{I} : t \leq i\}$. The proof proceeds in three
  steps:
  \begin{enumerate}

  \item First, we prove that it holds $i' = i_{\rm prev}$. By
    definition of $i'$, it holds $i' \in \mathcal{I}$ and $(i' \dd
    i] \cap \mathcal{I} = \emptyset$. Thus, for every $t \in (i' \dd
    i]$, $\BWT{\Text}[t-1] = \BWT{\Text}[t]$.  Consequently,
    by \cref{lm:lf-within-run}, for every $t \in [i' \dd i]$, it holds
    $\LF{\Text}[t] = \LF{\Text}[t-1]+1 = \LF{\Text}[t-2]+2 = \dots
    = \LF{\Text}[i'] + (t-i')$.  In other words, the sequence of
    positions $\LF{\Text}[i'], \LF{\Text}[i'+1], \dots, \LF{\Text}[i]$
    is increasing, and forms a contiguous block.  On the other hand,
    note that $[i' \dd i] \cap \mathcal{I} = \{i'\}$ implies that
    $[\LF{\Text}[i'] \dd \LF{\Text}[i]] \cap \mathcal{J}
    = \{\LF{\Text}[i']\}$. Recalling that $\LF{\Text}[i] = j$, we
    therefore obtain $j_{\rm prev} = \max\{t \in \mathcal{J} : t \leq
    j\} = \LF{\Text}[i']$. This implies that $\ILF{\Text}[j_{\rm
    prev}] = i'$, or equivalently, $i_{\rm prev} = i'$.

  \item Next, we prove that $j - j_{\rm prev} = i - i_{\rm
    prev}$. Above we proved that, for every $t \in [i_{\rm prev} \dd
    i]$, it holds $\LF{\Text}[t] = \LF{\Text}[i_{\rm prev}] + (t -
    i_{\rm prev})$.  For $t = i$, we thus obtain $\LF{\Text}[i]
    = \LF{\Text}[i_{\rm prev}] + (i - i_{\rm prev})$. Substituting
    $\LF{\Text}[i_{\rm prev}] = j_{\rm prev}$ and $\LF{\Text}[i] = j$,
    we thus obtain $j = j_{\rm prev} + (i - i_{\rm prev})$, or
    equivalently, $j - j_{\rm prev} = i - i_{\rm prev}$.

  \item Putting together the two facts proved above, we obtain
    \[
      \ILF{\Text}[j]
        = i
        = i_{\rm prev} + (j - j_{\rm prev})
        = \ILF{\Text}[j_{\rm prev}] + (j - j_{\rm prev}).
       \qedhere
    \]
  \end{enumerate}
\end{proof}

\begin{lemma}\label{lm:ilf-for-string-without-dollar}
  Let $\Text \in \IntegerAlphabet^{\Textlen}$.
  Denote
  \[
    \Text' = \Big(\textstyle\bigodot_{i=1}^{\Textlen} (\Text[i] + 1) \Big) \cdot \zero \in [0 \dd \AlphabetSize+1)^{\Textlen+1}.
  \]
  Let also $i_{\rm first} = \ISA{\Text}[1]$ (\cref{def:isa}) and $i_{\rm last} = \ISA{\Text}[\Textlen]$.
  Then, for every $i \in [1 \dd \Textlen]$, it holds (see \cref{def:ilf})
  \[
    \ILF{\Text}[i] =
      \begin{cases}
        \ILF{\Text'}[i+1] - 1 & \text{if }i \neq i_{\rm last},\\
        i_{\rm first} & \text{otherwise}.\\
      \end{cases}
  \]
\end{lemma}
\begin{proof}
  We proceed in three steps:
  \begin{enumerate}

  \item First, we characterize $\SA{\Text'}[1 \dd |\Text'|]$. More precisely,
    we show that, for every $i \in [1 \dd \Textlen]$, it holds
    $\SA{\Text}[i] = \SA{\Text'}[i+1]$. To see this, it
    suffices to note that $\SA{\Text'}[1] = \Textlen + 1$ and, for every $i,j \in [1 \dd n]$,
    $\Text[i \dd |\Text|] \prec \Text[j \dd |\Text|]$ holds if and only if
    $\Text'[i \dd |\Text'|] \prec \Text'[j \dd |\Text'|]$.

  \item Next, we observe that by the above characterization of $\SA{\Text'}$, we
    also obtain that, for every $j \in [1 \dd \Textlen]$, $\ISA{\Text}[j] = \ISA{\Text'}[j]-1$.

  \item We are now ready to prove the main claim. The equality
    $\ILF{\Text}[i_{\rm last}] = i_{\rm first}$ follows by definition (see \cref{def:ilf}).
    Let us thus consider any $i \in [1 \dd \Textlen] \setminus \{i_{\rm last}\}$.
    By \cref{def:ilf}, we then have
    \begin{align*}
      \ILF{\Text}[i]
        &= \ISA{\Text}[\SA{\Text}[i] + 1]\\
        &= \ISA{\Text'}[\SA{\Text}[i] + 1] - 1\\
        &= \ISA{\Text'}[\SA{\Text'}[i+1] + 1] - 1\\
        &= \ILF{\Text'}[i+1] - 1,
    \end{align*}
    where in the second equality we used that $i \neq i_{\rm last}$
    implies that $\SA{\Text}[i] + 1 \in [1 \dd \Textlen]$.
    \qedhere
  \end{enumerate}
\end{proof}

\begin{lemma}\label{lm:number-of-runs-after-appending-dollar}
  Let $\Text \in \IntegerAlphabet^{\Textlen}$.
  Denote
  \[
    \Text' = \Big(\textstyle\bigodot_{i=1}^{\Textlen} (\Text[i] + 1) \Big) \cdot \zero \in [0 \dd \AlphabetSize+1)^{\Textlen+1}.
  \]
  Then, it holds $\RLBWTSize{\Text'} \leq \RLBWTSize{\Text} + 3$ (\cref{def:bwt}).
\end{lemma}
\begin{proof}
  Observe that, for every $i,j \in [1 \dd \Textlen]$, $\Text[i \dd \Textlen] \prec \Text[j \dd \Textlen]$ holds
  if and only if $\Text'[i \dd \Textlen + 1] \prec \Text'[j \dd \Textlen + 1]$. On the other hand, it holds
  $\SA{\Text'}[1] = \Textlen + 1$. This implies that, for every $i \in [2 \dd \Textlen + 1]$, it holds
  $\SA{\Text'}[i] = \SA{\Text}[i-1]$. Consequently, by \cref{def:bwt}, we obtain
  \[
    \BWT{\Text'}[i] =
      \begin{cases}
        \Text[\Textlen] + 1 & \text{if }i = 1,\\
        \BWT{\Text}[i-1]+1 & \text{if }i > 1\text{ and }i-1\neq \ISA{\Text}[1],\\
        \zero & \text{if }i > 1\text{ and }i-1 = \ISA{\Text}[1].
      \end{cases}
  \]
  In other words, $\BWT{\Text'}$ is obtained from $\BWT{\Text}$ by
  \begin{enumerate}
  \item increasing all symbols by one (which does not increase the number of runs),
  \item replacing the symbol at index $\ISA{\Text}[1]$ with $\zero$ (which can increase the number of runs by at most two), and
  \item prepending the BWT with symbol $\Text[\Textlen] + 1$ (which can increase the number of runs by at most one).
  \end{enumerate}
  Thus, $\RLBWTSize{\Text'} \leq \RLBWTSize{\Text} + 3$.
\end{proof}

\begin{theorem}\label{th:ilf-upper-bound-integer-alphabet-with-dollar}
  For every text $\Text \in \IntegerAlphabet^{\Textlen}$ such that $\Text[\Textlen] = \zero$, and $\zero$
  does not occur in $\Text[1 \dd \Textlen)$,
  there exists a data structure of size $\bigO(\RLBWTSize{\Text})$ (see \cref{def:bwt})
  that answers inverse LF queries on $\Text$ (that, given any $i \in [1 \dd \Textlen]$, return
  $\ILF{\Text}[i]$; see \cref{def:ilf}) in $\bigO(\log \log \Textlen)$ time.
\end{theorem}
\begin{proof}

  Let $\mathcal{I} = \{i \in [2 \dd \Textlen] : \BWT{\Text}[i-1] \neq \BWT{\Text}[i]\} \cup \{1\}$
  (\cref{def:bwt}) and
  $\mathcal{J} = \{\LF{\Text}[i] : i \in \mathcal{I}\}$ (\cref{def:lf}).
  Denote $m = |\mathcal{J}|$
  and let $(p_j)_{j \in [1 \dd m]}$ be an increasing sequence containing all elements of $\mathcal{J}$, i.e.,
  such that $\{p_1, \dots, p_m\} = \mathcal{J}$ and $p_1 < p_2 < \dots < p_m$. We also set $p_0 = -\infty$.
  Let $A_{\mathcal{J}}[0 \dd m]$ be an array defined by $A_{\mathcal{J}}[i] = p_i$ for $i \in [0 \dd m]$.
  Finally, let $A_{\rm ILF}[1 \dd m]$ be an array defined by
  $A_{\rm ILF}[i] = \ILF{\Text}[p_i]$ for $i \in [1 \dd m]$.

  The data structure to answer inverse LF queries on $\Text$ consists of the following components:
  \begin{enumerate}
  \item The predecessor data structure from \cref{th:predecessor-yfast-trie} for the sequence $(p_i)_{i \in [0 \dd m]}$.
    Since for every $i \in [1 \dd m]$, it holds $p_i \in [0 \dd \Textlen]$,
    the structure answers queries
    in $\bigO(\log \log \Textlen)$ time. To bound its space, observe that
    $m = |\mathcal{J}| = |\mathcal{I}| = \RLBWTSize{\Text}$
    (see \cref{def:bwt}).
    Thus, the structure needs $\bigO(m) = \bigO(\RLBWTSize{\Text})$ space.
  \item The array $A_{\mathcal{J}}[0 \dd m]$ stored in plain form. It needs $\bigO(m) = \bigO(\RLBWTSize{\Text})$ space.
  \item The array $A_{\rm ILF}[1 \dd m]$ stored in plain form. It needs $\bigO(m) = \bigO(\RLBWTSize{\Text})$ space.
  \end{enumerate}
  In total, the structure needs $\bigO(\RLBWTSize{\Text})$ space.

  Let $j \in [1 \dd \Textlen]$. To compute $\ILF{\Text}[j]$ using the above data structure, we proceed as follows:
  \begin{enumerate}
  \item Using the structure from \cref{th:predecessor-yfast-trie},
    in $\bigO(\log \log \Textlen)$ time, we compute the index $k \in [0 \dd m]$ satisfying
    $p_k = \Predecessor{\mathcal{J}}{j}$ (\cref{def:predecessor}).
    If $k + 1 \leq m$ and $p_{k+1} = j$, then we set $j_{\rm prev} = A_{\mathcal{J}}[k+1]$
    and $x = A_{\rm ILF}[k+1]$.
    Otherwise, we set $j_{\rm prev} = A_{\mathcal{J}}[k]$ and $x = A_{\rm ILF}[k]$.
    Note that at this point, we have $j_{\rm prev} = \max\{t \in \mathcal{J} : t \leq j\}$ and $x = \ILF{\Text}[j_{\rm prev}]$.
  \item In $\bigO(1)$ time, we set $y = x + (j - j_{\rm prev})$. By \cref{lm:ilf-upper-bound}, it holds
    $y = \ILF{\Text}[j]$. Thus, we return $y$ as the answer.
  \end{enumerate}
  In total, the query takes $\bigO(\log \log \Textlen)$ time.
\end{proof}

\begin{theorem}\label{th:ilf-upper-bound-integer-alphabet}
  For every nonempty text $\Text \in \IntegerAlphabet^{\Textlen}$,
  there exists a data structure of size $\bigO(\RLBWTSize{\Text})$ (see \cref{def:bwt})
  that answers inverse LF queries on $\Text$ (that, given any $i \in [1 \dd \Textlen]$, return
  $\ILF{\Text}[i]$; see \cref{def:ilf}) in $\bigO(\log \log \Textlen)$ time.
\end{theorem}
\begin{proof}

  Let $\Text' = (\bigodot_{i=1}^{\Textlen}(\Text[i] + 1)) \cdot \zero \in
  [0 \dd \AlphabetSize + 1)^{\Textlen + 1}$. Let also
  $i_{\rm first} = \ISA{\Text}[1]$ and
  $i_{\rm last} = \ISA{\Text}[\Textlen]$ (\cref{def:isa}).

  The data structure to answer inverse LF queries on $\Text$ consists of the
  following components:
  \begin{enumerate}
  \item The integers $i_{\rm first}$ and $i_{\rm last}$ stored in $\bigO(1)$ space.
  \item The data structure from \cref{th:ilf-upper-bound-integer-alphabet-with-dollar}
    for text $\Text'$. Note that, by \cref{lm:number-of-runs-after-appending-dollar}, it holds
    $\RLBWTSize{\Text'} \leq \RLBWTSize{\Text} + 3$. Thus,
    the structure needs
    $\bigO(\RLBWTSize{\Text'}) = \bigO(\RLBWTSize{\Text})$ space.
  \end{enumerate}
  In total, the structure needs $\bigO(\RLBWTSize{\Text})$ space.

  Let $i \in [1 \dd \Textlen]$.
  To compute $\ILF{\Text}[i]$ using the above data structure, we proceed as follows:
  \begin{enumerate}
  \item If $i = i_{\rm last}$, then in $\bigO(1)$ time
    we return $\ILF{\Text}[i] = i_{\rm first}$ (this holds by \cref{def:ilf}),
    and conclude the query algorithm. Let us thus assume that $i \neq i_{\rm last}$.
  \item Using the structure from \cref{th:ilf-upper-bound-integer-alphabet-with-dollar},
    in $\bigO(\log \log \Textlen)$ time, we compute
    $x = \ILF{\Text'}[i+1]$.
  \item In $\bigO(1)$ time, we set $y = x - 1$. By \cref{lm:ilf-for-string-without-dollar},
    it holds $y = \ILF{\Text}[i]$. Thus, we return $y$ as the answer.
  \end{enumerate}
  In total, the query takes $\bigO(\log \log \Textlen)$ time.
\end{proof}

\begin{theorem}\label{th:ilf-upper-bound-in-terms-of-r}
  For every nonempty text $\Text \in \Sigma^{\Textlen}$,
  there exists a data structure of size $\bigO(\RLBWTSize{\Text})$ (see \cref{def:bwt})
  that answers inverse LF queries on $\Text$ (that, given any $i \in [1 \dd \Textlen]$, return
  $\ILF{\Text}[i]$; see \cref{def:ilf}) in $\bigO(\log \log \Textlen)$ time.
\end{theorem}
\begin{proof}

  Let $\Sigma_{\Text}$ denote the effective alphabet of $\Text$, i.e., $\Sigma_{\Text} = \{\Text[i] : i \in [1 \dd \Textlen]\}$.
  Let $\Text_{\rm rank}[1 \dd \Textlen]$ denote a string defined so that, for every $i \in [1 \dd \Textlen]$,
  $\Text_{\rm rank}[i]$ is the rank of $\Text[i]$ in $\Sigma_{\Text}$,
  i.e., $\Text_{\rm rank}[i] = |\{c \in \Sigma_{\Text} : c \prec \Text[i]\}|$.

  The data structure to answer inverse LF queries on $\Text$ consists of a single
  component: the structure from \cref{th:ilf-upper-bound-integer-alphabet} for text $\Text_{\rm rank}$.
  Since $\RLBWTSize{\Text} = \RLBWTSize{\Text_{\rm rank}}$, the structure from \cref{th:ilf-upper-bound-integer-alphabet}
  (and hence the whole structure) needs $\bigO(\RLBWTSize{\Text})$ space.

  Let $i \in [1 \dd \Textlen]$.
  To compute $\ILF{\Text}[i]$ using the above data structure, we simply compute
  and return $\ILF{\Text_{\rm rank}}[i] = \ILF{\Text}[i]$ in $\bigO(\log \log \Textlen)$ using
  the structure from \cref{th:ilf-upper-bound-integer-alphabet}.
\end{proof}

\begin{theorem}\label{th:ilf-upper-bound-in-terms-of-delta}
  For every nonempty text $\Text \in \Sigma^{\Textlen}$,
  there exists a data structure of size $\bigO(\SubstringComplexity{\Text} \log^2 \Textlen)$
  (see \cref{def:delta})
  that answers inverse LF queries on $\Text$ (that, given any $i \in [1 \dd \Textlen]$, return
  $\ILF{\Text}[i]$; see \cref{def:ilf}) in $\bigO(\log \log \Textlen)$ time.
\end{theorem}
\begin{proof}
  The claim follows by combining \cref{th:ilf-upper-bound-in-terms-of-r} and
  \cref{th:rlbwt-size} (see also \cref{rm:rlbwt-size}).
\end{proof}

\subsection{\boldmath Lexicographical Predecessor (\texorpdfstring{$\Phi$}{Φ}) Queries}\label{sec:phi}

\subsubsection{Problem Definition}\label{sec:phi-problem-def}
\vspace{-1.5ex}

\begin{framed}
  \noindent
  \probname{Indexing for Lexicographical Predecessor ($\Phi$) Queries}
  \begin{bfdescription}
  \item[Input:]
    A string $\Text \in \Sigma^{\Textlen}$.
  \item[Output:]
    A data structure that, given any index $j \in [1 \dd \Textlen]$,
    returns $\PhiArray{\Text}[j]$ (\cref{def:phi-array}).
  \end{bfdescription}
\end{framed}
\vspace{2ex}

\subsubsection{Lower Bound}\label{sec:phi-lower-bound}

\begin{lemma}\label{lm:reduce-predecessor-to-phi}
  Let $A \subseteq [1 \dd m^2]$ be a nonempty set of size $|A| = m$.
  Denote $A = \{a_1, a_2, \dots, a_m\}$, where $a_1 < a_2 < \dots < a_m$, and
  let $a_{0} = -\infty$.
  Let
  \[
    \Text =
      \Big( \textstyle\bigodot_{i=1}^{m} (\zero^{a_i}\one^{m^2-a_i+2}) \Big) \cdot
      \Big( \zero^{m^2+1}\one \Big) \cdot
      \Big( \zero^{m^2}\one^{m^2+2} \Big) \in
      \BinaryAlphabet^{*}
  \]
  (brackets added for clarity). Then, for every $x \in [1 \dd m^2]$, it holds
  (see \cref{def:predecessor})
  \[
    \Predecessor{A}{x} = a_i,
  \]
  where $\Delta = |\Text| - 2(m^2 + 1)$
  and $i = \left\lceil \tfrac{\PhiArray{\Text}[\Delta + m^2 - x + 1]}{m^2+2} \right\rceil - 1$
  (\cref{def:phi-array}).
\end{lemma}
\begin{proof}
  Denote $j = \Delta + m^2 - x + 1$ and let $a_{m+1} = m^2 + 1$. Let
  $p = \min\{t \in [1 \dd m+1] : a_t \geq x\}$.
  Note that that $p$ is well-defined, since $a_{m+1} = m^2 + 1$.
  For every $t \in [p \dd m+1]$, denote $s_t = (t-1)(m^2+2) + (a_{t}-x+1)$.
  By definition of $\Text$, it holds $|\OccTwo{\zero^{x}\one}{\Text}| = n - p + 3$ and, moreover,
  $\OccTwo{\zero^{v}\one}{\Text} = \{s_t\}_{t \in [p \dd n+1]} \cup \{j\}$.
  Next, observe that:
  \begin{itemize}
  \item $\Text[j \dd |\Text|]$ has $\zero^{x}\one^{m^2+2}$ as a prefix and
  \item for every $t \in [p \dd m+1]$, $\Text[s_t \dd |\Text|]$
    has $\zero^{x}\one^{m^2-a_t+2}\zero$ as a prefix.
  \end{itemize}
  This implies that it holds
  $\Text[s_{m+1} \dd |\Text|] \prec
    \Text[s_{m} \dd |\Text|] \prec \dots \prec
    \Text[s_{p} \dd |\Text|] \prec
    \Text[j \dd |\Text|]$.
  Consequently, it holds $\PhiArray{\Text}[j] = s_{p}$. It remains to note that
  for every $t \in [p \dd m+1]$, it holds $\lceil s_t/(m^2+2) \rceil = t$.
  Combining this with $\Predecessor{A}{x} = a_{p-1}$ (following from
  the definition of $p$), we obtain
  \begin{align*}
    i
      &= \left\lceil \frac{\PhiArray{\Text}[\Delta + m^2 - x + 1]}{m^2+2} \right\rceil - 1
      = \left\lceil \frac{\PhiArray{\Text}[j]}{m^2+2} \right\rceil - 1
      = \left\lceil \frac{s_p}{m^2+2} \right\rceil - 1
      = p - 1.
  \end{align*}
  Thus, indeed we have $\Predecessor{A}{x} = a_{p-1} = a_{i}$.
\end{proof}

\begin{theorem}\label{th:phi-lower-bound}
  There is no data structure that, for every text $\Text \in \BinaryAlphabet^{\Textlen}$,
  uses $\bigO(\SubstringComplexity{\Text} \log^{\bigO(1)}
  \Textlen)$ space and answers $\Phi$ queries on $\Text$ (that, given
  any $j \in [1 \dd \Textlen]$, return $\PhiArray{\Text}[j]$; see
  \cref{def:phi-array}) in $o(\log \log \Textlen)$ time.
\end{theorem}
\begin{proof}

  Suppose that the claim does not hold. Let $D_{\Phi}$ denote the
  hypothetical data structure answering $\Phi$ queries, and let $c =
  \bigO(1)$ be a positive constant such that, that for
  a text $\Text \in \BinaryAlphabet^{\Textlen}$, $D_{\Phi}$
  uses $\bigO(\SubstringComplexity{\Text}
  \log^c \Textlen)$ space, and answers $\Phi$ queries on $\Text$ in
  $t_{\Phi}(|\Text|) = o(\log \log |\Text|)$ time.  We will prove
  that this implies that there exists a data structure answering
  predecessor queries,
  that contradicts \cref{th:predecessor-lower-bound}.

  Let $m \geq 1$ and let $A \subseteq [1 \dd m^2]$ be a nonempty set
  of size $m$.  Denote $A = \{a_1, a_2, \dots, a_m\}$, where $a_1 <
  a_2 < \dots < a_m$.  Let $\Text_{A} =
  (\bigodot_{i=1}^{m}(\zero^{a_i}\one^{m^2-a_i+2})) \cdot
  (\zero^{m^2+1}\one) \cdot (\zero^{m^2}\one^{m^2+2}) \in
  \BinaryAlphabet^{*}$ be a text defined as in
  \cref{lm:reduce-predecessor-to-phi}.  Denote $\Delta = |\Text_{A}|
  - 2(m^2 + 1)$. Let $A_{\rm val}[0 \dd m]$ be an array such that
  $A_{\rm val}[0] = -\infty$ and, for every $i \in [1 \dd m]$, $A_{\rm
  val}[i] = a_i$.

  Let $D_{\rm pred}$ denote the data structure consisting of the
  following components:
  \begin{enumerate}
  \item The array $A_{\rm val}[0 \dd m]$ stored in plain form. It
    needs $\bigO(m)$ space.
  \item The integer $\Delta$ stored in $\bigO(1)$ space.
  \item The data structure $D_{\Phi}$ for the string $\Text_{A}$.
    Note that
    \[
      |\Text_{A}|
        = m^3 + 3m^2 + 2m + 3
        \in \bigO(m^3).
    \]
    Moreover, note that $|\RL{\Text_{A}}| = 2(m+2)$. By applying
    \cref{ob:rl} and
    \cref{lm:z-and-r-upper-bound}\eqref{lm:z-and-r-upper-bound-it-1},
    we thus have $\SubstringComplexity{\Text_{A}} = \bigO(m)$.
    Consequently, $D_{\Phi}$ for $\Text_{A}$ needs
    $\bigO(\SubstringComplexity{\Text_{A}} \log^c |\Text_{A}|) =
    \bigO(m \log^c m)$ space.
  \end{enumerate}
  In total, $D_{\rm pred}$ needs $\bigO(m \log^c m)$ space.

  Given any $x \in \Z$, we compute $\Predecessor{A}{x}$ as follows:
  \begin{enumerate}
  \item If $x < 1$ (resp.\ $x > m^2$), we return $\Predecessor{A}{x} =
    -\infty$ (resp.\ $\Predecessor{A}{x} = A_{\rm val}[m]$) in
    $\bigO(1)$ time, and conclude the query algorithm. Let us thus
    assume that $x \in [1 \dd m^2]$.
  \item Using $D_{\Phi}$, in $\bigO(t_{\Phi}(|\Text_{A}|))$
    time we compute $j = \PhiArray{\Text_{A}}[\Delta + m^2 - x + 1]$.
    We then let $i = \lceil j/(m^2+2) \rceil - 1$, and return $p = A_{\rm
    val}[i]$ as the answer.  By
    \cref{lm:reduce-predecessor-to-phi}, it holds $p =
    \Predecessor{A}{x}$.
  \end{enumerate}
  In total, the query takes $\bigO(t_{\Phi}(|\Text_{A}|)) = o(\log
  \log |\Text_{A}|) = o(\log \log (m^3)) = o(\log \log m)$ time.

  We have thus proved that there exists a data structure that,
  for every set $A \subseteq [1 \dd m^2]$ of size $|A| = m$, uses
  $\bigO(m \log^c m) = \bigO(m \log^{\bigO(1)} m)$ space, and answers
  predecessor queries on $A$ in $o(\log \log m)$ time. This
  contradicts \cref{th:predecessor-lower-bound}.
\end{proof}

\subsection{\boldmath Inverse Lexicographical Predecessor (\texorpdfstring{$\Phi^{-1}$}{Φ⁻¹}) Queries}\label{sec:iphi}

\subsubsection{Problem Definition}\label{sec:iphi-problem-def}
\vspace{-1.5ex}

\begin{framed}
  \noindent
  \probname{Indexing for Inverse Lexicographical Predecessor ($\Phi^{-1}$) Queries}
  \begin{bfdescription}
  \item[Input:]
    A string $\Text \in \Sigma^{\Textlen}$.
  \item[Output:]
    A data structure that, given any index $j \in [1 \dd \Textlen]$,
    returns $\InvPhiArray{\Text}[j]$ (\cref{def:inv-phi-array}).
  \end{bfdescription}
\end{framed}
\vspace{2ex}

\subsubsection{Lower Bound}\label{sec:iphi-lower-bound}

\begin{lemma}\label{lm:invert-phi}
  Let $\AlphabetSize \geq 1$ and $\Text \in \IntegerAlphabet^{\Textlen}$ be a nonempty text.
  Let
  \[
    \Text' = \left(
          \textstyle\bigodot_{i=1}^{\Textlen}
            \Big(
              \zero\zero\one \cdot (\AlphabetSize - 1 - \Text[i]) \cdot \one
            \Big)
        \right) \cdot \one \in \IntegerAlphabet^{*}
  \]
  (brackets added for clarity). Then:
  \begin{itemize}
  \item For every $j \in [1 \dd \Textlen] \setminus \{\SA{\Text}[1]\}$,
    it holds (see \cref{def:phi-array,def:inv-phi-array})
    \[
      \PhiArray{\Text}[j] = \tfrac{1}{5} \Big( \InvPhiArray{\Text'}[1 + 5(j-1)] - 1 \Big) + 1.
    \]
  \item For every $j \in [1 \dd \Textlen] \setminus \{\SA{\Text}[\Textlen]\}$,
    it holds
    \[
      \InvPhiArray{\Text}[j] = \tfrac{1}{5} \Big( \PhiArray{\Text'}[1 + 5(j-1)] - 1 \Big) + 1.
    \]
  \end{itemize}
\end{lemma}
\begin{proof}
  Let $\Textlen' = |\Text'| = 5\Textlen + 1$.
  Let $(p_j)_{j \in [1 \dd \Textlen]}$ be a sequence defined by
  $p_j = 1 + 5(j-1)$. For every $a \in \IntegerAlphabet$, denote
  $s(a) = \zero\zero\one \cdot (\AlphabetSize - 1 - a) \cdot \one \in \IntegerAlphabet^{*}$. Observe
  that for every $j \in [1 \dd \Textlen]$, we then have
  \[
    \Text'[p_j \dd \Textlen'] = s(\Text[j])s(\Text[j+1]) \cdots s(\Text[\Textlen]) \cdot \one.
  \]
  Finally, let $\alpha = \RangeBegTwo{\zero\zero}{\Text'}$.

  The proof proceeds in three steps:
  \begin{enumerate}

  \item First, we prove that, for every
    $j_1, j_2 \in [1 \dd \Textlen]$,
    $\Text[j_1 \dd \Textlen] \prec \Text[j_2 \dd \Textlen]$ implies that
    $\Text'[p_{j_1} \dd \Textlen'] \succ \Text'[p_{j_2} \dd \Textlen']$.
    We consider two cases:
    \begin{itemize}
    \item First, assume that $\Text[j_1 \dd \Textlen]$ is a proper
      prefix of $\Text[j_2 \dd \Textlen]$. Note that, letting $\ell = \Textlen - j_1 + 1$, we then
      have $j_2 < j_1$
      and
      \begin{align*}
        \Text'[p_{j_1} \dd p_{j_1} + 5\ell)
          &= s(\Text[j_1])s(\Text[j_1+1]) \cdots s(\Text[\Textlen])\\
          &= s(\Text[j_2])s(\Text[j_2+1]) \cdots s(\Text[j_2+\ell-1])\\
          &= \Text'[p_{j_2} \dd p_{j_2} + 5\ell).
      \end{align*}
      Moreover, by
      $p_{j_1} + 5\ell = 1 + 5(j_1 - 1) + 5(\Textlen - j_1 + 1) = 5\Textlen + 1 = \Textlen'$,
      we then have $\Text'[p_{j_1} + 5\ell] = \one$. On the other hand,
      $p_{j_2} + 5\ell = 1 + 5(j_2-1) + 5(\Textlen - j_1 + 1) =
      5\Textlen + 1 - 5(j_1-j_2) = \Textlen' - 5(j_1-j_2) < \Textlen'$ implies that
      $\Text'[p_{j_2} + 5\ell] = \zero$ (this follows since, for every $a \in \IntegerAlphabet$, the first
      symbol of $s(a)$ is $\zero$). Thus, we have
      $\Text'[p_{j_1} \dd \Textlen'] \succ \Text'[p_{j_2} \dd \Textlen']$.
    \item Let us now assume that there exists $\ell \in [0 \dd \min(\Textlen - j_1 + 1, \Textlen - j_2 + 1))$
      such that $\Text[j_1 \dd j_1 + \ell) = \Text[j_2 \dd j_2 + \ell)$ and $\Text[j_1 + \ell] \prec \Text[j_2 + \ell]$.
      By definition of $\Text'$, we then
      have
      \begin{align*}
        \Text'[p_{j_1} \dd p_{j_1} + 5\ell)
          &= s(\Text[j_1])s(\Text[j_1+1]) \cdots s(\Text[j_1+\ell-1])\\
          &= s(\Text[j_2])s(\Text[j_2+1]) \cdots s(\Text[j_2+\ell-1])\\
          &= \Text'[p_{j_2} \dd p_{j_2} + 5\ell).
       \end{align*}
       Moreover, by $\Text[j_1 + \ell] \prec \Text[j_2 + \ell]$, we then have
       \begin{align*}
         \Text'[p_{j_1} + 5\ell \dd p_{j_1} + 5(\ell+1))
           &= s(\Text[j_1 + \ell])\\
           &\succ s(\Text[j_2 + \ell])\\
           &=\Text'[p_{j_2} + 5\ell \dd p_{j_2} + 5(\ell+1)).
       \end{align*}
       Thus, we obtain $\Text'[p_{j_1} \dd \Textlen'] \succ \Text'[p_{j_2} \dd \Textlen']$.
    \end{itemize}

  \item In the second step, we prove that, for every $i \in [1 \dd \Textlen]$, it holds
    \[
      \SA{\Text'}[\alpha + i] = p_{\SA{\Text}[(\Textlen - i) + 1]}.
    \]
    Recall that, by definition of the suffix array, it holds
    $\Text[\SA{\Text}[1] \dd \Textlen] \prec \cdots \prec \Text[\SA{\Text}[\Textlen] \dd \Textlen]$.
    By the above step, we thus have
    \[
      \Text'[p_{\SA{\Text}[\Textlen]} \dd \Textlen'] \prec \cdots \prec \Text'[p_{\SA{\Text}[1]} \dd \Textlen'].
    \]
    It remains to observe that $\OccTwo{\zero\zero}{\Text'} = \{p_1, p_2, \dots, p_{\Textlen}\}$.
    Recalling that $\alpha = \RangeBegTwo{\zero\zero}{\Text'}$, we thus have
    $\SA{\Text'}[\alpha + i] = p_{\SA{\Text}[(\Textlen - i) + 1]}$ for every $i \in [1 \dd \Textlen]$.

  \item We are now ready to show the claims. Let $j \in [1 \dd \Textlen] \setminus \{\SA{\Text}[1]\}$.
    Let $i = \ISA{\Text}[j]$. By \cref{def:phi-array}, we then have $i > 1$ and
    $\PhiArray{\Text}[j] = \SA{\Text}[i-1]$. In the previous step, we proved
    that $\InvPhiArray{\Text'}[p_{\SA{\Text}[i]}] = p_{\SA{\Text}[i-1]}$.
    Substituting $\SA{\Text}[i]$ with $j$ and $\SA{\Text}[i-1]$ with $\PhiArray{\Text}[j]$, we thus 
    obtain that $\InvPhiArray{\Text'}[p_{j}] = p_{\PhiArray{\Text}[j]}$. Expanding the definition
    of $p_{j}$ and $p_{\PhiArray{\Text}[j]}$, we therefore have
    \[
      \InvPhiArray{\Text'}[1 + 5(j-1)] = 1 + 5 \cdot (\PhiArray{\Text}[j] - 1).
    \]
    By rewriting for $\PhiArray{\Text}[j]$, we obtain
    $\PhiArray{\Text}[j] = \tfrac{1}{5}(\InvPhiArray{\Text'}[1 + 5(j-1)] - 1) + 1$, i.e., the first claim.
    Let us now consider $j \in [1 \dd \Textlen] \setminus \{\SA{\Text}[\Textlen]\}$.
    Let again $i = \ISA{\Text}[j]$. By \cref{def:inv-phi-array}, we have $i < \Textlen$ and
    $\InvPhiArray{\Text}[j] = \SA{\Text}[i+1]$. In the previous step, we proved that
    $\PhiArray{\Text'}[p_{\SA{\Text}[i]}] = p_{\SA{\Text}[i+1]}$.
    Substituting $\SA{\Text}[i]$ with $j$ and $\SA{\Text}[i+1]$ with $\InvPhiArray{\Text}[j]$,
    we thus obtain that $\PhiArray{\Text'}[p_{j}] = p_{\InvPhiArray{\Text}[j]}$. Expanding the definition
    of $p_{j}$ and $p_{\InvPhiArray{\Text}[j]}$, we therefore have
    \[
      \PhiArray{\Text'}[1 + 5(j-1)] = 1 + 5 \cdot (\InvPhiArray{\Text}[j] - 1).
    \]
    By rewriting for $\InvPhiArray{\Text}[j]$, we obtain
    $\InvPhiArray{\Text}[j] = \tfrac{1}{5}(\PhiArray{\Text'}[1 + 5(j-1)] - 1) + 1$, i.e., the second claim.
    \qedhere
  \end{enumerate}
\end{proof}

\begin{lemma}\label{lm:delta-for-morphism}
  Let $\Sigma$ be a nonempty set, and let $f : \Sigma \rightarrow \Sigma^{k}$, where $k \in \Zp$, be any function.
  Then, for every nonempty text $\Text \in \Sigma^{\Textlen}$, it holds (see \cref{def:delta})
  \[
    \SubstringComplexity{\Text'} = \bigO(k \cdot \SubstringComplexity{\Text} \cdot \log \Textlen),
  \]
  where $\Text' = \bigodot_{i=1}^{\Textlen} f(\Text[i])$.
\end{lemma}
\begin{proof}
  The proof proceeds in two steps:
  \begin{enumerate}
  \item First, we show that $\LZSize{\Text'} \leq k \cdot \LZSize{\Text}$. To this end, we observe that
    $\Text'$ has an LZ77-like factorization consisting of $k \cdot \LZSize{\Text}$ phrases, which is constructed as follows:
    \begin{itemize}
    \item If $\Text[i]$ in the LZ77 factorization is encoded as a literal phrase,
      then in the LZ77-like parsing of $\Text'$, we encode symbols in $\Text'(k(i-1) \dd ki]$ as $k$ literal phrases.
    \item If the substring $\Text(i \dd j]$ is encoded as a repeat phrase
      in the LZ77 factorization of $\Text$,
      with a source starting at $i' < i$, then in the LZ77-like parsing
      of $T'$, we encode the substring $\Text(ki \dd kj]$ as a repeat phrase,
      with a source starting at position $1 + k(i'-1)$.
    \end{itemize}
    It remains to apply \cref{th:lz77-size} to obtain $\LZSize{\Text'} \leq k \cdot \LZSize{\Text}$.
  \item We are now ready to complete the proof. By \cref{lm:lz-upper-bound} (see also \cref{rm:lz-upper-bound}),
    it holds $\LZSize{\Text} = \bigO(\SubstringComplexity{\Text} \cdot \log \Textlen)$. On the other hand,
    by \cref{lm:z-and-r-upper-bound}\eqref{lm:z-and-r-upper-bound-it-1}, we have
    $\SubstringComplexity{\Text'} = \bigO(\LZSize{\Text'})$. Putting these two observations together
    with the previous step, we thus obtain
    \[
      \SubstringComplexity{\Text'}
        = \bigO(\LZSize{\Text'})
        = \bigO(k \cdot \LZSize{\Text})
        = \bigO(k \cdot \SubstringComplexity{\Text} \cdot \log \Textlen).
        \qedhere
    \]
  \end{enumerate}
\end{proof}

\begin{theorem}\label{th:phi-and-invphi-equvalence}
  Let $\AlphabetSize \geq 1$ and $\Text \in \IntegerAlphabet^{\Textlen}$ be a nonempty text.
  The $\PhiArray{\Text}[j]$ and $\InvPhiArray{\Text}[j]$ queries (\cref{def:phi-array,def:inv-phi-array})
  are related in the following sense:
  If there exists a data structure using $\bigO(\SubstringComplexity{\Text} \log^c \Textlen)$ space
  (where $c = \bigO(1)$ is a positive constant) that answers one of these two types of queries
  in $t_{\rm query}(\Textlen)$ time, then there exists a
  data structure answering the other type of queries also in $t_{\rm query}(5\Textlen + 1)$ time and using
  $\bigO(\SubstringComplexity{\Text} \log^{c+1} \Textlen)$ space.
\end{theorem}
\begin{proof}

  For concreteness, we prove the equivalence in one direction. The proof in the opposite direction
  follows symmetrically (relying on the symmetry of \cref{lm:invert-phi}).
  Let $D_{\Phi}$ be a data structure that, for every $\Text \in \IntegerAlphabet^{\Textlen}$,
  uses $\bigO(\SubstringComplexity{\Text} \log^{c} \Textlen)$ space (where $c = \bigO(1)$ is a positive constant)
  and, given any $j \in [1 \dd \Textlen]$, returns $\PhiArray{\Text}[j]$ in $t_{\rm query}(\Textlen)$ time. We
  will present a data structure that uses $\bigO(\SubstringComplexity{\Text} \log^{c+1} \Textlen)$ space
  and, given any $j \in [1 \dd \Textlen]$, returns $\InvPhiArray{\Text}[j]$ in $t_{\rm query}(5\Textlen + 1)$ time.

  Let $\Text \in \IntegerAlphabet^{\Textlen}$ and $\Text' =
  (\bigodot_{i=1}^{\Textlen} (\zero\zero\one \cdot (\AlphabetSize - 1 - \Text[i]) \cdot \one))
  \cdot \one \in \IntegerAlphabet^{*}$ be a text defined as in \cref{lm:invert-phi}. Let
  $j_{\rm lexfirst} = \SA{\Text}[1]$ and
  $j_{\rm lexlast} = \SA{\Text}[\Textlen]$.

  Let $D_{\Phi^{-1}}$ denote a data structure consisting of the following components:
  \begin{enumerate}
  \item The integers $j_{\rm lexfirst}$ and $j_{\rm lexlast}$ using $\bigO(1)$ space.
  \item The data structure $D_{\Phi}$ for the string $\Text'$. Note that
    $|\Text'| = 5\Textlen + 1$. To show an upper bound for $\SubstringComplexity{\Text'}$,
    we define $\Text'_{\rm pref}$ as a string satisfying
    $\Text' = \Text'_{\rm pref} \cdot \one$. By \cref{lm:delta-for-morphism},
    we have $\SubstringComplexity{\Text'_{\rm pref}}
    = \bigO(\SubstringComplexity{\Text} \log \Textlen)$. On the other hand,
    by \cref{lm:delta-after-appending}, it holds $\SubstringComplexity{\Text'} =
    \bigO(\SubstringComplexity{\Text'_{\rm pref}})$. Thus, we obtain
    $\SubstringComplexity{\Text'} = \bigO(\SubstringComplexity{\Text} \log \Textlen)$.
    Consequently, $D_{\Phi}$ for $\Text'$ needs
    $\bigO(\SubstringComplexity{\Text'} \log^{c} |\Text'|) =
    \bigO((\SubstringComplexity{\Text} \log \Textlen) \cdot \log^{c} |\Text'|) =
    \bigO((\SubstringComplexity{\Text} \log \Textlen) \cdot \log^{c} \Textlen) =
    \bigO(\SubstringComplexity{\Text} \log^{c+1} \Textlen)$ space.
  \end{enumerate}
  In total, $D_{\Phi^{-1}}$ needs $\bigO(\SubstringComplexity{\Text} \log^{c+1} \Textlen)$ space.

  Given any $j \in [1 \dd \Textlen]$, we compute $\InvPhiArray{\Text}[j]$ as follows:
  \begin{enumerate}
  \item If $j = j_{\rm lexlast}$, then in $\bigO(1)$ time we return $\InvPhiArray{\Text}[j] = j_{\rm lexfirst}$.
    Let us now assume that $j \neq j_{\rm lexlast}$.
  \item Using $D_{\Phi}$, in $t_{\rm query}(|\Text'|) = t_{\rm query}(5\Textlen + 1)$ time we
    compute $j' = \PhiArray{\Text'}[1 + 5(j-1)]$. We then set $j'' = \tfrac{1}{5}(j' - 1) + 1$.
    By \cref{lm:invert-phi}, it holds $\InvPhiArray{\Text}[j] = j''$. Thus, we return
    $j''$ as the answer.
  \end{enumerate}
  In total, the query takes $\bigO(1) + t_{\rm query}(5\Textlen + 1) = t_{\rm query}(5\Textlen + 1)$ time.
\end{proof}

\begin{theorem}\label{th:iphi-lower-bound}
  There is no data structure that, for every text $\Text \in \BinaryAlphabet^{\Textlen}$,
  uses $\bigO(\SubstringComplexity{\Text} \log^{\bigO(1)}
  \Textlen)$ space and answers $\Phi^{-1}$ queries on $\Text$ (that, given
  any $j \in [1 \dd \Textlen]$, return $\InvPhiArray{\Text}[j]$; see
  \cref{def:inv-phi-array}) in $o(\log \log \Textlen)$ time.
\end{theorem}
\begin{proof}
  Suppose that the claim does not hold. Let $D_{\Phi^{-1}}$ denote the
  hypothetical data structure answering $\Phi^{-1}$ queries, and let $c =
  \bigO(1)$ be a positive constant such that, that for
  a text $\Text \in \BinaryAlphabet^{\Textlen}$, $D_{\Phi^{-1}}$
  uses $\bigO(\SubstringComplexity{\Text}
  \log^c \Textlen)$ space, and answers $\Phi^{-1}$ queries on $\Text$ in
  $t_{\Phi^{-1}}(|\Text|) = o(\log \log |\Text|)$ time. By
  \cref{th:phi-and-invphi-equvalence}, this implies that,
  for every text $\Text \in \BinaryAlphabet^{\Textlen}$,
  there exists a data structure of size $\bigO(\SubstringComplexity{\Text} \log^{c+1} \Textlen) =
  \bigO(\SubstringComplexity{\Text} \log^{\bigO(1)} \Textlen)$ that answers
  $\Phi$ queries on $\Text$ in $t_{\Phi^{-1}}(5|\Text| + 1) = o(\log \log (5\Textlen + 1))
  = o(\log \log \Textlen)$ time.
  This contradicts \cref{th:phi-lower-bound}.
\end{proof}

\appendix

\section{Stronger Lower Bound for Inverse Suffix Array
  (\texorpdfstring{$\text{SA}^{-1}$}{SA⁻¹}) Queries}\label{sec:isa-lower-bound-2}

In this section, we present a reduction from range counting queries to inverse
suffix array queries. While this does not change the main result of the paper
--- we can already prove an $\Omega\left(\tfrac{\log \Textlen}{\log \log \Textlen}\right)$
lower bound for inverse suffix array queries by reducing from random access queries
(which in turn are proved via a reduction from range counting \emph{parity}~\cite{VerbinY13}) ---
the reduction below (which is of independent interest) establishes a strictly
stronger result: namely, that inverse suffix array can be used to answer
full range counting queries.

\begin{lemma}\label{lm:reduce-range-count-to-isa}
  Let $A[1 \dd n]$ be a nonempty array containing a permutation
  of $\{1, 2, \dots, n\}$. Let
  \[
    \Text = \Big( \textstyle\bigodot_{i=1}^{n} (\zero^{A[i]} \one^{i}) \Big) \cdot
            \zero^{n+1}\one^{n+1} \cdot
            \Big( \textstyle\bigodot_{i=1}^{n+1} (\zero^{n+1} \one^{i}) \Big)
    \in \BinaryAlphabet^{*}
  \]
  (brackets added for clarity). Then, for every $j \in [0 \dd n]$ and every
  $v \in [1 \dd n]$, it holds (see \cref{def:range-count-and-select,def:isa})
  \[
    \TwoSidedRangeCount{A}{j}{v} = \ISA{\Text}[j'] - (b + j + 1),
  \]
  where
  \begin{itemize}
  \item $b = \RangeBegTwo{\zero^{v}\one}{\Text}$ (\cref{def:occ}),
  \item $\ell_1 = n \cdot (n+1)$,
  \item $\ell_2 = 2 \cdot (n+1)$,
  \item $\delta = j \cdot (n + 1 + \tfrac{j+1}{2}) + (n+2-v)$, and
  \item $j' = \ell_1 + \ell_2 + \delta$.
  \end{itemize}
\end{lemma}
\begin{proof}
  Denote $r = \TwoSidedRangeCount{A}{j}{v}$ and $k = \TwoSidedRangeCount{A}{n}{v}$.
  Let $(a_i)_{i \in [1 \dd k]}$
  be a sequence such that, for every $i \in [1 \dd k]$,
  $a_i = \RangeSelect{A}{r}{v}$.
  By definition, we then have $a_1 < a_2 < \dots < a_{r} \leq j < a_{r+1} < \dots < a_k$.
  Next, let $(s_i)_{i \in [1 \dd k]}$
  be a sequence defined such that for every $i \in [1 \dd k]$,
  \[
    s_i = \Big( \textstyle\sum_{t=1}^{a_i-1} (A[t]+t) \Big) +
    \Big(A[a_i]-v+1 \Big).
  \]
  Let also
  \[
    u = \ell_1 + (n+2-v).
  \]
  Next, denote $q = n + 1$, and let $(t_i)_{i \in [1 \dd q]}$
  be a sequence defined so that for every $i \in [1 \dd q]$,
  \[
    t_i = \ell_1 + \ell_2 + i \cdot (n + 1 + \tfrac{i+1}{2}) + (n+2-v).
  \]
  Observe that by definition of $\Text$, it follows that
  $|\OccTwo{\zero^v\one}{\Text}| = k + 1 + q$ and, moreover,
  $\OccTwo{\zero^v\one}{\Text}$ is a disjoint union
  \[
    \OccTwo{\zero^v\one}{\Text} =
    \{s_i\}_{i \in [1 \dd k]} \cup \{u\} \cup \{t_i\}_{i \in [1 \dd q]}.
  \]
  Finally, note that it holds
  \[
    \Text[j' \dd |\Text|] = \zero^v \one^{j+1} \cdot
    \Big(\textstyle\bigodot_{i=j+2}^{n+1} (\zero^{n+1}\one^{i})\Big).
  \]
  The rest of the proof proceeds in four steps:
  \begin{enumerate}

  \item First, we prove that $|\{i \in [1 \dd k] : \Text[s_i \dd
    |\Text|] \prec \Text[j' \dd |\Text|]\}| = r$.
    Observe that for every $i \in [1 \dd k]$,
    the suffix $\Text[s_i \dd |\Text|]$ has the string
    $\zero^{v}\one^{a_1}\zero$ as a prefix. By
    $a_1 < \dots < a_{r} \leq j$ and
    since $\Text[j' \dd |\Text|]$ has $\zero^{v}\one^{j+1}$ as a prefix, we obtain
    $\Text[s_1 \dd |\Text|] \prec \dots \prec
    \Text[s_r \dd |\Text|] \prec
    \Text[j' \dd |\Text|]$.
    We now show that for every $i \in [r+1 \dd k]$, it holds
    $\Text[j' \dd |\Text|] \prec \Text[s_i \dd |\Text|]$.
    Consider any $i \in [r+1 \dd k]$. Note that the existence of
    such $i$ implies that $j < n$, and hence also that 
    $\Text[j' \dd |\Text|]$ has
    $\zero^{v}\one^{j+1}\zero^{n+1}\one$ as a prefix. Consider two cases.
    \begin{itemize}
    \item First, assume that $i \in [r+2 \dd k]$.
      Recall that $\Text[s_i \dd |\Text|]$ has $\zero^{v}\one^{a_i}\zero$
      as a prefix. By $j+1 \leq a_{r+1} < a_i$, we thus
      immediately obtain $\Text[j' \dd |\Text|] \prec \Text[s_i \dd |\Text|]$.
    \item Let us now consider the case $i = r + 1$. Recall that by definition
      of $r$, we have $j+1 \leq a_{r+1}$. If $j+1 < a_{r+1}$, then we immediately
      obtain $\Text[j' \dd |\Text|] \prec \Text[s_{r+1} \dd |\Text|]$, since
      $\Text[j' \dd |\Text|]$ (resp.\ $\Text[s_{r+1} \dd |\Text|]$) has
      $\zero^{v}\one^{j+1}\zero$ (resp.\ $\zero^{v}\one^{a_{r+1}}\zero$) as a prefix.
      Let us thus assume that $j+1 = a_{r+1}$. If $a_{r+1} < n$, then again
      we obtain $\Text[j' \dd |\Text|] \prec \Text[s_{r+1} \dd |\Text|]$, since
      then $\Text[j' \dd |\Text|]$ (resp.\ $\Text[s_{r+1} \dd |\Text|]$)
      has $\zero^{v}\one^{j+1}\zero^{n+1}\one$ (resp.\
      $\zero^{v}\one^{a_{r+1}}\zero^{A[a_{r+1}+1]}\one =
      \zero^{v}\one^{j+1}\zero^{A[a_{r+1}+1]}\one$)
      as a prefix (recall that $A[t] \leq n$ holds for all $t \in [1 \dd n]$).
      Let us finally consider the remaining case $j+1 = a_{r+1} = n$.
      In this case, we have $\Text[j' \dd |\Text|] = \zero^{v}\one^{n}\zero^{n+1}\one^{n+1}$
      and $\Text[s_{r+1} \dd |\Text|]$ has $\zero^{v}\one^{a_{r+1}}\zero^{n+1}\one^{n+1}\zero
      = \zero^{v}\one^{n}\zero^{n+1}\one^{n+1}\zero$ as a prefix. Thus,
      $\Text[j' \dd |\Text|]$ is a prefix of $\Text[s_{r+1} \dd |\Text|]$, and
      hence $\Text[j' \dd |\Text|] \prec \Text[s_{r+1} \dd |\Text|]$.
    \end{itemize}

  \item Second, we show that $\Text[j' \dd |\Text|] \prec \Text[u \dd |\Text|]$.
    Consider two cases:
    \begin{itemize}
    \item First, assume that $j < n$. Then, $\Text[j' \dd |\Text|]$ has
      $\zero^{v}\one^{j+1}\zero$ as a prefix. Since, on the other hand,
      $\Text[u \dd |\Text|]$ has $\zero^{v}\one^{n+1}$ as a prefix, we thus
      obtain $\Text[j' \dd |\Text|] \prec \Text[u \dd |\Text|]$.
    \item Let us now assume that $j = n$. Note that then $\Text[j' \dd |\Text|]
      = \zero^{v}\one^{n+1}$. Since $\Text[u \dd |\Text|]$ has
      $\zero^{v}\one^{n+1}\zero$ as a prefix, we thus again obtain
      $\Text[j' \dd |\Text|] \prec \Text[u \dd |\Text|]$.
    \end{itemize}

  \item Next, we prove that $|\{i \in [1 \dd q] : \Text[t_i \dd
    |\Text|] \prec \Text[j' \dd |\Text|]\}| = j$. We proceed in three
    steps.
    \begin{itemize}
    \item We start by observing
      that, for every $i \in [1 \dd q-1]$, the suffix $\Text[t_i \dd |\Text|]$
      has the string $\zero^{v}\one^{i}\zero$ as a prefix. Since
      $\Text[j' \dd |\Text|]$ has $\zero^{v}\one^{j+1}$ as a prefix, we obtain
      that for all $i \in [1 \dd j]$, it holds
      $\Text[t_i \dd |\Text|] \prec \Text[j' \dd |\Text|]$.
    \item Next, observe that $j' = t_{j+1}$. Thus, it does not hold that
      $\Text[t_{j+1} \dd |\Text|] \prec \Text[j' \dd |\Text|]$.
    \item Finally, let us consider any $i \in [j+2 \dd q]$. Note that
      the existence of such $i$ implies that $j+1 < j+2 \leq q = n + 1$.
      Thus, $\Text[j' \dd |\Text|]$ then has $\zero^{v}\one^{j+1}\zero$ as
      a prefix. On the other hand, $\Text[t_{i} \dd |\Text|]$
      has $\zero^{v}\one^{i}$ as a prefix. By $j+1 < i$, we thus obtain
      $\Text[j' \dd |\Text|] \prec \Text[t_i \dd |\Text|]$.
    \end{itemize}

  \item We now put everything together. First, note that
    by the three facts proved above, and the earlier
    observation that $|\OccTwo{\zero^{v}\one}{\Text}| = k + 1 + q$
    and $\OccTwo{\zero^v\one}{\Text} = \{s_i\}_{i \in [1 \dd k]} \cup \{u\} \cup
    \{t_i\}_{i \in [1 \dd q]}$, we obtain that
    \[
      |\{t \in \OccTwo{\zero^{v}\one}{\Text} :
        \Text[t \dd |\Text|] \prec \Text[j' \dd |\Text|]\}| = j + r.
    \]
    By $j' \in \OccTwo{\zero^{v}\one}{\Text}$, it follows that
    $\RangeBegTwo{\Text[j' \dd |\Text|]}{\Text} = \RangeBegTwo{\zero^{v}\one}{\Text} +
    |\{t \in \OccTwo{\zero^v\one}{\Text} : \Text[t \dd |\Text|] \prec \Text[j' \dd |\Text|]\}|$.
    Consequently, by definition of $\ISA{\Text}[j']$, we have
    \begin{align*}
      \ISA{\Text}[j']
        &= 1 + \RangeBegTwo{\Text[j' \dd |\Text|]}{\Text}\\
        &= 1 + \RangeBegTwo{\zero^{v}\one}{\Text} +
               |\{t \in \OccTwo{\zero^{v}\one}{\Text} :
                 \Text[t \dd |\Text|] \prec \Text[j' \dd |\Text|]\}|\\
        &= 1 + b + j + r\\
        &= (1 + b + j) + \TwoSidedRangeCount{A}{j}{v},
    \end{align*}
    which is equivalent to the claim.
    \qedhere
  \end{enumerate}
\end{proof}

\begin{theorem}\label{th:isa-lower-bound-2}
  There is no data structure that, for every text $\Text \in \BinaryAlphabet^{\Textlen}$,
  uses $\bigO(\SubstringComplexity{\Text} \log^{\bigO(1)}
  \Textlen)$ space and answers inverse SA queries on $\Text$ (that,
  given any $i \in [1 \dd \Textlen]$, return $\ISA{\Text}[i]$; see
  \cref{def:isa}) in $o(\tfrac{\log \Textlen}{\log \log \Textlen})$
  time.
\end{theorem}
\begin{proof}

  Suppose that the claim does not hold. Let $D_{\rm ISA}$ denote the
  hypothetical data structure answering inverse SA queries, and let
  $c = \bigO(1)$ be a positive constant such that, that for
  a text $\Text \in \BinaryAlphabet^{\Textlen}$, $D_{\rm ISA}$ uses
  $\bigO(\SubstringComplexity{\Text} \log^c \Textlen)$ space, and
  answers inverse SA queries on $\Text$ in $t_{\rm ISA}(|\Text|) =
  o(\tfrac{\log |\Text|}{\log \log |\Text|})$ time.  We will prove
  that this implies that there exists a data structure answering range
  counting queries that contradicts
  \cref{th:range-count-lower-bound}.

  Consider any array $A[1 \dd n]$ containing a permutation of
  $\{1, 2, \dots, n\}$, where $n \geq 1$.
  Let $\Text_{A} = (\bigodot_{i=1}^{n} (\zero^{A[i]} \one^{i})) \cdot
  \zero^{n+1}\one^{n+1} \cdot (\bigodot_{i=1}^{n+1} (\zero^{n+1} \one^{i}))
  \in \BinaryAlphabet^{*}$ be a text defined as in
  \cref{lm:reduce-range-count-to-isa}. Let $R[1 \dd n]$ be
  an array defined so that, for every $v \in [1 \dd n]$,
  $R[i] = \RangeBegTwo{\zero^{i}\one}{\Text_{A}}$ (\cref{def:occ}).

  Let $D_{\rm count}$ denote the data structure consisting of the
  following components:
  \begin{enumerate}
  \item The array $R[1 \dd n]$ stored in plain form. It uses
    $\bigO(n)$ space.
  \item The data structure $D_{\rm ISA}$ for the string $\Text_{A}$.
    Note that
    \begin{align*}
      |\Text_{A}|
        &= (\textstyle\sum_{i=1}^{n}(A[i]+i)) +
           2(n+1) +
           (\textstyle\sum_{i=1}^{n+1}(n+1+i))\\
        &= (\tfrac{5}{2}n+4)(n+1) \in \bigO(n^2).
    \end{align*}
    Moreover, note that $|\RL{\Text_{A}}| = 4(n+1)$. By applying
    \cref{ob:rl} and
    \cref{lm:z-and-r-upper-bound}\eqref{lm:z-and-r-upper-bound-it-1},
    we thus obtain $\SubstringComplexity{\Text_{A}} = \bigO(n)$.
    Consequently, $D_{\rm ISA}$ for $\Text_{A}$ needs
    $\bigO(\SubstringComplexity{\Text_{A}} \log^c |\Text_{A}|) =
    \bigO(n \log^c n)$ space.
  \end{enumerate}
  In total, $D_{\rm count}$ needs $\bigO(n \log^c n)$ space.

  Given any $j \in [0 \dd n]$ and $v \geq 0$, we compute
  $\TwoSidedRangeCount{A}{j}{v}$ as follows:
  \begin{enumerate}
  \item If $v < 1$ (resp.\ $v > n$), then in $\bigO(1)$ time we return that
    $\TwoSidedRangeCount{A}{j}{v} = j$ (resp.\ $\TwoSidedRangeCount{A}{j}{v} = 0$),
    and conclude the query algorithm. Let us thus assume that $v \in [1 \dd n]$.
  \item Using array $R$, in $\bigO(1)$ time we compute
    $b = \RangeBegTwo{\zero^{v}\one}{\Text_{A}} = R[v]$.
    In $\bigO(1)$ time we also compute
    $\ell_1 = n \cdot (n + 1)$,
    $\ell_2 = 2 \cdot (n + 1)$
    $\delta = j \cdot (n + 1 + \tfrac{j+1}{2}) + (n + 2 - v)$, and
    $j' = \ell_1 + \ell_2 + \delta$.
  \item Using $D_{\rm ISA}$, in $\bigO(t_{\rm ISA}(|\Text_{A}|))$
    time we compute $x = \ISA{\Text_{A}}[j']$.
  \item In $\bigO(1)$ time we set $y = x - (b + j + 1)$.
    By \cref{lm:reduce-range-count-to-isa}, it holds
    $y = \TwoSidedRangeCount{A}{j}{v}$. Thus, we return $y$ as the answer.
  \end{enumerate}
  In total, the query time is
  \[
    \bigO(t_{\rm ISA}(|\Text_{A}|))
     = o\Big(\tfrac{\log |\Text_{A}|}{\log \log |\Text_{A}|}\Big)
     = o\Big(\tfrac{\log (n^2)}{\log \log (n^2)}\Big)
     = o\Big(\tfrac{\log n}{\log \log n}\Big).
  \]

  We have thus proved that there exists a data structure that,
  for every array $A[1 \dd n]$ containing a permutation of $\{1, \dots, n\}$,
  uses $\bigO(n \log^c n) = \bigO(n \log^{\bigO(1)} n)$ space, and answers
  range counting queries on $A$ in $o(\tfrac{\log n}{\log \log n})$ time.
  This contradicts \cref{th:range-count-lower-bound}.
\end{proof}

\section{\boldmath Faster Longest Common Prefix Range Minimum Queries (LCP RMQ)}\label{sec:lcp-rmq}

\subsection{Problem Definition}\label{sec:lcp-rmq-problem-def}
\vspace{-1.5ex}

\begin{framed}
  \noindent
  \probname{Indexing for Longest Common Prefix Range Minimum Queries (LCP RMQ)}
  \begin{bfdescription}
  \item[Input:]
    A string $\Text \in \Sigma^{\Textlen}$.
  \item[Output:]
    A data structure that, given any $b,e \in [0 \dd \Textlen]$
    such that $b < e$, returns the position
    $\argmin_{i \in (b \dd e]} \LCP{\Text}[i]$
    (see \cref{def:lcp-array,rm:argmin}).
  \end{bfdescription}
\end{framed}
\vspace{2ex}

\subsection{Upper Bound}\label{sec:lcp-rmq-upper-bound}

In this section, we present the modifications of the data structure of
Gagie, Navarro, and Prezza~\cite{Gagie2020}, that enable answering LCP
RMQ queries in $\bigO(\tfrac{\log \Textlen}{\log \log \Textlen})$ time
using $\bigO(\RLBWTSize{\Text} \log^{2+\epsilon} \Textlen)$ space, for
any constant $\epsilon \in (0,1)$ and any text $\Text \in
\Sigma^{\Textlen}$ (see \cref{def:bwt}). These improvements build upon
the original structure described in~\cite[Section~6.3]{Gagie2020},
which achieves $\bigO(\RLBWTSize{\Text} \log \Textlen)$ space and
$\bigO(\log \Textlen)$ query time.

\begin{theorem}[{\cite[Lemma~6.6]{Gagie2020}}]\label{th:diff-lcp-slp}
  Let $\Text \in \Sigma^{\Textlen}$ be a nonempty text. Let $A[1 \dd \Textlen]$ be
  an array defined by setting $A[1] = \LCP{\Text}[1]$ and, for every
  $i \in [2 \dd \Textlen]$, $A[i] = \LCP{\Text}[i] - \LCP{\Text}[i-1]$ (see \cref{def:lcp-array}).
  There exists an SLP $G$ (\cref{def:slp}) of size $|G| = \bigO(\RLBWTSize{\Text} \log^2 \Textlen)$
  (see \cref{def:bwt}) and height $h = \bigO(\log \Textlen)$ (\cref{def:slg-height})
  that expands to $A$, i.e., $\Lang{G} = \{A\}$.
\end{theorem}

\begin{remark}\label{rm:diff-lcp-slp}
  We remark that~\cite[Lemma~6.6]{Gagie2020} proves a stronger result than above, i.e., it
  describes the construction of a so-called \emph{run-length SLP (RLSLP)} $G'$
  of size $|G'| = \bigO(\RLBWTSize{\Text} \log \Textlen)$ that expands to the differentially encoded LCP array $A$.
  However, this implies the existence of an equivalent (i.e., expanding to the same string) SLP of size
  $\bigO(\RLBWTSize{\Text} \log^2 \Textlen)$. This expansion may increase the height
  of the grammar, but it can be rebalanced to height $\bigO(\log \Textlen)$
  without asymptotically increasing its size by using~\cite{balancing}.
\end{remark}

\begin{corollary}\label{cr:diff-lcp-slg}
  Let $\epsilon \in (0, 1)$ be a constant and $\Text \in \Sigma^{\Textlen}$ be a nonempty text. Let $A[1 \dd \Textlen]$ be
  an array defined by setting $A[1] = \LCP{\Text}[1]$ and, for every
  $i \in [2 \dd \Textlen]$, $A[i] = \LCP{\Text}[i] - \LCP{\Text}[i-1]$ (see \cref{def:lcp-array}).
  There exists an SLG $G = (V, \Sigma, R, S)$ (\cref{def:slg}) of size $|G| = \bigO(\RLBWTSize{\Text} \log^{2+\epsilon} \Textlen)$
  (see \cref{def:bwt}) and height $h = \bigO\big(\tfrac{\log \Textlen}{\log \log \Textlen}\big)$ (\cref{def:slg-height})
  that expands to $A$, i.e., $\Lang{G} = \{A\}$, and such that for some $\ell = \bigO(\log^{\epsilon} \Textlen)$ it holds
  $|\Rhs{G}{N}| \leq \ell$ for all $N \in V$.
\end{corollary}
\begin{proof}
  Let $G_{\rm in} = (V, \Sigma, R_{\rm in}, S)$ be an SLP from \cref{th:diff-lcp-slp}. Based on $G_{\rm in}$,
  we will construct an SLG $G = (V, \Sigma, R, S)$ with the required properties.
  Our construction follows the technique described in~\cite[Theorem~2]{BelazzouguiCPT15}.
  Let $k = \epsilon \cdot \log \log \Textlen$.
  For every $N \in V$, we define $\Rhs{G}{N}$ to be the string obtained by starting
  with $\Rhs{G_{\rm in}}{N}$, and iteratively expanding all variables $k$ times (or less, if the expansion reaches
  a terminal symbol). Since $G_{\rm in}$ is an SLP, each step of the expansion at most doubles the length of the current string,
  and hence it holds $|\Rhs{G}{N}| \leq \ell$, where
  $\ell = 2 \cdot 2^{k} = 2 \cdot \log^{\epsilon} \Textlen = \bigO(\log^{\epsilon} \Textlen)$.
  The resulting SLG $G$ expands to the same string as $G_{\rm in}$, the size of $G$ satisfies
  $|G| = \bigO(|G_{\rm in}| \cdot \ell) = \bigO(\RLBWTSize{\Text} \log^{2+\epsilon} \Textlen)$, and
  its height is equal to height of $G_{\rm in}$ divided by $k$, i.e.,
  $\bigO(\tfrac{\log \Textlen}{\log \log \Textlen})$.
\end{proof}

\begin{lemma}[Based on {\cite[Section~6.3]{Gagie2020}}]\label{lm:slg-nonterminal-prefix}
  Denote $\Sigma = \Z$ and let
  $G = (V, \Sigma, R, S)$ be an SLG of size $g$ (\cref{def:slg}) and
  height $h$ (\cref{def:slg-height}) such that, for every $N \in V$,
  it holds $|\Rhs{G}{V}| \leq \ell$. Denote $n = |\Exp{G}{S}|$,
  $k = |V|$, and $V = \{N_1, \dots, N_k\}$.
  In the word RAM model with word size $w = \Omega(\log n)$, there
  exists a data structure of size $\bigO(g)$ that, given any
  $x \in [1 \dd k]$ and $p \in [1 \dd |\Exp{G}{N_x}|]$, in
  $\bigO(h \cdot (1 + \log_{w} \ell))$ time returns three values:
  \begin{itemize}
  \item $B[1] + \dots + B[p]$,
  \item $\min_{t \in [1 \dd p]} (B[1] + \dots + B[t])$, and
  \item $\argmin_{t \in [1 \dd p]} (B[1] + \dots + B[t])$ (see \cref{rm:argmin}),
  \end{itemize}
  where $B = \Exp{G}{N_x}$.
\end{lemma}
\begin{proof}

  We use the following definitions.
  For every $i \in [1 \dd k]$, let
  $\ell_i = |\Rhs{G}{N_i}|$, and let $R_i[1 \dd \ell_i]$ and $R'_i[1 \dd \ell_i]$
  denote arrays defined such that for every $j \in [1 \dd \ell_i]$, 
  $R_i[j] = \Rhs{G}{N_i}[j]$ and $R'_i[j] = \zero$ holds if $\Rhs{G}{N_i}[j] \in \Sigma$, and
  $R_i[j] = j'$ and $R'_i[j] = \one$ holds if $\Rhs{G}{N_i}[j] \in V$ (and $j' \in \{1,\dots,k\}$ is
  such that $\Rhs{G}{N_i}[j] = N_{j'}$).
  For every $i \in [1 \dd k]$, we define
  $P_i^{\rm len}[1 \dd \ell_i]$,
  $P_i^{\rm sum}[1 \dd \ell_i]$,
  $P_i^{\rm min}[1 \dd \ell_i]$, and
  $P_i^{\rm pos}[1 \dd \ell_i]$,
  so that, for every $\delta \in [1 \dd \ell_i]$, letting $B = \Rhs{G}{N_i}[1 \dd \delta)$ and $E = \Exp{G}{B}$,
  it holds
  \begin{align*}
    P_i^{\rm len}[\delta] &= |E|,\\
    P_i^{\rm sum}[\delta] &= E[1] + \dots + E[|E|],\\
    P_i^{\rm min}[\delta] &= \min \{E[1] + \dots + E[t] : t \in [1 \dd |E|]\} \cup \{\infty\}.
  \end{align*}
  If $P_i^{\rm min}[\delta] \neq \infty$, then we also define
  \[
    P_i^{\rm pos}[\delta] = \textstyle\argmin_{t \in [1 \dd |E|]} (E[1] + \dots + E[t]).
  \]
  Otherwise (i.e., if $\delta = 1$), we leave $P_i^{\rm pos}[\delta]$ undefined.
  For every $i \in [1 \dd k]$, we define a sequence $(b_{i,j})_{j \in [0 \dd \ell_i]}$
  so that $b_{i,0} = -\infty$ and, for every $j \in [1 \dd \ell_i]$, it holds
  \[
    b_{i,j} = |\Exp{G}{\Rhs{G}{N_i}[1 \dd j)}| = P_i^{\rm len}[j].
  \]

  \DSComponents
  The data structure consists of the following components:
  \begin{enumerate}
  \item For every $i \in [1 \dd k]$, we store the predecessor data structure from \cref{th:predecessor-fusion-tree}
    for the sequence $(b_{i,j})_{j \in [0 \dd \ell_i]}$. The structure for $i \in [1 \dd k]$ needs
    $\bigO(\ell_i)$ space (recall that $\ell_i \geq 1$),
    and hence in total all structures need $\bigO(\sum_{i \in [1 \dd k]} \ell_i)
    = \bigO(g)$ space.
  \item For every $i \in [1 \dd k]$, we store the arrays
    $P_i^{\rm len}[1 \dd \ell_i]$,
    $P_i^{\rm sum}[1 \dd \ell_i]$,
    $P_i^{\rm min}[1 \dd \ell_i]$,
    $P_i^{\rm pos}[1 \dd \ell_i]$,
    $R_i[1 \dd \ell_i]$, and
    $R'_i[1 \dd \ell_i]$ in plain form.
    In total, they need $\bigO(\sum_{i \in [1 \dd k]} \ell_i) = \bigO(g)$ space.
  \end{enumerate}
  In total, the structure uses $\bigO(g)$ space.

  \DSQueries
  Let $x \in [1 \dd k]$, $B = \Exp{G}{N_x}$, and $p \in [1 \dd |B|]$.
  We now show how, given $x$ and $p$, to compute
  the following three values:
  \begin{itemize}
  \item $s_{x,p} = B[1] + \dots + B[p]$,
  \item $v_{x,p} = \min_{t \in [1 \dd p]} (B[1] + \dots + B[t])$, and
  \item $\delta_{x,p} = \argmin_{t \in [1 \dd p]} (B[1] + \dots + B[t])$
  \end{itemize}
  in $\bigO(h \cdot (1 + \log_{w} \ell))$ time.
  Let $(a_i)_{i \in [0 \dd q]}$ be the sequence of symbol identifiers corresponding to the nodes
  in the parse tree $\mathcal{T}_{G}(N_x)$ (see \cref{def:parse-tree}) on the path from the root to the $p$th leftmost leaf,
  and $(\delta_i)_{i \in [1 \dd q]}$ is such that, for every $i \in [1 \dd q]$, node $i$ on the path is the $\delta_i$th
  leftmost child of node $i-1$ on the path. Formally,
  $(a_i)_{i \in [0 \dd q]} \in \Z^{q+1}$
  and $(\delta_i)_{i \in [1 \dd q]} \in \Zp^{q}$, where $q \geq 1$, are the unique sequences satisfying
  the following conditions:
  \begin{itemize}
  \item $a_0 = x$,
  \item for every $i \in [1 \dd q)$, $\Rhs{G}{N_{a_{i-1}}}[\delta_i] = N_{a_{i}}$,
  \item $\Rhs{G}{N_{a_{q-1}}}[\delta_q] = a_q \in \Sigma$,
  \item $\big(\sum_{i=1}^{q} |\Exp{G}{\Rhs{G}{N_{a_{i-1}}}[1 \dd \delta_i)}|\big) + 1 = p$.
  \end{itemize}

  Note that then it holds:
  \[
    \Exp{G}{N_x}[1 \dd p] =
      \Big(\textstyle\bigodot_{i=1}^{q} \Exp{G}{\Rhs{G}{N_{a_{i-1}}}[1 \dd \delta_i)} \Big) \cdot a_q.
  \]

  By definition of arrays $P_i^{\rm sum}$, we thus have:
  \[
    s_{x,p} = \Big( \textstyle\sum_{j \in [1 \dd q]} P_{a_{j-1}}^{\rm sum}[\delta_j] \Big) + a_q.
  \]

  Moreover, by definition of arrays $P_i^{\rm sum}$ and $P_i^{\rm min}$, we also have:
  \[
    v_{x,p} =
      \min
        \Big\{
          \min_{i \in [1 \dd q]}
            \Big( \textstyle\sum_{j \in [1 \dd i)} P_{a_{j-1}}^{\rm sum}[\delta_j] \Big) + P_{a_{i-1}}^{\rm min}[\delta_i],
            s_{x,p}
        \Big\}.
  \]

  Finally, to compute the smallest prefix length $\delta_{x,p}$ that minimizes the prefix sum, i.e., the value
  $\argmin_{t \in [1 \dd p]} (B[1] + \dots + B[t])$, we observe
  that:
  \begin{itemize}
  \item If $s_{x,p} < \min_{i \in [1 \dd q]} (\sum_{j \in [1 \dd i)} P_{a_{j-1}}^{\rm sum}[\delta_j]) + P_{a_{i-1}}^{\rm min}[\delta_i]$,
    then $\delta_{x,p} = p$.
  \item Otherwise, letting $i'$ be the smallest index in $[1 \dd q]$ such that
    $(\sum_{j \in [1 \dd i')} P_{a_{j-1}}^{\rm sum}[\delta_j]) + P_{a_{i'-1}}^{\rm min}[\delta_{i'}] = v_{x,p}$,
    it holds by definition of arrays $P_i^{\rm len}$ and $P_i^{\rm pos}$ that
    \[
      \delta_{x,p} =
        \Big(\textstyle\sum_{j \in [1 \dd i')} P_{a_{j-1}}^{\rm len}[\delta_j] \Big) + P_{a_{i'-1}}^{\rm pos}[\delta_{i'}].
    \]
  \end{itemize}

  We are now ready to present the query algorithm. Given $x$ and $p$, we proceed as follows:
  \begin{enumerate}

  \item In the first step, we compute the arrays $a[0 \dd q]$ and $\delta[1 \dd q]$ containing the sequences
    $(a_i)_{i \in [0 \dd q]}$ and $(\delta_i)_{i \in [1 \dd q]}$ defined above. We initialize the current
    position $p_{\rm cur} := p$, current nonterminal number $x_{\rm cur} := x$, and current depth
    $d_{\rm cur} := 0$ in the parse tree $\mathcal{T}_{G}(N_x)$.
    We also set $a[d_{\rm cur}] := x_{\rm cur}$. We then repeat the following sequence of steps:
    \begin{enumerate}
    \item In $\bigO(1)$ time, we set $d_{\rm cur} := d_{\rm cur} + 1$.
    \item Denote $i = x_{\rm cur}$.
      Using the structure from \cref{th:predecessor-fusion-tree} for the sequence
      $(b_{i,j})_{j \in [0 \dd \ell_i]}$,
      in $\bigO(1 + \log_{w} \ell_i) = \bigO(1 + \log_{w} \ell)$ time
      we compute $\delta \in [1 \dd \ell_i]$ satisfying $b_{i,\delta} = \Predecessor{C}{p_{\rm cur}}$,
      where $C = \{b_{i,j}\}_{j \in [1 \dd \ell_i]}$
      (note that $\delta \geq 1$ follows by $p_{\rm cur} \geq 1$ and $b_{i,1} = 0$).
      Note that the $p_{\rm cur}$th leftmost leaf in the parse tree $\mathcal{T}_{G}(N_{x_{\rm cur}})$
      is then in the subtree rooted in the $\delta$th leftmost child of the
      root of $\mathcal{T}_{G}(N_{x_{\rm cur}})$.
    \item In $\bigO(1)$ time, we set $\delta[d_{\rm cur}] := \delta$.
    \item In $\bigO(1)$ time, we determine if $\Rhs{G}{N_{x_{\rm cur}}}[\delta] \in \Sigma$ by
      checking if $R'_{x_{\rm cur}}[\delta] = \zero$. We then consider two cases:
      \begin{itemize}
      \item If $R'_{x_{\rm cur}}[\delta] = \zero$, then in $\bigO(1)$ time we first compute
        $c := \Rhs{G}{N_{x_{\rm cur}}}[\delta] = R_{x_{\rm cur}}[\delta]$ and set $a[d_{\rm cur}] := c$. We
        then terminate the query algorithm with $q = d_{\rm cur}$.
      \item Otherwise, in $\bigO(1)$ time we compute the index $x_{\rm next}$ satisfying
        $\Rhs{G}{N_{x_{\rm cur}}}[\delta] = N_{x_{\rm next}}$ by looking up $x_{\rm next} = R_{x_{\rm cur}}[\delta]$.
        We then set $a[d_{\rm cur}] := x_{\rm next}$. Finally, in preparation for the next iteration, we set
        $p_{\rm cur} := p_{\rm cur} - P_{x_{\rm cur}}^{\rm len}[\delta]$ and $x_{\rm cur} := x_{\rm next}$.
      \end{itemize}
    \end{enumerate}
    Each of the iterations takes $\bigO(1 + \log_{w} \ell)$. In total we thus spend
    $\bigO(q \cdot (1 + \log_{w} \ell))$ time.

  \item In the second step, we compute the value $s_{x,p}$, i.e., $B[1] + \dots + B[p]$.
    \begin{enumerate}
    \item First, we compute an array $A_{\rm sum}[0 \dd q]$ defined by
      $A_{\rm sum}[i] = \sum_{j \in [1 \dd i]} P_{a_{j-1}}^{\rm sum}[\delta_j]$. This is easily done in
      $\bigO(q)$ time: first set $A_{\rm sum}[0] = 0$, and then, for every $j \in [1 \dd q]$,
      $A_{\rm sum}[j] = A_{\rm sum}[j-1] + P_{a[j-1]}^{\rm sum}[\delta[j]]$.
    \item We then return that $s_{x,p} = A_{\rm sum}[q] + a[q]$.
    \end{enumerate}
    In total, this takes $\bigO(q)$ time.

  \item In the third step, we compute $\min_{t \in [1 \dd p]} (B[1] + \dots + B[t])$. To this end, in
    $\bigO(q)$ time we compute the value of the expression
    \[
      \min \Big\{ A_{\rm sum}[0] + P_{a[0]}^{\rm min}[\delta[1]],
                            \dots,
                            A_{\rm sum}[q-1] + P_{a[q-1]}^{\rm min}[\delta[q]],
                            s_{x,p} \Big\}.
    \]
    By the above discussion, the result is equal to $v_{x,p}$, i.e., $\min_{t \in [1 \dd p]} (B[1] + \dots + B[t])$.
    Computing $v_{x,p}$ thus takes $\bigO(q)$ time.

  \item In the fourth step, we compute $\delta_{x,p}$, i.e., $\argmin_{t \in [1 \dd p]} (B[1] + \dots + B[t])$.
    First, in $\bigO(q)$ time we compute
    $v = \min \{ A_{\rm sum}[0] + P_{a[0]}^{\rm min}[\delta[1]],
    \dots, A_{\rm sum}[q-1] + P_{a[q-1]}^{\rm min}[\delta[q]]\}$ (where $A_{\rm sum}$ is the array computed above).
    \begin{itemize}
    \item If $v > s_{x,p}$, then by the above discussion we
      have $\argmin_{t \in [1 \dd p]} (B[1] + \dots + B[t]) = p$.
    \item Let us now assume that $v \leq s_{x,p}$. In $\bigO(q)$ time,
      we compute the smallest index $i' \in [1 \dd q]$ that
      satisfies $A_{\rm sum}[i'-1] + P_{a[i'-1]}^{\rm min}[\delta[i']] = v_{x,p}$ (the latter value was computed above).
      In $\bigO(q)$ time, we then obtain $\delta_{x,p} =
      \big( \sum_{j \in [1 \dd i')} P_{a[j-1]}^{\rm len}[\delta[j]] \big) +
      P_{a[i'-1]}^{\rm pos}[\delta[i']]$. The correctness of this computation follows by the above discussion.
    \end{itemize}
    In total, we spend $\bigO(q)$ time.
  \end{enumerate}
  In total, the computation of $s_{x,p}$, $v_{x,p}$, and $\delta_{x,p}$
  takes $\bigO(q \cdot (1 + \log_{w} \ell)) = \bigO(h \cdot (1 + \log_{w} \ell))$ time.
\end{proof}

\begin{remark}
  The next result shows how to compute the largest prefix sum on the suffix
  of the expansion of the nonterminal in a grammar. Note that although the
  data structure is similar to the one in \cref{lm:slg-nonterminal-prefix},
  it is not symmetric. This is because this time we are investigating
  the \emph{suffix} of a nonterminal expansion, but still ask about the
  largest \emph{prefix} sum of that substring, making the details of the
  query slightly different.
\end{remark}

\begin{lemma}[Based on {\cite[Section~6.3]{Gagie2020}}]\label{lm:slg-nonterminal-suffix}
  Denote $\Sigma = \Z$ and let
  $G = (V, \Sigma, R, S)$ be an SLG of size $g$ (\cref{def:slg}) and
  height $h$ (\cref{def:slg-height}) such that, for every $N \in V$,
  it holds $|\Rhs{G}{V}| \leq \ell$. Denote $n = |\Exp{G}{S}|$,
  $k = |V|$, and $V = \{N_1, \dots, N_k\}$.
  In the word RAM model with word size $w = \Omega(\log n)$, there
  exists a data structure of size $\bigO(g)$ that, given any
  $x \in [1 \dd k]$ and $p \in [1 \dd m]$ (where $m = |\Exp{G}{N_x}|$),
  in $\bigO(h \cdot (1 + \log_{w} \ell))$ time returns three values:
  \begin{itemize}
  \item $B[m-p+1] + \dots + B[m]$,
  \item $\min_{t \in [1 \dd p]} (B[m-p+1] + \dots + B[m-p+t])$, and
  \item $\argmin_{t \in [1 \dd p]} (B[m-p+1] + \dots + B[m-p+t])$ (see \cref{rm:argmin}),
  \end{itemize}
  where $B = \Exp{G}{N_x}$.
\end{lemma}
\begin{proof}

  We use the following definitions.
  For every $i \in [1 \dd k]$, let
  $\ell_i = |\Rhs{G}{N_i}|$, and let $R_i[1 \dd \ell_i]$ and $R'_i[1 \dd \ell_i]$
  denote arrays defined such that for every $j \in [1 \dd \ell_i]$, 
  $R_i[j] = \Rhs{G}{N_i}[j]$ and $R'_i[j] = \zero$ holds if $\Rhs{G}{N_i}[j] \in \Sigma$, and
  $R_i[j] = j'$ and $R'_i[j] = \one$ holds if $\Rhs{G}{N_i}[j] \in V$ (and $j' \in \{1,\dots,k\}$ is
  such that $\Rhs{G}{N_i}[j] = N_{j'}$).
  For every $i \in [1 \dd k]$, we define
  $S_i^{\rm len}[1 \dd \ell_i]$,
  $S_i^{\rm sum}[1 \dd \ell_i]$,
  $S_i^{\rm min}[1 \dd \ell_i]$, and
  $S_i^{\rm pos}[1 \dd \ell_i]$,
  so that, for every $\delta \in [1 \dd \ell_i]$,
  letting $B = \Rhs{G}{N_i}(\delta \dd |\Rhs{G}{N_i}|]$ and $E = \Exp{G}{B}$,
  it holds
  \begin{align*}
    S_i^{\rm len}[\delta] &= |E|,\\
    S_i^{\rm sum}[\delta] &= E[1] + \dots + E[|E|],\\
    S_i^{\rm min}[\delta] &= \min \{E[1] + \dots + E[t] : t \in [1 \dd |E|]\} \cup \{\infty\}.
  \end{align*}
  If $S_i^{\rm min}[\delta] \neq \infty$, then we also define
  \[
    S_i^{\rm pos}[\delta] = \textstyle\argmin_{t \in [1 \dd |E|]} (E[1] + \dots + E[t]).
  \]
  Otherwise (i.e., if $\delta = |\Rhs{G}{N_i}|$), we leave $S_i^{\rm pos}[\delta]$ undefined.
  For every $i \in [1 \dd k]$, we define a sequence $(b_{i,j})_{j \in [0 \dd \ell_i]}$
  so that $b_{i,0} = -\infty$ and, for every $j \in [1 \dd \ell_i]$, it holds
  \[
    b_{i,j} = |\Exp{G}{\Rhs{G}{N_i}[1 \dd j)}| = S_i^{\rm len}[j].
  \]
  We define $A_{\rm explen}[1 \dd k]$ so that,
  for every $i \in [1 \dd k]$, $A_{\rm explen}[i] = |\Exp{G}{N_i}|$.

  \DSComponents
  The data structure consists of the following components:
  \begin{enumerate}
  \item For every $i \in [1 \dd k]$, we store the predecessor data structure from \cref{th:predecessor-fusion-tree}
    for the sequence $(b_{i,j})_{j \in [0 \dd \ell_i]}$. The structure for $i \in [1 \dd k]$ needs
    $\bigO(\ell_i)$ space (recall that $\ell_i \geq 1$),
    and hence in total all structures need $\bigO(\sum_{i \in [1 \dd k]} \ell_i)
    = \bigO(g)$ space.
  \item For every $i \in [1 \dd k]$, we store the arrays
    $S_i^{\rm len}[1 \dd \ell_i]$,
    $S_i^{\rm sum}[1 \dd \ell_i]$,
    $S_i^{\rm min}[1 \dd \ell_i]$,
    $S_i^{\rm pos}[1 \dd \ell_i]$,
    $R_i[1 \dd \ell_i]$, and
    $R'_i[1 \dd \ell_i]$ in plain form.
    In total, they need $\bigO(\sum_{i \in [1 \dd k]} \ell_i) = \bigO(g)$ space.
  \item We store the array $A_{\rm explen}[1 \dd k]$ in plan form using $\bigO(k) = \bigO(g)$ space.
  \end{enumerate}
  In total, the structure uses $\bigO(g)$ space.

  \DSQueries
  Let $x \in [1 \dd k]$, $B = \Exp{G}{N_x}$, $m = |B|$, and $p \in [1 \dd m]$.
  We now show how, given $x$ and $p$, to compute
  the following three values:
  \begin{itemize}
  \item $s_{x,p} = B[m-p+1] + \dots + B[m]$,
  \item $v_{x,p} = \min_{t \in [1 \dd p]} (B[m-p+1] + \dots + B[m-p+t])$, and
  \item $\delta_{x,p} = \argmin_{t \in [1 \dd p]} (B[m-p+1] + \dots + B[m-p+t])$
  \end{itemize}
  in $\bigO(h \cdot (1 + \log_{w} \ell))$ time.
  Let $(a_i)_{i \in [0 \dd q]}$ be the sequence of symbol identifiers corresponding to the nodes
  in the parse tree $\mathcal{T}_{G}(N_x)$ (see \cref{def:parse-tree}) on the path from the root to the $p$th rightmost leaf,
  and $(\delta_i)_{i \in [1 \dd q]}$ is such that, for every $i \in [1 \dd q]$, node $i$ on the path is the $\delta_i$th
  leftmost child of node $i-1$ on the path. Formally,
  $(a_i)_{i \in [0 \dd q]} \in \Z^{q+1}$
  and $(\delta_i)_{i \in [1 \dd q]} \in \Zp^{q}$, where $q \geq 1$, are the unique sequences satisfying
  the following conditions:
  \begin{itemize}
  \item $a_0 = x$,
  \item for every $i \in [1 \dd q)$, $\Rhs{G}{N_{a_{i-1}}}[\delta_i] = N_{a_{i}}$,
  \item $\Rhs{G}{N_{a_{q-1}}}[\delta_q] = a_q \in \Sigma$,
  \item $\big(\sum_{i=1}^{q} |\Exp{G}{\Rhs{G}{N_{a_{i-1}}}(\delta_i \dd |\Rhs{G}{N_{a_{i-1}}}|]}|\big) + 1 = p$.
  \end{itemize}

  Note that then it holds:
  \[
    \Exp{G}{N_x}(m-p \dd m] =
      a_q \cdot \Big(\textstyle\bigodot_{i=q,\dots,1} \Exp{G}{\Rhs{G}{N_{a_{i-1}}}(\delta_i \dd |\Rhs{G}{N_{a_{i-1}}}|]} \Big).
  \]

  By definition of arrays $S_i^{\rm sum}$, we thus have:
  \[
    s_{x,p} = \Big( \textstyle\sum_{j \in [1 \dd q]} S_{a_{j-1}}^{\rm sum}[\delta_j] \Big) + a_q.
  \]

  Moreover, by definition of arrays $S_i^{\rm sum}$ and $S_i^{\rm min}$, we also have:
  \[
    v_{x,p} =
      \min
        \Big\{
          a_{q},
          \min_{i \in [1 \dd q]}
            \Big(a_q + \textstyle\sum_{j \in (i \dd q]} S_{a_{j-1}}^{\rm sum}[\delta_j] \Big) + S_{a_{i-1}}^{\rm min}[\delta_i]
        \Big\}.
  \]

  Finally, to compute the smallest prefix length $\delta_{x,p}$ that minimizes the prefix sum, i.e., the value
  $\argmin_{t \in [1 \dd p]} (B[m-p+1] + \dots + B[m-p+t])$, we observe
  that:
  \begin{itemize}
  \item If $a_q < \min_{i \in [1 \dd q]}
    (a_q + \sum_{j \in (i \dd q]} S_{a_{j-1}}^{\rm sum}[\delta_j]) + S_{a_{i-1}}^{\rm min}[\delta_i]$,
     then $\delta_{x,p} = 1$.
  \item Otherwise, letting $i'$ be the largest index in $[1 \dd q]$ such that
    $(a_q + \sum_{j \in (i \dd q]} S_{a_{j-1}}^{\rm sum}[\delta_j]) + S_{a_{i-1}}^{\rm min}[\delta_i] = v_{x,p}$,
    it holds by definition of arrays $S_i^{\rm len}$ and $S_i^{\rm pos}$ that
    \[
      \delta_{x,p} =
        1 + \Big(\textstyle\sum_{j \in (i' \dd q]} S_{a_{j-1}}^{\rm len}[\delta_j] \Big) + S_{a_{i'-1}}^{\rm pos}[\delta_{i'}].
    \]
  \end{itemize}

  We are now ready to present the query algorithm. Given $x$ and $p$, we proceed as follows:
  \begin{enumerate}

  \item In the first step, we compute the arrays $a[0 \dd q]$ and $\delta[1 \dd q]$ containing the sequences
    $(a_i)_{i \in [0 \dd q]}$ and $(\delta_i)_{i \in [1 \dd q]}$ defined above. The query
    algorithm proceeds the same as in the proof of \cref{lm:slg-nonterminal-suffix}, except
    at the beginning of the algorithm, we initialize $p_{\rm cur} := A_{\rm explen}[x] - p + 1$.
    The computation of the two arrays takes $\bigO(q \cdot (1 + \log_{w} \ell))$ time.

  \item In the second step, we compute the value $s_{x,p}$, i.e., $B[m-p+1] + \dots + B[m]$.
    \begin{enumerate}
    \item First, we compute an array $A_{\rm sum}[0 \dd q]$ defined by
      $A_{\rm sum}[i] = \sum_{j \in (i \dd q]} S_{a_{j-1}}^{\rm sum}[\delta_j]$. This is easily done in
      $\bigO(q)$ time: first set $A_{\rm sum}[q] = 0$, and then, for every $i=q-1,\dots,0$, set
      $A_{\rm sum}[i] = A_{\rm sum}[i+1] + S_{a[i]}^{\rm sum}[\delta[i+1]]$.
    \item We then return that $s_{x,p} = A_{\rm sum}[0] + a[q]$.
    \end{enumerate}
    In total, this takes $\bigO(q)$ time.

  \item In the third step, we compute $\min_{t \in [1 \dd p]} (B[m-p+1] + \dots + B[m-p+t])$. To this end, in
    $\bigO(q)$ time we compute the value of the expression
    \[
      \min \Big\{ a[q],
                  a[q] + A_{\rm sum}[1] + S_{a[0]}^{\rm min}[\delta[1]],
                  \dots,
                  a[q] + A_{\rm sum}[q] + S_{a[q-1]}^{\rm min}[\delta[q]],
           \Big\}.
    \]
    By the above discussion, the result is equal to $v_{x,p}$, i.e., $\min_{t \in [1 \dd p]} (B[m-p+1] + \dots + B[m-p+t])$.
    Computing $v_{x,p}$ thus takes $\bigO(q)$ time.

  \item In the fourth step, we compute $\delta_{x,p}$, i.e., $\argmin_{t \in [1 \dd p]} (B[m-p+1] + \dots + B[m-p+t])$.
    First, in $\bigO(q)$ time we compute
    $v = \min \{ a[q] + A_{\rm sum}[1] + S_{a[0]}^{\rm min}[\delta[1]],
    \dots, a[q] + A_{\rm sum}[q] + S_{a[q-1]}^{\rm min}[\delta[q]]\}$ (where $A_{\rm sum}$ is the array computed above).
    \begin{itemize}
    \item If $a[q] < v$, then by the above discussion we
      have $\argmin_{t \in [1 \dd p]} (B[m-p+1] + \dots + B[m-p+t]) = 1$.
    \item Let us now assume that $a[q] \geq v$. In $\bigO(q)$ time,
      we compute the largest index $i' \in [1 \dd q]$ that
      satisfies $a[q] + A_{\rm sum}[i'] + S_{a[i'-1]}^{\rm min}[\delta[i']] = v_{x,p}$
      (the latter value was computed above).
      In $\bigO(q)$ time, we then obtain $\delta_{x,p} =
      1 + \big( \sum_{j \in (i' \dd q]} S_{a[j-1]}^{\rm len}[\delta[j]] \big) +
      S_{a[i'-1]}^{\rm pos}[\delta[i']]$.
      The correctness of this computation follows by the above discussion.
    \end{itemize}
    In total, we spend $\bigO(q)$ time.
  \end{enumerate}
  In total, the computation of $s_{x,p}$, $v_{x,p}$, and $\delta_{x,p}$
  takes $\bigO(q \cdot (1 + \log_{w} \ell)) = \bigO(h \cdot (1 + \log_{w} \ell))$ time.
\end{proof}

\begin{lemma}[Based on {\cite[Section~6.3]{Gagie2020}}]\label{lm:min-prefix-sum-slg}
  Denote $\Sigma = \Z$ and let
  $G = (V, \Sigma, R, S)$ be an SLG of size $g$ (\cref{def:slg}) and
  height $h$ (\cref{def:slg-height}) such that, for every $N \in V$,
  it holds $|\Rhs{G}{V}| \leq \ell$. Denote $A = \Exp{G}{S}$ and $n = |A|$.
  In the word RAM model with word size $w = \Omega(\log n)$, there
  exists a data structure of size $\bigO(g)$ that, given any
  $b, e \in [0 \dd n]$ satisfying $b < e$, in
  $\bigO(h \cdot (1 + \log_{w} \ell))$ time returns (see \cref{rm:argmin})
  \[
    \argmin_{i \in (b \dd e]}\, (A[1] + \dots + A[i]).
  \]
\end{lemma}
\begin{proof}

  We use the following definitions.
  Denote $k = |V|$ and $V = \{N_1, \dots, N_k\}$.
  For every $i \in [1 \dd k]$, let
  $\ell_i = |\Rhs{G}{N_i}|$, and let $R_i[1 \dd \ell_i]$
  denote an array defined such that for every $j \in [1 \dd \ell_i]$, 
  $R_i[j] = \Rhs{G}{N_i}[j]$ holds if $\Rhs{G}{N_i}[j] \in \Sigma$, and
  $R_i[j] = j'$ if $\Rhs{G}{N_i}[j] \in V$ (and $j' \in \{1,\dots,k\}$ is
  such that $\Rhs{G}{N_i}[j] = N_{j'}$).
  For every $i \in [1 \dd k]$, we define
  $P_i^{\rm len}[1 \dd \ell_i+1]$ and $P_i^{\rm sum}[1 \dd \ell_i+1]$
  so that, for every $\delta \in [1 \dd \ell_i+1]$,
  letting $B = \Rhs{G}{N_i}[1 \dd \delta)$ and $E = \Exp{G}{B}$,
  it holds:
  \begin{align*}
    P_i^{\rm len}[\delta] &= |E|,\\
    P_i^{\rm sum}[\delta] &= E[1] + \dots + E[|E|].
  \end{align*}
  For every $i \in [1 \dd k]$, we define a sequence $(b_{i,j})_{j \in [0 \dd \ell_i+1]}$
  so that $b_{i,0} = -\infty$ and, for every $j \in [1 \dd \ell_i+1]$, it holds
  \[
    b_{i,j} = |\Exp{G}{\Rhs{G}{N_i}[1 \dd j)}| = P_i^{\rm len}[j].
  \]
  For every $i \in [1 \dd k]$, we define
  $M_i^{\rm min}[1 \dd \ell_i]$ and $M_i^{\rm pos}[1 \dd \ell_i]$ so that,
  for every $\delta \in [1 \dd \ell_i]$,
  letting $E = \Exp{G}{N_i}$, $\ell_{\rm beg} = b_{i,\delta}$, and $\ell_{\rm end} = b_{i,\delta+1}$,
  it holds:
  \begin{align*}
    M_i^{\rm min}[\delta] &= \textstyle\min_{t \in (\ell_{\rm beg} \dd \ell_{\rm end}]} E[1] + \dots + E[t],\\
    M_i^{\rm pos}[\delta] &= \textstyle\argmin_{t \in (\ell_{\rm beg} \dd \ell_{\rm end}]} E[1] + \dots + E[t].
  \end{align*}
  Let $s \in [1 \dd k]$ be such that $S = N_{s}$ is the start symbol of $G$.

  \DSComponents
  The data structure consists of the following components:
  \begin{enumerate}
  \item For every $i \in [1 \dd k]$, we store the predecessor data structure
    from \cref{th:predecessor-fusion-tree} for the sequence $(b_{i,j})_{j \in [0 \dd \ell_i+1]}$.
    The structure for $i \in [1 \dd k]$ needs $\bigO(\ell_i)$ space (recall that $\ell_i \geq 1$), and
    hence in total all structures need $\bigO(\sum_{i=1}^{k} \ell_i) = \bigO(g)$ space.
  \item For every $i \in [1 \dd k]$, we store the RMQ data structure
    from \cref{th:rmq} for the array $M_i^{\rm min}[1 \dd \ell_i]$. The structure for $i \in [1 \dd k]$
    needs $\bigO(\ell_i)$ space, and hence in total all
    structures need $\bigO(\sum_{i=1}^{k} \ell_i) = \bigO(g)$ space.
  \item For every $i \in [1 \dd k]$, we store the arrays
    $R_i[1 \dd \ell_i]$,
    $P_i^{\rm len}[1 \dd \ell_i + 1]$,
    $P_i^{\rm sum}[1 \dd \ell_i + 1]$,
    $M_i^{\rm min}[1 \dd \ell_i]$, and
    $M_i^{\rm pos}[1 \dd \ell_i]$
    in plain form. In total, they need $\bigO(\sum_{i=1}^{k} \ell_i) = \bigO(g)$ space.
  \item The data structure from \cref{lm:slg-nonterminal-prefix} for the SLG $G$.
    It uses $\bigO(g)$ space.
  \item The data structure from \cref{lm:slg-nonterminal-suffix} for the SLG $G$.
    It also uses $\bigO(g)$ space.
  \end{enumerate}
  In total, the structure uses $\bigO(g)$ space.

  \DSQueries
  Let $b,e \in [0 \dd n]$ be such that $b < e$. We now show how, given $b$ and $e$, to compute
  the position $\argmin_{i \in (b \dd e]}\, (A[1] + \dots + A[i])$ in $\bigO(h \cdot (1 + \log_{w} \ell))$ time.
  We proceed as follows:
  \begin{enumerate}

  \item In the first step, we perform a traversal in the parse tree of $G$ to decompose the input interval $(b \dd e]$ into
    simpler form. We initialize $b' := b$, $e' := e$, and $x := s$. We then repeat the following sequence
    of steps:
    \begin{enumerate}

    \item Using \cref{th:predecessor-fusion-tree}
      for the sequence $(b_{x,t})_{t \in [1 \dd \ell_x+1]}$,
      in $\bigO(1 + \log_{w} \ell_x) = \bigO(1 + \log_{w} \ell)$
      time we compute $i \in [0 \dd \ell_x + 1]$ satisfying
      $b_{x,i} = \Predecessor{C}{b'}$,
      where $C = \{b_{x,t}\}_{t \in [1 \dd \ell_x+1]}$.

    \item Analogously,
      in $\bigO(1 + \log_{w} \ell)$ time
      compute $j \in [0 \dd \ell_x + 1]$ satisfying
      $b_{x,j} = \Predecessor{C}{e'+1}$.

    \item We then consider two cases:
      \begin{itemize}
      \item If $i = j$, then it holds
        $b_{x,i} < b' < e' < b_{x,i+1}$ and hence
        we continue the traversal. Note that this implies tat $i \geq 1$
        and $\Rhs{G}{N_x}[i] \in V$.
        In preparation for the next iteration, we set
        $b' := b' - b_{x,i} = b' - P_{x}^{\rm len}[i]$,
        $e' := e' - b_{x,i} = e' - P_{x}^{\rm len}[i]$,
        and $x := R_{x}[i]$.
      \item If $i < j$, we stop the downward traversal.
        Observe that at the end of the traversal, the variables
        $b'$, $e'$, $x$, $i$, and $j$ satisfy the following conditions:
        \begin{itemize}
        \item $0 \leq b' < e' \leq |\Exp{G}{N_x}|$,
        \item $\Exp{G}{N_x}(b' \dd e'] = A(b \dd e]$,
        \item $0 \leq i < j \leq \ell_x + 1$,
        \item $b_{x,i} < b' \leq b_{x,i+1}$,
        \item $b_{x,j} \leq e' < b_{x,j+1}$.
        \end{itemize}
      \end{itemize}
    \end{enumerate}
    In each step, we spend $\bigO(1 + \log_{w} \ell)$ time, and descend one level
    in the parse tree of $G$. Thus, the above traversal takes
    $\bigO(h \cdot (1 + \log_{w} \ell))$ time.

  \item In the next step, we compute $\argmin_{i \in (b \dd e]}\,(A[1] + \dots + A[i])$.
    The central properties that we will use are:
    \begin{itemize}
    \item First, we observe that by $\Exp{G}{N_x}(b' \dd e'] = A(b \dd e]$, letting $B = \Exp{G}{N_x}(b' \dd e']$, it holds
      \[
        \argmin_{t \in (b \dd e]}\,(A[1] + \dots + A[t]) = b + \argmin_{t \in [1 \dd e'-b']}\,(B[1] + \dots + B[t]).
      \]
      This holds because each of the prefix sums on the right can be mapped to exactly one prefix sum on the left such
      that they differ by exactly $A[1] + \dots + A[b]$. Thus, the position of the minimum is not affected.
    \item Second, note that we can write the interval $(b' \dd e']$ as a disjoint union (with either
      of the three subintervals possibly empty)
      \[
        (b' \dd e'] = (b' \dd b_{x,i+1}] \cup (b_{x,i+1} \dd b_{x,j}] \cup (b_{x,j} \dd e'].
      \]
    \end{itemize}
    The query proceeds as follows:
    \begin{enumerate}

    \item In $\bigO(1)$ time we initialize
      the minimum prefix sum to $v_{\rm cur} = \infty$,
      the length of the prefix corresponding to this prefix sum to $\delta_{\rm cur} = 0$,
      and the sum and length of the prefix examined so far to $s_{\rm cur} = 0$ and $p_{\rm cur} = 0$.
      We also set
      $p_L := P_{x}^{\rm len}[i+1] - b'$,
      $p_M := P_{x}^{\rm len}[j] - P_{x}^{\rm len}[i+1]$, and
      $p_R := e' - P_{x}^{\rm len}[j]$
      to be lengths of the three subintervals in the above decomposition.

    \item If $p_L = 0$, then we skip this step. Let us thus assume that $p_L > 0$.
      Observe that since above we noted that $b_{x,i} < b' \leq b_{x,i+1}$, it follows by $p_L > 0$ that
      $|\Exp{G}{N_x}[i]| = b_{x,i+1}-b_{x,i} = (b_{x,i+1}-b') + (b'-b_{x,i}) = p_{L} + (b'-b_{x,i}) \geq 2$,
      and hence $\Rhs{G}{N_x}[i] \in V$.
      \begin{enumerate}
      \item In $\bigO(1)$ time we compute $x_{L} := R_{x}[i]$. We then have $\Rhs{G}{N_x}[i] = N_{x_L}$.
        Denote $B_L = \Exp{G}{N_{x_L}}$ and $m_L = |B_L|$.
      \item Using \cref{lm:slg-nonterminal-suffix} and $x_{L}$ and $p_{L}$ and input,
        in $\bigO(h \cdot (1 + \log_{w} \ell))$ time we compute
        three values:
        \begin{itemize}
        \item $s_L = B_L[m_L - p_L + 1] + \dots + B_L[m_L]$,
        \item $v_L = \min_{t \in [1 \dd p_L]} B_L[m_L - p_L + 1] + \dots + B_L[m_L - p_L + t]$,
        \item $\delta_L = \argmin_{t \in [1 \dd p_L]} B_L[m_L - p_L + 1] + \dots + B_L[m_L - p_L + t]$.
        \end{itemize}
      \item We then update the values recording the currently smallest prefix sum and the auxiliary statistics
        by setting $v_{\rm cur} := v_{L}$, $\delta_{\rm cur} = \delta_{L}$,
        $s_{\rm cur} := s_{L}$, and $p_{\rm cur} := p_{L}$.
      \end{enumerate}
      In total, above steps take $\bigO(h \cdot (1 + \log_{w} \ell))$ time.

    \item If $p_{M} = 0$, then we skip this step. Let us thus assume that $p_{M} > 0$.
      Note that this implies that $i < j-1$.
      \begin{enumerate}
      \item Using \cref{th:rmq}, in $\bigO(1)$ time we compute
        $t_{\rm child} = \argmin_{t \in (i \dd j-1]} M_x^{\rm min}[t]$.
      \item In $\bigO(1)$ time we set
        $s_M := P_x^{\rm sum}[j] - P_x^{\rm sum}[i+1]$,
        $v_M := M_x^{\rm min}[t_{\rm child}] - P_x^{\rm sum}[i+1]$, and
        $\delta_M := M_x^{\rm pos}[t_{\rm child}] - P_x^{\rm len}[i+1]$.
        Denoting $B_M = \Exp{G}{\Rhs{G}{N_x}(i \dd j-1]}$ and $m_M = |B_M|$, we then have:
        \begin{itemize}
        \item $s_M = B_M[1] + \dots + B_M[m_M]$,
        \item $v_M = \min_{t \in [1 \dd m_M]} (B_M[1] + \dots + B_M[t])$,
        \item $\delta_M = \argmin_{t \in [1 \dd m_M]} (B_M[1] + \dots + B_M[t])$.
        \end{itemize}
      \item In $\bigO(1)$ time we update the values corresponding
        to the smallest prefix.
        More precisely, if $s_{\rm cur} + v_{M} < v_{\rm cur}$, then we set
        $v_{\rm cur} := s_{\rm cur} + v_{M}$ and $\delta_{\rm cur} := p_{\rm cur} + \delta_{M}$.
      \item We update the remaining variables by setting
        $s_{\rm cur} := s_{\rm cur} + s_{M}$ and $p_{\rm cur} := p_{\rm cur} + p_{M}$.
      \end{enumerate}
      In total, above steps take $\bigO(1)$ time.

    \item If $p_R = 0$, then we skip this step. Let us thus assume that $p_R > 0$.
      Observe that since above we noted that $b_{x,j} \leq e' < b_{x,j+1}$, it follows by $p_R > 0$ that
      $|\Exp{G}{N_x}[j]| = b_{x,j+1}-b_{x,j} = (b_{x,j+1}-e') + (e'-b_{x,j}) = (b_{x,j+1}-e') + p_{R} \geq 2$,
      and hence $\Rhs{G}{N_x}[j] \in V$.
      \begin{enumerate}
      \item In $\bigO(1)$ time we compute $x_{R} := R_{x}[j]$. We then have $\Rhs{G}{N_x}[j] = N_{x_R}$.
        Denote $B_R = \Exp{G}{N_{x_R}}$ and $m_R = |B_R|$.
      \item Using \cref{lm:slg-nonterminal-prefix} and $x_{R}$ and $p_{R}$ and input,
        in $\bigO(h \cdot (1 + \log_{w} \ell))$ time we compute
        three values:
        \begin{itemize}
        \item $s_R = B_R[1] + \dots + B_R[p_{R}]$,
        \item $v_R = \min_{t \in [1 \dd p_R]} B_R[1] + \dots + B_R[t]$,
        \item $\delta_R = \argmin_{t \in [1 \dd p_R]} B_R[1] + \dots + B_R[t]$.
        \end{itemize}
      \item In $\bigO(1)$ time we update the values corresponding
        to the smallest prefix.
        More precisely, if $s_{\rm cur} + v_{R} < v_{\rm cur}$, then we set
        $v_{\rm cur} := s_{\rm cur} + v_{R}$ and $\delta_{\rm cur} := p_{\rm cur} + \delta_{R}$.
      \end{enumerate}
      In total, above steps take $\bigO(h \cdot (1 + \log_{w} \ell))$ time.

    \item At this point, we have $\delta_{\rm cur} = \argmin_{t \in [1 \dd e'-b']}\,(B[1] + \dots + B[t])$,
      where $B$ is defined as above, i.e., $B = \Exp{G}{N_x}(b' \dd e']$. Thus, by the above discussion, in
      $\bigO(1)$ we return that $\argmin_{t \in (b \dd e]} (A[1] + \dots + A[t]) = b + \delta_{\rm cur}$.
    \end{enumerate}
    In total, the above steps take $\bigO(h \cdot (1 + \log_{w} \ell))$ time.
  \end{enumerate}
  In total, the whole query algorithm takes $\bigO(h \cdot (1 + \log_{w} \ell))$ time.
\end{proof}

\begin{theorem}[Based on~{\cite[Section~6.3]{Gagie2020}}]\label{th:lcp-rmq-upper-bound}
  Let $\epsilon \in (0, 1)$ be a constant.
  For every nonempty set $\Text \in \Sigma^{\Textlen}$, there exists a data
  structure of size $\bigO(\RLBWTSize{\Text} \log^{2+\epsilon})$ (see \cref{def:bwt})
  that answers LCP RMQ queries on $\Text$ (that, given any $b,e \in [0 \dd \Textlen]$ satisfying
  $b < e$, return $\argmin_{i \in (b \dd e]} \LCP{\Text}[i]$; see \cref{def:lcp-array,rm:argmin}) in
  $\bigO\big(\tfrac{\log \Textlen}{\log \log \Textlen}\big)$ time.
\end{theorem}
\begin{proof}

  We use the following definitions. Let $A[1 \dd \Textlen]$ be an array
  defined such that $A[1] = \LCP{\Text}[1]$ and, for every $i \in [2 \dd \Textlen]$,
  $A[i] = \LCP{\Text}[i] - \LCP{\Text}[i-1]$.
  Let $G = (V,\Sigma,R,S)$ denote the SLG from \cref{cr:diff-lcp-slg}. Recall
  that $G$ has size $g = \bigO(\RLBWTSize{\Text} \log^{2+\epsilon} \Textlen)$,
  it has height $h = \bigO(\tfrac{\log \Textlen}{\log \log \Textlen})$,
  it expands to $A$ (i.e., $\Lang{G} = \{A\}$),
  and there exists $\ell = \bigO(\log^{\epsilon} \Textlen)$ such that, for every
  $N \in V$, it holds $|\Rhs{G}{V}| \leq \ell$.

  \DSComponents
  The data structure to answer LCP RMQ queries consists of a single component: the data
  structure from \cref{lm:min-prefix-sum-slg} for the SLG $G$. The data structure
  from \cref{lm:min-prefix-sum-slg} (and hence the entire structure for LCP RMQ queries)
  needs $\bigO(g) = \bigO(\RLBWTSize{\Text} \log^{2+\epsilon} \Textlen)$ space.

  \DSQueries
  Let $b,e \in [0 \dd \Textlen]$ be such that $b < e$. To compute
  $\argmin_{i \in (b \dd e]} \LCP{\Text}[i]$, given the indexes $b$ and $e$,
  we use \cref{lm:min-prefix-sum-slg} to compute $i_{\rm min} := \argmin_{i \in (b \dd e]} (A[1] + \dots + A[i])$
  in
  \[
    \bigO(h \cdot (1 + \log_{w} \ell)) =
    \bigO(h \cdot (1 + \tfrac{\log \ell}{\log w})) =
    \bigO(h) =
    \bigO(\tfrac{\log \Textlen}{\log \log \Textlen})
  \]
  time. We then return $i_{\rm min}$ as the answer. The correctness follows by observing that, for every
  $i \in [1 \dd \Textlen]$, $\sum_{j=1}^{i} A[j] = \LCP{\Text}[1] + (\LCP{\Text}[2] - \LCP{\Text}[1]) + \dots +
  (\LCP{\Text}[i] - \LCP{\Text}[i-1]) = \LCP{\Text}[i]$.
  Thus, $i_{\rm min} = \argmin_{i \in (b \dd e]} \LCP{\Text}[i]$.
\end{proof}

\bibliographystyle{alphaurl}
\bibliography{paper}

\end{document}